\documentclass[a4paper,11pt]{article}
\usepackage[margin=1.2in]{geometry}\usepackage[utf8]{inputenc}
\usepackage{jheppub} 
\usepackage{amsmath}
\usepackage{bbold}
\usepackage{slashed}
\usepackage{graphicx}
\usepackage{subfig}
\usepackage{booktabs}
\usepackage{epstopdf}
\usepackage{cancel}

\def\lvec#1{\setbox0=\hbox{$#1$}
    \setbox1=\hbox{$\scriptstyle\leftarrow$}
    #1\kern-\wd0\smash{
    \raise\ht0\hbox{$\raise1pt\hbox{$\scriptstyle\leftarrow$}$}}
    \kern-\wd1\kern\wd0}
\def\rvec#1{\setbox0=\hbox{$#1$}
    \setbox1=\hbox{$\scriptstyle\rightarrow$}
    #1\kern-\wd0\smash{
    \raise\ht0\hbox{$\raise1pt\hbox{$\scriptstyle\rightarrow$}$}}
    \kern-\wd1\kern\wd0}

\newcommand{\mr}{m_{\rm R}}
\newcommand{\gr}{g_{\rm R}}
\newcommand{\cT}{c_{\rm T}}
\newcommand{\ca}{c_{\rm A}}
\newcommand{\cv}{c_{\rm V}}
\newcommand{\zf}{z_f}
\newcommand{\mcr}{m_{\rm cr}}

\newcommand{\unit}{1\kern-.25em {\rm l}}
\newcommand{\Tr}{{\rm Tr}}
\newcommand{\tr}{{\rm tr}}

\newcommand{\Real}{\relax{\mathsf{\Gamma\kern-.35em R}}}
\newcommand{\Int}{\relax{\mathsf{Z\kern-.40em Z}}}
\newcommand{\half}{{\tfrac{1}{2}}}
\newcommand{\diag}{{\rm diag}}

\newcommand{\Nf}{N_{\rm f}}

\newcommand{\bfp}{{\bf p}}

\newcommand{\bfk}{{\bf k}}
\newcommand{\bfx}{{\bf x}}
\newcommand{\bfy}{{\bf y}}
\newcommand{\bfz}{{\bf z}}

\newcommand{\rmd}{{\rm d}}
\newcommand{\rme}{{\rm e}}
\newcommand{\rmO}{{\rm O}}

\newcommand{\psibar}{\bar{\psi}}

\newcommand{\zetabar}{\bar{\zeta}}
\newcommand{\zetaprime}{\zeta\kern1pt'}
\newcommand{\zetabarprime}{\zetabar\kern1pt'}

\newcommand{\msbar}{{\rm \overline{MS\kern-0.05em}\kern0.05em}}

\def\cx{c_{\rm X}}
\def\ca{c_{\rm A}}
\def\cv{c_{\rm V}}
\def\csw{c_{\rm sw}}

\def\ct{c_{\rm t}}

\def\ctildet{\tilde{c}_{\rm t}}
\def\ctt{\ctildet}

\def\Cf{C_{\rm F}}

\def\Za{Z_{\rm A}}
\def\Zv{Z_{\rm V}}
\def\Zp{Z_{\rm P}}
\def\Zs{Z_{\rm S}}
\def\Zt{Z_{\rm T}}
\def\Ztt{Z_{\rm \widetilde{T}}}
\def\Zx{Z_{\rm X}}

\def\fv{f_{\rm V}}
\def\fa{f_{\rm A}}
\def\fs{f_{\rm S}}
\def\fp{f_{\rm P}}
\def\fx{f_{\rm X}}

\def\gv{g_{\rm V}}
\def\gvt{g_{\rm \widetilde{V}}}
\def\ga{g_{\rm A}}
\def\gs{g_{\rm S}}
\def\gp{g_{\rm P}}
\def\gx{g_{\rm X}}

\def\ka{k_{\rm A}}
\def\kv{k_{\rm V}}
\def\kt{k_{\rm T}}
\def\ktt{k_{\rm \widetilde{T}}}
\def\ky{k_{\rm Y}}

\def\la{l_{\rm A}}
\def\lv{l_{\rm V}}
\def\lvt{l_{\rm \widetilde{V}}}
\def\lt{l_{\rm T}}
\def\ltt{l_{\rm \widetilde{T}}}
\def\ly{l_{\rm Y}}

\def\xSF{\chi{\rm SF}}

\def\rmd{{\rm d}}

\def\rme{{\rm e}}
\def\rmO{{\rm O}}

\def\proof{\noindent{\sl Proof:}\kern0.6em}

\def\dual{\mathstrut^*\kern-0.1em}

\def\lvec#1{\setbox0=\hbox{$#1$}
    \setbox1=\hbox{$\scriptstyle\leftarrow$}
    #1\kern-\wd0\smash{
    \raise\ht0\hbox{$\raise1pt\hbox{$\scriptstyle\leftarrow$}$}}
    \kern-\wd1\kern\wd0}
\def\rvec#1{\setbox0=\hbox{$#1$}
    \setbox1=\hbox{$\scriptstyle\rightarrow$}
    #1\kern-\wd0\smash{
    \raise\ht0\hbox{$\raise1pt\hbox{$\scriptstyle\rightarrow$}$}}
    \kern-\wd1\kern\wd0}
\def\slash#1{\setbox0=\hbox{$#1$}\setbox1=\hbox{$\kern1pt/$}
    #1\kern-\wd0\kern1pt/\kern-\wd1\kern\wd0}

\def\nabstar#1{{\nabla\kern0.5pt\smash{\raise 4.5pt\hbox{$\ast$}}
               \kern-5.5pt_{#1}}}

\def\drvstar#1{{\partial\kern0.5pt\smash{\raise 4.5pt\hbox{$\ast$}}
               \kern-6.0pt_{#1}}}

\def\ldrvstar#1{{\lvec{\,\partial}\kern-0.5pt\smash{\raise 4.5pt\hbox{$\ast$}}
               \kern-5.0pt_{#1}}}

\def\MSbar{\overline{\rm MS\kern-0.5pt}\kern0.5pt}
\def\Nf{{N_{\rm f}}}

\def\psibar{\bar{\psi}}

\def\zetabar{\bar{\zeta}}
\def\zetaprime{\zeta\kern1pt'}
\def\zetabarprime{\zetabar\kern1pt'}

\def\dirac#1{\gamma_{#1}}
\def\diracstar#1#2{
    \setbox0=\hbox{$\gamma$}\setbox1=\hbox{$\gamma_{#1}$}
    \gamma_{#1}\kern-\wd1\kern\wd0
    \smash{\raise4.5pt\hbox{$\scriptstyle#2$}}}

\def\tr{{\rm tr}}
\def\Tr{{\rm Tr}}

\def\Ds{D_{\rm s}}
\def\DsdagDs{\Ds{\Ds}^{\kern-1pt\dagger}}

\def\avg#1{{\kern1.0pt\overline{\kern-1.0pt#1\kern-1.0pt}\kern1.0pt}}

\title{\boldmath The chirally rotated Schr\"{o}dinger functional: theoretical expectations
                 and perturbative tests}

\author[a]{Mattia Dalla Brida,}
\author[b]{Stefan Sint}
\author[c]{and Pol Vilaseca}

\affiliation[a]{NIC, DESY, Platanenallee 6, 15738 Zeuthen, Germany}
\affiliation[b]{School of Mathematics, Trinity College Dublin, Dublin 2, Ireland}
\affiliation[c]{Istituto Nazionale di Fisica Nucleare, Sezione di Roma, P.le A. Moro 2,
		I-00185, Roma, Italy}

\emailAdd{mattia.dalla.brida@desy.de}
\emailAdd{sint@maths.tcd.ie}
\emailAdd{pol.vilaseca.mainar@roma1.infn.it}

\abstract{
The chirally rotated Schr\"odinger functional ($\chi$SF) with massless  Wilson-type 
fermions provides an alternative lattice regularization of the Schr\"odinger functional (SF), 
with different lattice symmetries and a common continuum limit expected from universality.
The explicit breaking of flavour and parity symmetries needs to be repaired
by tuning the bare fermion mass  and the coefficient of a
dimension 3 boundary counterterm. Once this is achieved one expects the mechanism of automatic
O($a$) improvement to be operational in the $\chi$SF, in contrast to the standard formulation
of the SF. This is expected to significantly improve the attainable precision for step-scaling
functions of some composite operators.  Furthermore, the $\chi$SF offers new strategies
to determine finite renormalization constants which are traditionally obtained from chiral Ward identities.
In this paper we consider a complete set of fermion bilinear operators, define corresponding
correlation functions and explain the relation to their standard SF counterparts.
We discuss renormalization and O($a$) improvement and then use this set-up to 
formulate the theoretical expectations which follow from universality.
Expanding the correlation functions to one-loop order of perturbation theory
we then perform a number of non-trivial checks. In the process we obtain the
action counterterm coefficients to one-loop order and reproduce some known perturbative
results for renormalization constants of fermion bilinears.
By confirming the theoretical expectations, this perturbative study lends further support to the soundness
of the $\chi$SF framework and prepares the ground for non-perturbative applications.
}

\preprint{
  \begin{flushright}
     DESY 16--040\\
     TCDMATH 16--02
  \end{flushright}
}

\begin{document} 
\maketitle
\flushbottom

\section{Introduction}

The chirally rotated Schr\"odinger functional ($\chi$SF)~\cite{Sint:2005qz,Sint:2010eh} provides a new tool to 
address renormalization and O($a$) improvement problems in lattice QCD and similar lattice
gauge theories with Wilson type fermions.
With an even number of massless fermion flavours it is formally related to the standard
Schr\"odinger functional (SF)~\cite{Luscher:1992an,Sint:1993un,Sint:1995rb} by a non-singlet chiral field rotation.
Such chirally rotated SF boundary conditions have first appeared with staggered
fermions~\cite{Miyazaki:1994nu} where the chiral rotation can be absorbed in the reconstruction of four-spinors
from the one-component staggered fermion field~\cite{Heller:1997pn,PerezRubio:2008yd}.
Similarly, with Ginsparg-Wilson or domain-wall fermions such boundary
conditions~\cite{Taniguchi:2004gf,Taniguchi:2006qw} can be
re-interpreted as standard SF boundary conditions, based on exact Ginsparg-Wilson-type
lattice symmetries~\cite{Sint:2007zz}.
With Wilson fermions the chiral field rotation does not correspond to a lattice symmetry, and
the $\chi$SF can thus be seen as an alternative lattice regularization of the SF.
The $\chi$SF has practical advantages when applied to non-perturbative renormalization problems.
In particular, the expected property of automatic O($a$) improvement~\cite{Frezzotti:2003ni,Sint:2010eh} 
will potentially be very helpful in reducing systematic errors in continuum extrapolations of step-scaling functions.
The theoretical framework for the $\chi$SF has been defined in ref.~\cite{Sint:2010eh} where also some
perturbative tests have been performed at tree-level. Here we would like to define
the framework for systematic tests and applications of the $\chi$SF. In particular we define
boundary-to-boundary correlation functions, as well as 
boundary-to-bulk correlation functions for a complete set of non-singlet fermion bilinear operators.
We then establish a dictionary translating them to their SF counterparts. This is reminiscent of 
twisted mass QCD~\cite{Frezzotti:2000nk}, except that we will here exclusively focus 
on the massless theory. From universality one then expects that
the same dictionary holds in terms of renormalized correlation functions up to cutoff effects.
We formulate various consequences of this expectation such as flavour and parity symmetry restoration,
the possibility to determine finite renormalization constants (otherwise obtainable by chiral Ward identities),
and scale dependent renormalization constants in SF schemes, together with their step-scaling functions.
We then use one-loop perturbation theory to perform non-trivial tests of these expectations.
Some elements of the set-up together with tests in quenched QCD have already appeared in ref.~\cite{Sint:2010xy},
and preliminary one-loop results have been given in refs.~\cite{Sint:2011gv,Sint:2012ae,Sint:2014wwa}.
For related non-perturbative applications of the $\chi$SF to quenched lattice QCD cf.~refs.~\cite{Lopez:2012as,Lopez:2012mc}.
Preliminary results for two-flavour lattice QCD can be found in~ref.~\cite{Brida:2014zwa}. 

The paper is organized as follows: in Section~2 we use a continuum language to discuss the connection
between the SF and the $\chi$SF and the respective correlation functions of interest.
The transcription to the lattice regularization is described in Section~3,
followed by a discussion of renormalization and Symanzik O($a$) improvement,
for both the standard SF and the $\chi$SF.
In Section~4 we summarize the theoretical expectations for the $\chi$SF.
The remainder of this paper discusses the perturbative expansion and one-loop
results for the action parameters (Section 5) and various ways to test and apply
the theoretical expectations in perturbation theory (Sects.~6 and 7). Section~8 
contains a discussion of a gluonic observable, the SF coupling, to one-loop order.
Conclusions are drawn in Section 9, and 3 appendices collect some definitions regarding
fermion bilinear fields (Appendix A), a few details on the calculation of the fermionic contribution
to the SF coupling at one-loop order (Appendix B) and  a comparison at weak coupling between
perturbation theory and Monte-Carlo simulations (Appendix C).

\section{Correlation functions and universality relations}
\label{sec:XSF}

In this section we recall the formal continuum relations between SF and $\chi$SF correlation
functions, which are obtained by a change of variables in the functional integral of
the formal continuum theory. We then establish a dictionary between specific SF and $\chi$SF correlation
functions of non-singlet fermion bilinear operators. On the lattice with Wilson type fermions
the chiral symmetry relating the SF and the $\chi$SF is broken explicitly.
Therefore the relation between both formulations is expected to assume the simple form of the continuum dictionary
only after appropriate renormalization and up to cutoff effects.

\subsection{Chiral rotations and correlation functions}

The continuum action for $\Nf$ massless fermions in an external gauge field\footnote{With fermions
in the fundamental representation of the gauge group SU($N$) we have $A_\mu = A_\mu^a T^a$,
where $T^a$ are the anti-hermitian generators in the fundamental representation,
a sum over $a=1,\ldots,N^2-1$ is implied, and we normalize the generators by $\tr(T^aT^b) = -\half \delta^{ab}$.
Generalizations to other representations are straightforward and do not affect the discussion of chiral and flavour symmetries.}
$A_\mu(x)$,
\begin{equation}
   S_f = \int \rmd^4 x\,\, \psibar(x) \gamma_\mu D_\mu \psi(x),\qquad D_\mu = \partial_\mu+ A_\mu,
\end{equation}
has exact flavour and chiral symmetries. The latter are
broken if one imposes the standard SF boundary conditions on the fermionic fields,
\begin{xalignat}{2}
     P_+\psi(x)\vert_{x_0=0} &=0,    & P_-\psi(x)\vert_{x_0=T} &=0,\nonumber\\
    \psibar(x)P_-\vert_{x_0=0} &= 0, & \psibar(x)P_+\vert_{x_0=T} &= 0,
\label{eq:stdbc}
\end{xalignat}
with the projectors $P_\pm=\half(1\pm\gamma_0)$.
Indeed, assuming $\Nf=2$ flavours, a chiral non-singlet transformation,
\begin{equation}
    \psi = R(\pi/2)\psi',\qquad \psibar = \psibar' R(\pi/2),
 \label{eq:chiral_rot}
\end{equation}
with  $R(\alpha)=\exp(i\alpha \gamma_5\tau^3/2)$, transforms Eqs.~(\ref{eq:stdbc}) to
\begin{xalignat}{2}
     \tilde{Q}_+\psi'(x)\vert_{x_0=0} &=0,
    &\tilde{Q}_-\psi'(x)\vert_{x_0=T} &=0,\nonumber\\
   \psibar'(x)\tilde{Q}_+\vert_{x_0=0} &= 0,
    &\psibar'(x)\tilde{Q}_-\vert_{x_0=T} &= 0\,,
\label{eq:rotbc}
\end{xalignat}
where
\begin{equation}
   \tilde{Q}_\pm = \half(1\pm i\gamma_0\gamma_5\tau^3),
\label{eq:Qtilde}
\end{equation}
and the Pauli matrix $\tau^3$ acts on the flavour indices.
If the field transformation is performed as a change of variables in the functional integral
one obtains relations between standard SF and $\chi$SF correlation functions,
\begin{eqnarray}
 \left\langle O[\psi,\psibar]\right\rangle_{(\tilde{Q}_{+})}&=&
 \left\langle O[R(-\pi/2)\psi,\psibar R(-\pi/2)]\right\rangle_{(P_+)},
 \label{eq:univ1}\\
 \left\langle O[\psi,\psibar]\right\rangle_{(P_{+})}&=&
 \left\langle O[R(\pi/2)\psi,\psibar R(\pi/2)]\right\rangle_{(\tilde{Q}_+)}.
 \label{eq:univ2}
\end{eqnarray}
Here, the subscript to the correlation function indicates the projector defining the Dirichlet component
of the fermion field at $x_0=0$. This notation unambiguously specifies the boundary conditions
for the integration variables in the functional integral, which we will always denote by $\psi$ and $\psibar$,
thereby removing the prime from the fields in the $\chi$SF.
Note also that the composite fields $O[\psi,\psibar]$ inside correlation functions
may include the boundary fermion fields,
\begin{xalignat}{2}
  \zeta(\bfx)         &=  \psi(0_+,\bfx),
  &\zetabar(\bfx)     &=  \psibar(0_+,\bfx),
  \label{eq:zetazetabar}\\
  \zetaprime(\bfx)    &=  \psi(T_-,\bfx),
&\zetabarprime(\bfx)  &=  \psibar(T_-,\bfx).
  \label{eq:zetazetabarprime}
\end{xalignat}
The time arguments $0_+$ or $T_-$ indicate that the fields are located infinitesimally
away from the boundaries at $x_0=0,T$. For later convenience we have omitted the projectors
$P_\pm$ or $\tilde{Q}_\pm$, in contrast to conventions used
in the literature~\cite{Luscher:1996sc,Sint:2010eh}. Instead we include
these projectors explicitly when defining the bilinear boundary source fields
for SF and $\chi$SF correlation functions (cf.~Subsections~\ref{subsect:SFcorr},\ref{subsect:chiSFcorr}).

\subsection{Flavour structure and symmetries}
\label{subsect:flavsym}

While the standard SF can be formulated for any number of flavours,
this is not straightforward for the $\chi$SF~\cite{Sint:2010eh}.
We will restrict attention to gauge theories with an even number of fermion flavours.
So far we have assumed $\Nf=2$, i.e.~a doublet structure,
\begin{equation}
   \psi = \begin{pmatrix}  \psi_u \\ \psi_d \end{pmatrix},
\end{equation}
with up and down type flavours. For the correlation functions defined below it will be
convenient to introduce more than a single up or down type flavour, such that flavour
non-singlet fermion bilinear fields can be formed with only up- or only down-type fermions.
We are thus led to consider the case $\Nf=4$ which we obtain by replicating the doublet structure,
\begin{equation}
  \psi = \begin{pmatrix}  \psi_u \\ \psi_d \\ \psi_{u'} \\ \psi_{d'} \end{pmatrix},
\end{equation}
i.e.~there are two up and two down type flavours. Obviously this implies that
the flavour matrix $\tau^3$ in Eqs.~(\ref{eq:chiral_rot}),(\ref{eq:Qtilde}) should be replaced by
\begin{equation}
   \tau^3 \rightarrow \unit_2 \otimes \tau^3 = \diag(1,-1,1,-1).
\end{equation}
It is often convenient to reduce the flavour structure of the projectors,
\begin{equation}
   \left.\tilde{Q}\right\vert_{\Nf=2} = \diag(Q_+,Q_-), \qquad
   \left.\tilde{Q}\right\vert_{\Nf=4} = \diag(Q_+,Q_-,Q_+,Q_-),
\end{equation}
with
\begin{equation}
   Q_\pm = \half(1\pm i\gamma_0\gamma_5).
\end{equation}
Although the $\chi$SF boundary conditions differ for up and down type flavours,
this does not mean that the SU($\Nf$) flavour
symmetry is broken. In fact, as discussed in ref.~\cite{Sint:2010eh},
the distinction between flavour and chiral symmetries in the absence of mass terms is conventional.
We here follow the convention used in ref.~\cite{Sint:2010eh} and define the flavour symmetry such
that the corresponding field transformations take their usual form in the standard SF basis.
In this basis, a flavour transformation for $\Nf=2$ flavours with
parameters $\omega^a$ ($a=1,2,3$), looks as usual,
\begin{equation}
    \psi \rightarrow \exp\left(i\sum_{a=1}^3 \omega^a \tau^a\right) \psi,\qquad
    \psibar \rightarrow  \psibar \exp\left(-i\sum_{a=1}^3\omega^a \tau^a\right).
\end{equation}
As the SF and $\chi$SF fields are related by the chiral rotation (\ref{eq:chiral_rot})
the same flavour symmetry transformation on the $\chi$SF fields takes the form
\begin{eqnarray}
    \psi' &\rightarrow&  R(-\pi/2) \exp\left(i\sum_{a=1}^3\omega^a \tau^a\right) R(\pi/2) \psi',\\
 \psibar' &\rightarrow&  \psibar'\, R(\pi/2)\exp\left(-i\sum_{a=1}^3\omega^a \tau^a\right) R(-\pi/2).
\end{eqnarray}
In particular, in the continuum the $\chi$SF shares all the symmetries with the standard SF,
i.e.~the full flavour symmetry, charge conjugation, spatial rotations and parity.
Of particular interest is the parity symmetry, which in the SF basis is realized by
\begin{equation}
   P: \begin{cases} \psi(x) \rightarrow \gamma_0\psi(\tilde x),\\
                    \psibar(x) \rightarrow \psibar(\tilde x)\gamma_0,
      \end{cases}           \qquad \tilde x = (x_0,-\bfx),
  \label{eq:P}
\end{equation}
whereas its covariantly rotated $\chi$SF version reads,
\begin{equation}
   P_5: \begin{cases} \psi'(x) \rightarrow i\gamma_0\gamma_5\tau^3\psi'(\tilde x),\\
                    \psibar'(x) \rightarrow -\psibar'(\tilde x) i\gamma_0\gamma_5\tau^3,
      \end{cases}            \qquad \tilde x = (x_0,-\bfx).
   \label{eq:P5}
\end{equation}
The $P_5$-symmetry plays an important r\^ole in the following, as it may be
used to classify lattice correlation functions and their approach to the continuum limit.
More precisely, in the lattice regularized $\chi$SF the $P_5$-even
correlation functions are automatically O($a$) improved in the bulk,
whereas their $P_5$-odd counterparts are pure lattice artefacts.
Hence, $P_5$ may be taken as a substitute for the $\gamma_5\tau^1$ symmetry
used in ref.~\cite{Sint:2010eh},
\begin{equation}
   \psi' \rightarrow \gamma_5\tau^1\psi',\qquad
   \psibar' \rightarrow -\psibar'\gamma_5\tau^1,
\end{equation}
which corresponds to a discrete flavour symmetry. The advantage of $P_5$
is that it is flavour diagonal and therefore more suitable
for $\chi$SF correlation functions with specific flavour assignments.

\subsection{SF correlation functions}

\label{subsect:SFcorr}
The SF correlation functions required for this work have previously
appeared in the literature, e.g.~in refs.~\cite{Luscher:1996sc,Sint:1997jx}.
When written in terms of fixed flavours, $f_1,f_2 \in \{u,d,u',d'\}$, with $f_1\ne f_2$,
they take the form
\begin{equation}
\fx(x_{0})=-{1\over2}\langle X^{f_{1}f_{2}}(x)\mathcal{O}_{5}^{f_{2}f_{1}}\rangle_{(P_+)},\qquad
\ky(x_{0})=-{1\over6}\sum_{k=1}^{3}\langle Y_{k}^{f_{1}f_{2}}(x)\mathcal{O}_{k}^{f_{2}f_{1}}\rangle_{(P_+)}.
\label{eq:bound_bulk}
\end{equation}
In the literature, the composite fields $X$ and $Y_{k}$ stand for
the fermion bilinears\footnote{cf.~Appendix~\ref{app:bilinear}
for our definitions and conventions.}
$X=A_{0},P$ and $Y_{k}=V_{k},T_{k0}$. Here we also include
$X=V_0,S$ and $Y_k=A_k,\widetilde{T}_{k0}$.
While these additional correlation functions are odd under parity 
(\ref{eq:P}) and thus vanish exactly, we will need them for the dictionary with
their $\chi$SF counterparts defined below.
Finally, the fermion bilinear source fields at the lower time boundary are defined by
\begin{align}
  {\cal O}_{5}^{f_1f_2} &=
  \int\rmd^3\bfy \rmd^3\bfz\,\, \zetabar_{f_1}(\bfy) P_+ \gamma_5 \zeta_{f_2}({\bfz}),
  \label{eq:O5}\\
   {\cal O}_{k}^{f_1 f_2} &= \int\rmd^3\bfy\rmd^3\bfz\,\,\zetabar_{f_1}(\bfy)P_+\gamma_k
   \zeta_{f_2}({\bfz}).
  \label{eq:Ok}
\end{align}
Note that the projector $P_+$ must be written explicitly as we did not include it
in the definition of the fermionic boundary fields $\zeta$ and $\zetabar$, Eq.~(\ref{eq:zetazetabar}).
Integrating over the fermion fields in the functional integral one obtains, for example,
\begin{equation}
   \fa(x_0) = \frac12\int\rmd^3\bfy \rmd^3\bfz
   \left\langle \tr\left\{S(x;0,\bfy)P_+\gamma_5 P_- S(0,\bfz;x)\gamma_0\gamma_5\right\}\right\rangle_G,
 \label{eq:facont}
\end{equation}
where $\langle \cdots\rangle_G$ denotes the gauge field average,
$S(x,y)$ the propagator for a single fermion flavour, and the trace is to be
taken over colour and Dirac indices. The SF boundary conditions in terms of the fermion propagator,
\begin{equation}
   P_+ S(x,y)\vert_{x_0=0} = 0= S(x,y)P_-\vert_{y_0=0},
\end{equation}
now imply that the correlation function vanishes if the projector in Eq.~(\ref{eq:O5})
is reverted, $P_+\rightarrow P_-$. In the lattice regularized theory this only
holds after taking the continuum limit and may thus be used as a check.
Finally, we also need the boundary-to-boundary correlators,
\begin{equation}
   f_{1}= - \frac12 \left\langle {\cal O}_{5}^{f_1f_2} {\cal O}_{5}^{'f_2f_1}\right\rangle_{(P_+)},\qquad
   k_{1}= - \frac16\sum_{k=1}^3\left\langle {\cal O}_{k}^{f_1f_2} {\cal O}_{k}^{'f_2f_1}\right\rangle_{(P_+)}\,,
\end{equation}
where the fermion bilinear source fields at the upper time boundary are defined by
\begin{eqnarray}
  {\cal O}_{5}^{'f_1f_2} &=& \int\rmd^3\bfy\rmd^3\bfz\,\,
  \zetabarprime_{f_1}(\bfy)P_- \gamma_5 \zetaprime_{f_2}({\bfz}),
\label{eq:Op5}\\
  {\cal O}_{k}^{'f_1 f_2} &=& \int\rmd^3\bfy\rmd^3\bfz \,\,
  \zetabarprime_{f_1}(\bfy)P_- \gamma_k \zetaprime_{f_2}({\bfz}).
\label{eq:Opk}
\end{eqnarray}

\subsection{$\chi$SF correlation functions}

\label{subsect:chiSFcorr}
To obtain correlation functions in the $\chi$SF we apply the
identities~(\ref{eq:univ1}),(\ref{eq:univ2}) to the standard SF correlation functions.
First we define the bilinear source fields ${\cal Q}_{5}^{f_1f_2}$ and ${\cal Q}_{k}^{f_1f_2}$ such that
they rotate into the standard SF sources (\ref{eq:O5}),(\ref{eq:Ok}), i.e.
\begin{equation}
  \langle O[R(\pi/2)\psi,\psibar R(\pi/2)] {\cal Q}_{5,k}^{f_1f_2}\rangle_{(\tilde{Q}_+)} =
  \langle O[\psi,\psibar] {\cal O}_{5,k}^{f_1f_2}\rangle_{(P_+)},
  \label{eq:chirot}
\end{equation}
and the same for the primed source fields at the upper time boundary. In this way one obtains,
for example,
\begin{eqnarray}
  {\cal Q}_{5}^{uu'} &=& \int \rmd^3\bfy \rmd^3\bfz\,\,
  \zetabar_u(\bfy)\gamma_0\gamma_5 Q_-\zeta_{u'}({\bfz}), 
\label{eq:Q5-uu}\\
  {\cal Q}_{5}^{du} &=& \int\rmd^3\bfy \rmd^3\bfz\,\,
  \zetabar_d(\bfy)\gamma_5 Q_-\zeta_{u}({\bfz}), 
  \label{eq:Q5-ud}
\end{eqnarray}
and the complete set of source fields can be found in Appendix~\ref{app:bilinear}.

We now define the correlation functions for fermion bilinears $X=V_0,A_0,S,P$, by
\begin{equation}
  \gx^{f_1f_2}(x_0) = -\frac12\left\langle X^{f_1f_2}(x){\cal Q}_{5}^{f_2f_1}\right\rangle_{(\tilde{Q}_+)},
\end{equation}
where we label the correlation functions by the flavour indices of the fermion bilinear operator in the bulk.
It is then straightforward to work out the relations~(\ref{eq:univ1}),(\ref{eq:univ2})
for these particular correlation functions:
 \begin{alignat}{4}
 \label{eq:dictfa}
 \fa\,  &=&\,  \ga^{uu'} &=&\,   \ga^{dd'} &=&\, -i\gv^{ud} &=\, i\gv^{du},\\
 \fp\,  &=&\, i\gs^{uu'} &=&\, -i\gs^{dd'} &=&\,   \gp^{ud} &=\,\phantom{i}\gp^{du},\\
 \fv\,  &=&\,  \gv^{uu'} &=&\,   \gv^{dd'} &=&\, -i\ga^{ud} &=\, i\ga^{du},
  \label{eq:dictfv}\\
 \fs\,  &=&\, i\gp^{uu'} &=&\, -i\gp^{dd'} &=&\,   \gs^{ud} &=\,\phantom{i}\gs^{du}.
  \label{eq:dictfs}
 \end{alignat}
Hence, by using the chirally covariant definition of the
boundary source fields, Eqs.~(\ref{eq:Q5-uu}) and (\ref{eq:Q5-ud}),
the properties of the correlation functions $\gx$ under chiral rotations are the same as for
the inserted fermion bilinear operators.

Proceeding similarly for the source fields with an open spatial vector index, Eq.~(\ref{eq:Ok}),
the correlation functions of the bilinear fields
$Y_k=A_k,V_k,T_{k0},\widetilde{T}_{k0}$ are defined by
\begin{equation}
  \ly^{f_1f_2}(x_0) = -\frac16\sum_{k=1}^3\left\langle Y_k^{f_1f_2}(x){\cal Q}_{k}^{f_2f_1}\right\rangle_{(\tilde{Q}_+)},
\end{equation}
and their relations to the standard SF correlation functions are found to be,
 \begin{alignat}{4}
  \kv  &=&\,  \lv^{uu'} &=&\,   \lv^{dd'} &=&\, -i\la^{ud} &= \, i\la^{du},\\
  \ka  &=&\,  \la^{uu'} &=&\,   \la^{dd'} &=&\, -i\lv^{ud} &= \, i\lv^{du},
   \label{eq:dictka}\\
  \kt  &=&\,i\ltt^{uu'} &=&\,-i\ltt^{dd'} &=&\,   \lt^{ud} &= \,\phantom{i} \lt^{du},\\
  \ktt &=&\, i\lt^{uu'} &=&\, -i\lt^{dd'} &=&\,  \ltt^{ud} &= \,\phantom{i} \ltt^{du}\,.
   \label{eq:dictktt}
 \end{alignat}
Finally, boundary-to-boundary correlators are defined by
\begin{eqnarray}
 g_{1}^{f_1f_2} &=&
 - \frac12 \left\langle {\cal Q}_{5}^{f_1f_2} {\cal Q}_{5}^{'f_2f_1}\right\rangle_{(\tilde{Q}_+)},\\
 l_{1}^{f_1f_2} &=& - \frac16\sum_{k=1}^3
           \left\langle {\cal Q}_{k}^{f_1f_2} {\cal Q}_{k}^{'f_2f_1}\right\rangle_{(\tilde{Q}_+)}.
\end{eqnarray}
Again, the primed sources at the upper time boundary are chirally
mapped to their standard SF counterparts, leading to rather simple entries for our
dictionary,
\begin{alignat}{4}
  f_1  &=&\,  g_1^{uu'} &=&\, g_1^{dd'} &=&\,  g_1^{ud} &=\, g_1^{du}\,,
  \label{eq:dictf1}\\
  k_1  &=&\,  l_1^{uu'} &=&\, l_1^{dd'} &=&\,  l_1^{ud} &=\, l_1^{du}.
  \label{eq:dictk1}
\end{alignat}
Note that, in the continuum, there are only 6 independent non-zero correlation functions,
namely $\fa,\fp,f_1$ and $\kv,\kt,k_1$ and the corresponding $\chi$SF correlation functions can be looked up in
the dictionary. As the standard SF correlation functions are real-valued, their $\chi$SF
counterparts must be either real or purely imaginary. While this dictionary is trivial in the formal continuum theory,
it does however lead to non-trivial consequences once the lattice regularization with Wilson-type fermions is in place,
due to the additional symmetry breaking by the Wilson term.

\section{Lattice set-up, renormalization and O($a$) improvement}
\label{sec:ren_XSF}

The lattice formulation of the standard Schr\"odinger functional on a lattice of spacing $a$ and
size $(T/a) \times (L/a)^3 $  is taken over from ref.~\cite{Luscher:1996sc}. 
The chirally rotated Schr\"odinger functional will be used in the form described 
in ref.~\cite{Sint:2010eh}. We refer to these references for unexplained notation.

\subsection{Lattice actions}

The lattice action, 
\begin{equation}
  S[U,\psi,\psibar] = S_g[U] + S_f[U,\psi,\psibar],
\end{equation}
consists of a pure gauge and a fermionic part.
For the former we choose Wilson's plaquette action~\cite{Luscher:1992an}, 
\begin{equation}
   S_g[U] = \dfrac{1}{g_0^2} \sum_p w(p)\tr \{1-U(p)\},
\end{equation}
where the sum is over all oriented plaquettes $p$, 
and $U(p)$ denotes the parallel transporter around $p$,
constructed from the link variables $U_\mu(x)$. We choose
$L$-periodic boundary conditions in all the spatial directions,
\begin{equation}
   U_\mu(x+L\hat\bfk) = U_\mu(x), \qquad k=1,2,3,
\end{equation}
where $\hat\bfk$ denotes a unit vector in direction $k$.
In the Euclidean time direction we choose homogeneous boundary conditions for the spatial 
gauge potential at $x_0=0,T$, i.e.~the spatial link variables at
the boundaries are set to unit matrices,
\begin{equation}
   U_k(0,\bfx)= \unit=U_k(T,\bfx),\quad k=1,2,3\,.
\label{eq:bclink}
\end{equation}
With these boundary conditions, the weight factors $w(p)$
take the values
\begin{equation}
   w(p) = \begin{cases}  \ct(g_0) & \text{if $p$ is a time like plaquette attached to a boundary plane,} \\
                          1       & \text{otherwise.}
          \end{cases}
\end{equation}
Here $\ct$ is an O($a$) boundary counterterm coefficient. Near the continuum limit it
is seen to multiply the dimension 4 operator $\tr\{F_{0k}F_{0k}\}$, where $F_{\mu\nu}$ denotes the gluonic field
strength tensor. Disregarding fermion fields, this operator is the only non-vanishing boundary counterterm 
at order $a$ given our choice of boundary conditions. 
Hence, all O($a$) effects in the pure gauge theory can be cancelled by choosing $\ct(g_0)$ appropriately.

The fermionic fields $\psi$ and $\psibar$ are taken to be $L$-periodic in space,
\begin{equation}
   \psi(x+L\hat\bfk) = \psi(x), \qquad \psibar(x+L\hat\bfk) = \psibar(x), \qquad k=1,2,3.
\end{equation}
Apart from the SU($N$) gauge field, the fermions are coupled to 
a constant U(1) background field $\lambda_\mu = \exp(ia\theta_\mu/L)$,
so that the covariant forward and backward derivatives are given by
\begin{eqnarray}
 \nabla_\mu^{}\psi(x) &=& 
    \frac{1}{a}\big[\lambda_\mu U_\mu(x)\psi(x+a\hat\mu)-\psi(x)\big], \\
 \nabla_\mu^\ast\psi(x)&=& 
    \frac{1}{a}\big[\psi(x)-\lambda_\mu^{-1} U_\mu(x-a\hat\mu)^{\dagger}\psi(x-a\hat\mu)\big].
\end{eqnarray}
We will always assume $\theta_0=0$ and $\theta_k=\theta$ ($k=1,2,3$), leaving $\theta$ as a single
parameter. On a lattice with infinite Euclidean time extent the Wilson-Dirac operator 
can be written as a finite difference operator in time,
\begin{equation}
   aD_W\psi(x) = -U_0(x)P_-\psi(x+a\hat{\bf 0}) + K\psi(x) - U_0(x-a\hat{\bf 0})^\dagger P_+ \psi(x-a\hat{\bf 0}),
\end{equation}
with the time diagonal operator $K$,
\begin{eqnarray}
 K\psi(x) &=& \left(1+ \frac12\sum_{k=1}^3\left\{ a(\nabla^{}_k+\nabla^\ast_k)\gamma_k
   -a^2\nabla^\ast_k\nabla^{}_k\right\}\right)\psi(x) \nonumber\\
 &&\mbox{} +\csw \frac{i}{4} a^2 \sum_{\mu,\nu=0}^3\sigma_{\mu\nu}\hat{F}_{\mu\nu}(x)\psi(x).
\end{eqnarray}
Here, the last term is the Sheikholeslami-Wohlert (SW) term~\cite{Sheikholeslami:1985ij}
in the notation of ref.~\cite{Luscher:1996sc}.
Using a continuum-like normalisation, the fermionic action for either the standard SF 
or the $\chi$SF takes the form,
\begin{equation}
   S_f[U,\psi,\psibar] = a^4\sum_x \psibar(x) \left({\cal D}_W + \delta {\cal D}_W + m_0\right) \psi(x)\,,
   \label{eq:Sferm}
\end{equation}
where ${\cal D}_W$ is the reduction of the Wilson-Dirac operator to the finite time interval between $x_0=0$
and $x_0=T$, which incorporates the respective boundary conditions, 
and $\delta {\cal D}_W$ arises due to the fermionic boundary counterterms.

In the case of the $\chi$SF, three different versions have been 
proposed in ref.~\cite{Sint:2010eh} and we here choose
\begin{equation}
 a{\cal D}_W\psi(x)= \begin{cases}        
         -U_0(x)P_-\psi(x+a\hat{\bf 0}) + (K\vert_{\csw=0}+i\gamma_5\tau^3P_-)\psi(x)
                                      &    \text{if $x_0=0$,}\\
                     aD_W\psi(x) & \text{if $0<x_0<T$,}\\
        (K\vert_{\csw=0}+i\gamma_5\tau^3P_+)\psi(x)- U_0(x-a\hat{\bf 0})^\dagger P_+ \psi(x-a\hat{\bf 0})
                                       &    \text{for $x_0=T$.}\\
                      \end{cases}
\label{eq:wilson_dirac_chiSF}
\end{equation}
Note that the dynamical field variables here include the fermion fields at
Euclidean times $x_0=0$ and $x_0=T$, i.e.~the sum over $x_0$ in Eq.~(\ref{eq:Sferm})
runs from $0$ to $T$. If the Sheikholeslami-Wohlert term is included we set it to zero at the boundaries,
even though the orbifold construction yields a different prescription~\cite{Sint:2010eh}. 
The difference in the action is of O($a^2$) and thus irrelevant.
The boundary counterterms for the $\chi$SF are included by setting
\begin{eqnarray}
 \delta {\cal D}_W \psi (x)  &=& \left(\delta_{x_0,0}+\delta_{x_0,T}\right)
\Bigl[\left(z_f-1\right)+ \left(d_s-1\right) a{\bf D}_s\Bigr]\psi(x),
\label{eq:deltaDW}\\
 {\bf D}_s &=& \frac12\sum_{k=1}^3\left\{ (\nabla^{}_k+\nabla^\ast_k)\gamma_k 
  -a\nabla^\ast_k\nabla^{}_k\right\},
 \label{eq:bfDs}
\end{eqnarray}
and the values for the two coefficients, $z_f$ and $d_s$ will be specified in Sect.~4. Note that
this definition of ${\bf D}_s$ differs from~\cite{Sint:2010eh} in that it also 
includes a second order derivative term\footnote{The motivation
is of purely technical origin as it led to a more transparent implementation 
of the counterterm in the Monte Carlo simulation programs.}.

The Wilson-Dirac operator for the standard SF in the same notation reads
\begin{equation}
 a{\cal D}_W\psi(x)= \begin{cases}        
         -U_0(x)P_-\psi(x+a\hat{\bf 0}) + K\psi(x)
                                      &    \text{if $x_0=a$\,,}\\
                     aD_W\psi(x) & \text{if $a<x_0<T-a$\,,}\\
        K\psi(x)- U_0(x-a\hat{\bf 0})^\dagger P_+ \psi(x-a\hat{\bf 0})
                                       &    \text{for $x_0=T-a$\,.}\\
                      \end{cases}
\label{eq:wilson_dirac}
\end{equation}
In contrast to our chosen set-up for the $\chi$SF the dynamical fermionic field variables
in the standard SF are restricted to Euclidean times $0<x_0<T$,
i.e. the sum over $x_0$ in Eq.~(\ref{eq:Sferm}) runs only from $a$ to $T-a$.
Finally, in the standard SF, the counterterm contribution is given by
\begin{equation}
  a\delta {\cal D}_W \psi (x)  = (\ctt-1)\left(\delta_{x_0,a}+\delta_{x_0,T-a}\right)\psi(x).
\label{eq:deltaDWSF}
\end{equation}

\subsection{Lattice correlation functions}

The correlation functions introduced in Sect.~2 can now easily be transcribed to
the lattice. One essentially needs to specify the boundary quark fields $\zeta$ and $\zetabar$
at time $x_0=0$ and $\zetaprime$ and $\zetabarprime$ at time $x_0=T$. 
As before we leave out the projectors here and the notation is
therefore the same for both the SF and the $\chi$SF, i.e.~in expectation values
one performs the replacement,
\begin{xalignat}{2}
 \zeta_f(\bfx)    &=  U_0(0,\bfx)\psi_f(a,\bfx),
 &\zetaprime_f(\bfx)    &=  U_0(T-a,\bfx)^\dagger\psi_f(T-a,\bfx),
 \label{eq:zeta}\\
 \zetabar_f(\bfx) &= \psibar_f(a,\bfx)U_0(0,\bfx)^\dagger,
 &\zetabarprime_f(\bfx) &= \psibar_f(T-a,\bfx) U_0(T-a,\bfx).
\label{eq:zetabar}
 \end{xalignat}
Note that this correspondence is incomplete if the Wick contractions include
two-point functions with source and sink at the same boundary~\cite{Luscher:1996sc,Sint:2010eh}.
Here we avoid this problem by our choice of flavour assignments in the
correlation functions of Sect.~2. Moreover, in the case of the $\chi$SF we have left out the O($a$)
counterterm proportional to $\bar{d}_s$~\cite{Sint:2010eh}, which can be included
by the replacement,
\begin{equation}
   \zeta_f(\bfx)  \rightarrow \left(1 + \bar{d}_s a{\bf D}_s\right)\zeta_f(\bfx),
   \label{eq:dsbar}
\end{equation}
and similarly for $\zetabar_f$ and $\zetaprime_f,\zetabarprime_f$.
As will be further explained in Section~4, these O($a$) counterterms produce
$P_5$-odd contributions to $P_5$-even observables affecting the latter only at O($a^2$).

With these conventions the fermion-bilinear boundary sources are obtained
from their continuum counterparts by replacing the integrals over space by lattice sums\footnote{In the standard SF
the rescaling by $\ctt$ combines with the $\ctt$-contribution to the Wilson-Dirac operator in Eq.~(\ref{eq:deltaDWSF}) to
form the O($a$) counterterm containing the time derivative~\cite{Luscher:1996sc}. Whether or
not the coefficient appears explicitly or is included in the definition of the fermion boundary fields
depends on the precise definition of the latter.}, e.g.
\begin{equation}
  {\cal O}_{5}^{f_1f_2} = a^6 \ctt^2\sum_{\bfy,\bfz}
  \zetabar_{f_1}(\bfy)P_+ \gamma_5 \zeta_{f_2}({\bfz}),\qquad
   {\cal Q}_{5}^{uu'} =  a^6 \sum_{\bfy,\bfz}
  \zetabar_u(\bfy)\gamma_0\gamma_5 Q_- \zeta_{u'}({\bfz}),
\end{equation}
and analogously for all other boundary source fields (cf.~Appendix~\ref{app:bilinear}).

Finally we mention that one may restrict attention to the flavour combinations
$ud$ and $uu'$ for all correlation functions, without loss of information.
This is due to an exact lattice symmetry, namely $P$-parity combined with up/down flavour exchange,
which may be used to show that
\begin{equation}
 \gx^{du} = \pm \gx^{ud},\qquad \gx^{dd'} = \pm \gx^{uu'},
\end{equation}
and analogously for $\ly$ and the boundary-to-boundary correlation functions.
Furthermore, combining this with charge conjugation,
some $\chi$SF correlation functions can be shown to vanish identically, namely
\begin{equation}
  \gs^{ud} = \gv^{uu'} = 0=\la^{uu'} = \ltt^{ud},
  \label{eq:exact0}
\end{equation}
in addition to the SF correlation functions $\fv,\fs$ and $\ka,\ktt$.

\subsection{Renormalization}

Renormalization requires the introduction of renormalized parameters,
\begin{equation}
   \gr^2 = Z_g(g_0^2,a\mu) g_0^2, \qquad \mr = Z_m(g_0^2,a\mu) \left(m_0 -\mcr(g_0^2)\right),
\end{equation}
and renormalized composite fields,
\begin{equation}
   [X^{f_1f_2}]_{\rm R} = Z_{\rm X}(g_0^2,a\mu) X^{f_1f_2},
\end{equation}
where $\mu$ denotes the renormalization scale and $X=A_\mu,V_\mu,P,S,T_{\mu\nu},\widetilde{T}_{\mu\nu}$.
In addition the boundary fermion fields $\zeta,\zetabar$ and $\zetaprime, \zetabarprime$
are multiplicatively renormalized by a common, scale dependent
renormalization constant, $Z_\zeta$~\cite{Luscher:1996sc,Sint:2010eh}.
This implies that renormalized SF correlation functions are of the form
\begin{equation}
   [\fx]_{\rm R}(x_0) = Z_\zeta^2 Z_{\rm X} \fx(x_0),\qquad [\ky]_{\rm R}(x_0) = Z_\zeta^2 Z_{\rm Y} \ky(x_0),
\end{equation}
and, for the boundary-to-boundary correlators,
\begin{equation}
   [f_1]_{\rm R} = Z_\zeta^4 f_1,\qquad [k_1]_{\rm R} = Z_\zeta^4 k_1\,.
\end{equation}
Provided the renormalization factors are chosen appropriately,
one expects that the continuum limit can be taken at fixed $\gr$ and $\mr$.
In this work we focus on the massless limit $\mr=0$, which implies that the bare mass, $m_0$,
is tuned to its critical value, $\mcr$. As usual, this can be
achieved by tuning to the point in parameter space where the non-singlet
axial current is conserved (see e.g. ref.~\cite{Luscher:1996sc}). In terms of the SF correlation
function one requires
\begin{equation}
   \tilde{\partial}_0 [\fa]_{\rm R}(x_0) = 0 \quad \Leftrightarrow \quad \tilde{\partial}_0 \fa(x_0) = 0\,,
   \label{eq:dA}
\end{equation}
for a chosen set of kinematical parameters $x_0$, $T/L$ and $\theta$. Note that the chiral limit
is special in that the renormalization constant of the axial current drops out in Eq.~(\ref{eq:dA}).

The renormalization of the $\chi$SF correlation functions is almost completely analogous, i.e.~one
defines renormalized $\chi$SF correlation functions,
\begin{align}
   \bigl[\gx^{f_1f_2}\bigr]_{\rm R}(x_0) &= Z_\zeta^2 Z_{\rm X} \gx^{f_1f_2}(x_0),
       &  \bigl[g_1^{f_1f_2}\bigr]_{\rm R} &= Z_\zeta^4 g_1^{f_1f_2}, \label{eq:ren_g}\\
  \bigl[\ly^{f_1f_2}\bigr]_{\rm R}(x_0) &= Z_\zeta^2 Z_{\rm Y} \ly^{f_1f_2}(x_0),
 & \bigl[l_1^{f_1f_2}\bigr]_{\rm R} &= Z_\zeta^4 l_1^{f_1f_2}\,,\label{eq:ren_l}
\end{align}
and one may again determine the massless limit by requiring,
\begin{equation}
   \tilde{\partial}_0 \ga^{f_1f_2}(x_0) = 0\,,
   \label{eq:cond_mc}
\end{equation}
for some choice of flavour indices and kinematical parameters.
However, with the $\chi$SF there is the additional complication
that the boundary conditions are not protected against renormalization~\cite{Sint:2010eh}.
In fact the scale-independent renormalization constant, $\zf(g_0)$ in (\ref{eq:deltaDW}), is 
required to ensure that the $\chi$SF boundary conditions and thus parity and flavour symmetry 
are restored up to cutoff effects. In order to determine $\zf$ one thus needs to require 
that some parity breaking correlation function vanishes exactly already at finite lattice 
spacing. 

From Section~2 we may choose any of the correlation functions on the RHS
of Eqs.~(\ref{eq:dictfv}), (\ref{eq:dictfs}) or Eqs.~(\ref{eq:dictka}),(\ref{eq:dictktt}),
which does not vanish exactly. An example would be to require
\begin{equation}
  \bigl[\ga^{ud}\bigr]_{\rm R}(x_0) = 0 \quad \Leftrightarrow \quad \ga^{ud}(x_0)=0,
\label{eq:cond_zf_gA}
\end{equation}
again with some choice for the kinematical parameters. Choosing $\ga^{ud}$ is in fact
appealing as it can be used to tune both the bare mass $m_0$ and $\zf$:
up to cutoff effects, the mass tuning renders $\ga^{ud}(x_0)$ independent of $x_0$,
whereas the tuning of $\zf$ shifts $\ga^{ud}(x_0)$  by an overall constant.

\subsection{Symanzik O($a$) improvement}\label{sec:symanzik}

On-shell O($a$) improvement in the chiral limit
requires the inclusion of the Sheikholeslami-Wohlert term in the action, with coefficient
$\csw$. Furthermore, there are 2 improvement coefficients, namely $\ct, \ctt$
in the case of the SF, and $\ct, d_s$ in the case of the $\chi$SF, which are required to
cancel O($a$) boundary effects.

To obtain O($a$) improved correlation functions one then needs to include
the counterterms that are required for the fermion bilinear operators $A_\mu,V_\mu$ and $T_\mu$
(cf. Appendix \ref{app:bilinear}), with coefficients $\ca,\cv$ and $\cT$, respectively.
Note that this affects the renormalization of the mass, as the mass determined from the
improved axial current depends on $\ca$. In terms of SF correlation functions
the condition of vanishing mass changes by an O($a$) offset,
$$
  \tilde{\partial}_0 \fa(x_0) = - \ca a \partial_0^\ast\partial_0^{} \fp(x_0),
$$
which directly translates to an O($a$) offset in the critical bare mass parameter.
In other words, to reduce the uncertainty in the renormalized mass to O($a^2$),
both $\csw$ and $\ca$ are required\footnote{Incidentally, this fact has been used
to obtain improvement conditions for the determination of both $\csw$ and $\ca$
in \cite{Luscher:1996sc}.}. 
For the SF correlation functions discussed here this exhausts the list
of required O($a$) improvement coefficients. For the $\chi$SF,
a further O($a$) boundary counterterm with coefficient $\bar{d}_s$ is needed to correct
the fermionic boundary fields $\zeta,\zetabar$ 
and $\zetaprime,\zetabarprime$, cf. ref.~\cite{Sint:2010eh} and Eq.~(\ref{eq:dsbar}).

\section{Theoretical expectations for the $\chi$SF} 
\label{sec:theory}

With the definitions made in the preceding sections we may now state
our theoretical expectations which will then be subjected to perturbative tests.
We assume that $m_0$ and, in the case of the $\chi$SF, also $\zf$ have
been determined as described in the previous section.

\subsection{Boundary conditions and symmetry restoration}

As discussed in ref.~\cite{Sint:2010eh}, the projectors $\tilde{Q}_\pm$ in the $\chi$SF
boundary conditions (\ref{eq:rotbc}) correspond to the special case $\alpha=\pi/2$ of
\begin{equation}
P_\pm(\alpha) = \frac12 \left(1 \pm \gamma_0 \rme^{i\alpha\gamma_5\tau^3}\right), \qquad P_\pm(\alpha=\pi/2) = \tilde{Q}_\pm.
\end{equation}
While parity protects the value $\alpha=0$ even for the lattice regularized SF,
there is no lattice symmetry protecting the value $\alpha=\pi/2$ in the case of the $\chi$SF.
Hence, restoring the $P_5$ symmetry, Eq.~(\ref{eq:P5}), on the lattice up to O($a$) effects,
through a condition like Eq.~(\ref{eq:cond_zf_gA}), is tantamount to implementing the correct $\chi$SF boundary conditions.
The boundary conditions, on the other hand, can be more directly checked by
reversing the projectors $Q_\pm \rightarrow Q_\mp$ in the boundary fermion bilinear
sources (cf. Appendix~\ref{app:bilinear}). Note that this reversal does not affect $P_5$-parity
as the projectors $Q_\pm$ commute with $P_5$. Denoting the thus obtained but otherwise unchanged 
correlation functions by a subscript ``$-$", one would like to check that
\begin{equation}
  \lim_{a\rightarrow 0} \left[g_{\rm X,-}^{f_1f_2}\right]_{\rm R}(x_0) = 0,
\end{equation}
and analogously for $\ly$, $g_1$ and $l_1$.
We focus on the $P_5$-even correlation functions and exclude those correlation functions which are
expected to vanish for being $P_5$-odd. In practice it is advantageous to cancel
the multiplicative renormalization constants by
forming ratios, i.e.
\begin{equation}
 R_{\rm X,-}^{g,f_1f_2}(x_0) = \dfrac{g_{\rm X,-}^{f_1f_2}(x_0)}{\gx^{f_1f_2}(x_0)}, \qquad
 R_{\rm Y,-}^{l,f_1f_2}(x_0) = \dfrac{l_{\rm Y,-}^{f_1f_2}(x_0)}{\ly^{f_1f_2}(x_0)}\,.
  \label{eq:rgxlyminus}
\end{equation}
While we expect these ratios to vanish in the continuum limit it is not immediately obvious
at which rate this should happen. We also note that the same question can be asked for the standard SF, although
in this case no tuning is required to ensure the correct boundary conditions are obtained in the continuum limit.

\subsection{Automatic O($a$) improvement}
\label{subsect:automatic}
Symanzik O($a$) improvement applies to both the $\chi$SF and the SF as discussed in the previous section.
However with massless Wilson fermions and $\chi$SF boundary conditions
there is a simplification due to automatic O($a$) improvement~\cite{Frezzotti:2003ni}, as explained
in \cite{Sint:2010eh}. It is convenient to distinguish between different kinds of O($a$) effects:
these may either arise from the bulk action and composite fields in the bulk, or due to the presence of the boundaries.
Bulk O($a$) counterterms contribute at O($a^2$) to $P_5$-even observables, and at O($a$) to $P_5$-odd
observables. In fact the latter are pure lattice artefacts and would vanish if parity
was exactly realized on the lattice. Since it is straightforward to classify observables by $P_5$
one may thus avoid O($a$) effects by restricting attention to $P_5$-even observables. This is
known as the mechanism of automatic O($a$) improvement~\cite{Frezzotti:2003ni}.
Unfortunately, this nice pattern in the bulk is distorted by boundary O($a$) effects,
which can be due to both $P_5$-even ($\ct,d_s$) and $P_5$-odd ($\bar{d}_s$) counterterm insertions.
Hence, those renormalized $\chi$SF correlation functions which translate to $\fa,\fp,f_1$ and $\kv,\kt,k_1$,
are expected to approach the continuum limit with bulk O($a^2$) and boundary O($a$) corrections;
the latter can be cancelled by appropriately tuning the boundary
improvement coefficients $\ct$ and $d_s$. This implies the possibility of using
unimproved Wilson fermions and omitting all O($a$) counterterms to the composite fields in the bulk.

Note that the tuning conditions for $\mcr$ and $\zf$ generally define these parameters
up to an O($a$) ambiguity, unless Symanzik O($a$) improvement is implemented. 
Hence, if $\zf$ is obtained from an alternative condition,
one generally expects the difference, $\Delta\zf$, to asymptotically vanish at a rate of O($a$),
and the same applies to the critical mass, $\mcr$. We emphasise that
these O($a$) ambiguities are not in conflict with automatic O($a$) improvement~\cite{Sint:2010eh}; 
for, treating any such O($a$) shift of $\zf$ or $\mcr$ as an insertion of the respective
$P_5$-odd counterterms into the $P_5$-even correlation function of interest, the result will
be of O($a$) and combine with the O($a$) coefficient to produce a total change of O($a^2$).

As mentioned above, $P_5$-odd correlation functions are expected to vanish in the continuum limit,
at a rate linear in the lattice spacing. If correctly O($a$) improved \`a la Symanzik,
this rate should change to O($a^2$). Conversely, this fact may be used to obtain alternative
O($a$) improvement conditions. This is potentially very interesting but will be left to future work.
Here we will only verify that $P_5$-odd observables vanish indeed at a rate proportional to $a$.
This includes the bulk O($a$) counterterm contributions to the $P_5$-even correlation functions,
$\ga^{uu'}$, $\lv^{uu'}$ and $\lt^{ud}$, namely
\begin{equation}
    \tilde\partial_0^{}\gp^{uu'}(x_0),\quad \,\tilde{\partial}_0\lt^{uu'}(x_0),
    \quad \tilde{\partial}_0 \lv^{ud}(x_0)\,.
\end{equation}
As these come with an explicit factor $a$, their contribution amounts to an O($a^2$) effect.

\subsection{Flavour symmetry restoration}
\label{subsect:flavour_restoration}

Focussing on the boundary-to-boundary correlation functions, Eqs.~(\ref{eq:dictf1}),(\ref{eq:dictk1}),
we expect that the chain of equalities on the RHS holds for renormalized correlation
functions, so that the ratios
\begin{equation}
  R_{g} = \dfrac{g_1^{uu'}}{g_1^{ud}},\qquad R_{l}=\dfrac{l_1^{uu'}}{l_1^{ud}},
  \label{eq:ratflav}
\end{equation}
should converge to 1 in the continuum limit, thereby demonstrating the restoration of
flavour symmetry.
Going a step further one may also show that the continuum limit is reached
with O($a^2$) corrections only: according to the above discussion of automatic
O($a$) improvement, the only O($a$) effects can be caused by the $P_5$-even boundary
counterterms with coefficients $\ct$ and $d_s$. In a Symanzik type analysis of
the cutoff effects  we may account for small changes $\Delta\ct$ and
$\Delta d_s$ in these coefficients by insertion of the respective counterterms.
Denoting these insertions by $g_{1;\ct}$ and $g_{1;d_s}$, we then obtain e.g.
\begin{equation}
   \bigl[g_1^{f_1f_2}\bigr]_{\rm R} = g_1^{f_1f_2} + a \left(\Delta\ct g_{1;\ct}^{f_1f_2}
   + \Delta d_s g_{1;d_s}^{f_1f_2}\right) + {\rm O} (a^2)\,,
\end{equation}
where the correlation functions on the RHS are calculated in Symanzik's effective continuum theory.
Expanding the first ratio, $R_g$, in Eq.~(\ref{eq:ratflav}), its expansion coefficient at O($a$) has 2 parts,
\begin{equation}
   \Delta\ct\left(\dfrac{g_{1;\ct}^{uu'}}{g_1^{uu'}} -  \dfrac{g_{1;\ct}^{ud}}{g_1^{ud}}\right)
   +  \Delta d_s \left(\dfrac{g_{1;d_s}^{uu'}}{g_1^{uu'}} -  \dfrac{g_{1;d_s}^{ud}}{g_1^{ud}}\right)\,.
\end{equation}
Due to $g_1^{uu'} = g_1^{ud}$, it remains to show that
\begin{equation}
  g^{uu'}_{1;\ct} = g^{ud}_{1;\ct},\qquad
  g^{uu'}_{1;d_s} = g^{ud}_{1;d_s}.
\end{equation}
This is straightforward: the counterterms are both invariant under chiral and flavour transformations, which
are the very symmetries of the continuum theory implying $g_1^{uu'} = g_1^{ud}$. Hence
the same relation must hold with the insertions of the counterterms.

\subsection{Scale-independent renormalization constants}
\label{subsec:rat_finiteZ}

We now apply the same universality argument to  correlation functions with fermion bilinear fields in the
bulk. Equating the right hand sides of Eq.~(\ref{eq:dictfa}), in terms of the renormalized
correlation functions, one obtains
\begin{equation}
  [\ga^{uu'}]_{\rm R}(x_0) = -i[\gv^{ud}]_{\rm R}(x_0)\,.
\end{equation}
Defining the ratio of bare correlation functions,
\begin{equation}
  R^{g}_{\rm AV}(g_0^2,a/L;x_0,\theta,T/L) = \dfrac{-i\gv^{ud}(x_0)}{\phantom{-i}\ga^{uu'}(x_0)}\,,
  \label{eq:RgAV}
\end{equation}
we expect that, at fixed renormalized parameters $\gr$ and $\mr=0$, and with fixed
kinematical parameters, for instance, $x_0=T/2$, $T=L$ and $\theta=0.5$,
\begin{equation}
     R^{g}_{\rm AV}\quad \mathop{\sim}_{a/L\rightarrow 0} \quad\dfrac{\Za}{\Zv} + \rmO(a^2)\,.
     \label{eq:RgAVas}
\end{equation}
Here, the renormalization constants $\Za$ and $\Zv$ are as required to restore the continuum
symmetries. We emphasize that these are the same continuum chiral and flavour symmetries
which are encoded in the corresponding Ward identities. Therefore, we expect that,
up to cutoff effects, $\Za$ and $\Zv$ or their ratio must coincide with
results obtained by imposing Ward identities as normalization
conditions~\cite{Bochicchio:1985xa,Luscher:1996jn}.

Why do we expect the cutoff effects to be of order $a^2$ in Eq.~(\ref{eq:RgAVas})? 
Firstly, automatic O($a$) improvement implies that $P_5$-odd O($a$) counterterms do not cause O($a$) effects
in these ratios of $P_5$-even correlation functions. Secondly, O($a$) corrections from
the $P_5$-even O($a$) boundary counterterms associated with $\ct$ and $d_s$ drop out
in the ratio for the same reason this happens in the ratios of boundary-to-boundary
correlation functions, Eq.~(\ref{eq:ratflav}). This corresponds with a
similar argument~\cite{Luscher:1996sc} regarding Ward identities: the external source fields 
localised outside the space-time region where the O($a$) improved Ward identity is probed need
not be O($a$) improved for the Ward identity to hold up to O($a^2$) effects (cf.~Section 6 of~\cite{Luscher:1996sc}).

At this point it is useful to recall that Wilson fermions in the bulk
actually enjoy exact lattice symmetries leading to the conserved vector currents,
\begin{equation}
 \widetilde{V}^{f_1f_2}_\mu(x)= \dfrac12
 \bigg[\psibar_{f_1}(x)(\dirac\mu-1)U_\mu(x)\psi_{f_2}(x+a\hat{\mu})+
 \psibar_{f_1}(x+a\hat{\mu})(\dirac\mu+1)U_\mu(x)^\dagger\psi_{f_2}(x)\bigg]\,.
 \label{eq:Vtilde}
\end{equation}
We recall that in our conventions (i.e.~the physical basis defined by standard SF boundary conditions, 
cf.~Subsect~\ref{subsect:flavsym}) the symmetries associated with these vector currents are interpreted
either as flavour or chiral symmetry, depending on the flavour assignments. In any case, since Noether currents associated 
with exact lattice symmetries are protected against 
renormalization, one may infer that $Z_{\rm \widetilde{V}}=1$, and, furthermore,
\begin{equation}
  \partial_0^\ast g^{f_1f_2}_{\rm \widetilde V}(x_0) = 0,\qquad a<x_0<T,
\end{equation}
{\em exactly}, i.e.~not just up to finite lattice spacing effects. Therefore one expects
\begin{equation}
     \label{eq:ZAgdef}
     R^g_{\rm A\widetilde{V}}\quad \mathop{\sim}_{a/L\rightarrow 0} \quad \Za  + \rmO(a^2),
\end{equation}
where this ratio is defined as in Eq.~(\ref{eq:RgAV}) but with the conserved current, Eq.~(\ref{eq:Vtilde}),
replacing the local current in the vector correlation function. Here we have again
assumed that the renormalized parameters and the kinematics have been chosen e.g.~as discussed
after Eq.~(\ref{eq:RgAV}). Having a conserved vector current also
allows for the determination of $\Zv$ for the non-conserved local current, 
simply by taking the ratio
\begin{equation}
  R^g_{\rm V\widetilde{V}}(x_0) = \dfrac{\gvt^{ud}(x_0)}{\gv^{ud}(x_0)}  \quad \mathop{\sim}_{a/L\rightarrow 0} \quad \Zv + \rmO(a^2).
  \label{eq:RgVVt}
\end{equation}
Alternative ratios for the current normalization constants $\Za$ and $\Zv$ can be formed with the $l$-correlation functions,
\begin{equation}
  R^l_{\rm A\widetilde{V}}(x_0) = \dfrac{i\lvt^{uu'}(x_0)}{\phantom{i}\la^{ud}(x_0)},\qquad
  R^l_{\rm V\widetilde{V}}(x_0) = \dfrac{\lvt^{uu'}(x_0)}{\lv^{uu'}(x_0)} \,.
\label{eq:ZlAV}
\end{equation}
Finally, one can also determine the finite ratios among
scale-dependent renormalization constants that belong to the same chiral
multiplet by considering the ratios,
\begin{equation}
  R^g_{\rm PS}(x_0) = \dfrac{i\gs^{uu'}(x_0)}{\phantom{i}\gp^{ud}(x_0)},\qquad
  R^l_{\rm T\widetilde{T}}(x_0) = \dfrac{i\ltt^{uu'}(x_0)}{\lt^{ud}(x_0)}.
\label{eq:ZSPTT}
\end{equation}
One then expects,
\begin{equation}
   R^g_{\rm PS}\quad \mathop{\sim}_{a/L\rightarrow 0} \quad \dfrac{\Zp}{\Zs}  + \rmO(a^2),
   \label{eq:ZSP}
\end{equation}
where we emphasize that both renormalization constants are associated with the flavour non-singlet 
operators. Regarding the tensor densities we expect
\begin{equation}
  R^l_{\rm T\widetilde{T}} = 1 + \rmO(a^2),
\end{equation}
since the operators $T_{\mu\nu}$ and $\widetilde{T}_{\mu\nu}$ are related by a lattice symmetry, 
cf.~Appendix~\ref{app:bilinear}.

\subsection{Scale-dependent renormalization constants}

\label{subsec:scaledep}

So far we have used the universality relations to the right hand sides
of our dictionary. A more direct comparison between renormalized correlation
functions calculated in the SF and in the $\chi$SF is rendered difficult by the fact
that the bare boundary source fields ${\cal O}_5$ and ${\cal Q}_5$
are not simply related to each other, due to the very different structure
of the lattice actions near the boundaries. This has to
be contrasted with bare composite fields in the bulk which can be chosen
to be the same independently of the boundary conditions.
Consequently, if we define $Z_\zeta$ through the respective ratios
\begin{equation}
  Z_\zeta^{\rm SF} = \left(f_1^{(0)}/f_1\right)^{\frac14},\qquad
  Z_\zeta^{\chi{\rm SF}} = \left(g_1^{(0)}/g_1\right)^{\frac14},
\end{equation}
the ratio of these $Z$-factors yields a scale independent constant
which only logarithmically approaches 1 in the continuum limit. Here,
the numerators are the lowest order perturbative expressions, e.g.
\begin{equation}
   f_1^{(0)} = f_1\vert_{g_0^2=0},
\end{equation}
such that the $Z$-factors are unity at leading order of perturbation theory.

Despite this limitation, we may compare scale-dependent renormalization constants
for bulk operators in SF renormalization schemes.
For instance, the SF scheme for the pseudo-scalar density
can be defined through~\cite{Sint:1998iq,Capitani:1998mq,Sint:2010xy},
\begin{eqnarray}
 \dfrac{[\fp]_{\rm R}(T/2)}{\sqrt{[f_1]_{\rm R}}} &=&
 \left.\dfrac{\fp(T/2)}{\sqrt{f_1}}\right\vert_{g_0=0},\\
 \dfrac{\bigl[\gp^{ud}\bigr]_{\rm R}(T/2)}{\sqrt{\bigl[g^{ud}_1\bigr]_{\rm R}}} &=&
 \left.\dfrac{\gp^{ud}(T/2)}{\sqrt{g^{ud}_1}}\right\vert_{g_0=0},
\end{eqnarray}
where at a given renormalization scale $\mu=L^{-1}$ (defined e.g. through the value of the renormalized
coupling) we require the renormalized matrix elements to be equal to their tree level values at
$g_0=0$. The boundary-to-boundary correlators $f_1$ and $g_1^{ud}$ are used to cancel the boundary
quark field renormalization factors $Z_\zeta$. The resulting expressions for the renormalization
constant of the pseudo-scalar density are then given by,
\begin{equation}
 Z^{\rm SF}_{\rm P}(g_0^2,L/a) = c(L/a) \dfrac{\sqrt{f_1}}{\fp(T/2)}\,,\quad
 Z^{\chi{\rm SF}}_{\rm P}(g_0^2,L/a) = c'(L/a) \dfrac{\sqrt{g_1^{ud}}}{\gp^{ud}(T/2)}\,,
 \label{eq:Zp}
\end{equation}
where the factors $c$ and $c'$ are chosen such that $Z^{\rm SF,\chi SF}_{\rm P}(0,L/a)=1$.
Note that the renormalization scale is fixed in terms of $L$,
the physical extent of the spatial volume. This implies that all dimensionful
parameters have to be scaled in a fixed proportion to $L$. Having set the mass to zero and $x_0=T/2$
one usually sets the aspect ratio $\rho=T/L=1$~\cite{Sint:1998iq}. Finally one needs to fix any
dimensionless parameters, e.g.~$\theta=0.5$,  in order to completely specify the SF scheme.

Similarly, one can define SF renormalization conditions for the tensor-density through,
\begin{equation}
 \label{eq:tensor_ren}
  Z^{\rm SF}_{\rm T}(g_0^2,L/a) = b(L/a) {\sqrt{k_1}\over \kt(T/2)},\qquad
  Z^{\xSF}_{\rm T}(g_0^2,L/a) = b'(L/a){\sqrt{l^{ud}_1}\over \lt^{ud}(T/2)},
\end{equation}
where again the factors $b$ and $b'$ are chosen such that $Z^{\rm SF,\chi SF}_{\rm T}(0,L/a)=1$ holds exactly
on a finite lattice with extent $L/a$. We note that the renormalization condition for the pseudo-scalar density can be
turned into a renormalization condition for the non-singlet scalar density by combining it with
an estimator of the ratio $\Zp/\Zs$, Eq.~(\ref{eq:ZSP}). We also remark that, by applying the same SF
renormalization procedure to scale-independent renormalization problems, one 
may define e.g.~a renormalized axial current in the SF scheme with corresponding renormalization
constants $\Za^{\rm SF}$ and $\Za^{\text{$\chi$SF}}$. However, we stress that such a renormalized
axial current is not canonically normalized, i.e.~it does not satisfy the axial Ward identities.

To conclude, we note that if O($a$) improved Wilson fermions are used in both the
SF and $\xSF$ determinations, one expects, for ${\rm X} = {\rm P,T,\ldots}$
\begin{equation}
   R_{\rm X}={Z^{\text{$\chi$SF}}_{\rm X}\over Z^{\rm SF}_{\rm X}} = 1 + {\rm O}(a^2),
 \label{eq:RatioZp}
\end{equation}
provided that the boundary improvement coefficients $\ct,\ctt$ for the SF and $\ct,d_s$ for the $\chi$SF
have been correctly tuned.
In the case of the ratio of $Z_{\rm T}$'s the SF computation also requires the necessary
O($a$) bulk counterterm for $T_{\mu\nu}$, otherwise uncancelled O($a$) effects are expected
in the ratio between the SF and $\xSF$ renormalization constants (\ref{eq:tensor_ren}).

The tensor density provides a first example where automatic O($a$) improvement
is advantageous in the calculation of the step-scaling function. On the lattice one defines
\begin{equation}
   \Sigma_{\rm T}(u,a/L) = \left.\dfrac{\Zt(g^2_0,2L/a)}{\Zt(g^2_0,L/a)}\right\vert_{u=\bar{g}^2(L)},
\end{equation}
with some renormalized coupling $\bar{g}^2(L)$ held fixed at the value $u$. Denoting the continuum step-scaling function by
$\sigma_{\rm T}(u)$ and with the correct choice for the boundary O($a$) improvement coefficients
$\ct$ and $d_s$ or $\ctt$, we expect, in the case of the $\chi$SF,
\begin{equation}
  \Sigma_{\rm T}(u,a/L) =\sigma_{\rm T}(u) + \rmO(a^2).
\end{equation}
In contrast, complete O($a$) improvement with the standard SF also requires the inclusion of
the bulk counterterm $\propto \cT$ (cf.~Appendix \ref{app:bilinear}).

\section{Perturbation theory}
\label{sec:PTsetup}
\subsection{Perturbative expansion of parameters and correlation functions}

The perturbative expansion of the renormalized correlation functions in
(\ref{eq:ren_g}) follows very closely the literature~\cite{Luscher:1992an,Luscher:1996vw}.
In particular, the gauge action remains the same, so that the gauge fixing
procedure can be taken over unchanged.

The coefficients in the action are functions of the bare coupling, and have a perturbative
expansion in $g_{0}^{2}$,
\begin{equation}
 c(g_{0})=c^{(0)}+g_{0}^{2}\,c^{(1)}+\rmO(g_{0}^{4}),
\end{equation}
where $c$ generically refers to $\mcr,z_{f},d_{s},\ct,\ctt$.
The tree-level values are given by~\cite{Luscher:1992an,Luscher:1996vw,Sint:2010eh},
\begin{equation}
  \mcr^{(0)}=0,  \quad z_{f}^{(0)}=1, \quad d_{s}^{(0)}=1/2, \quad \ct^{(0)}=1, \quad \ctt^{(0)}=1\,,
 \label{eq:acttree}
\end{equation}
and the one-loop coefficients $\mcr^{(1)}$, $z_{f}^{(1)}$ and $d_{s}^{(1)}$ and $\ct^{(1)}$
are given below.
Renormalization factors are expanded similarly,
\begin{equation}
  Z(g_0^2,L/a) =1+g_{0}^{2}\,Z^{(1)}(L/a)+\rmO(g_{0}^{4}),\qquad
  \label{eq:exp_ZX_ZY}
\end{equation}
where $Z$ stands for $Z_\zeta$ or $Z_{\rm X}$ in the case of fermion bilinear fields $X^{f_1f_2}$.
We distinguish between renormalization scale-independent and scale-dependent renormalization factors.
Among the former are $\Za$, $\Zv$ and ratios such as $\Zp/\Zs$, whereas $\Zp$, $\Zs$ and
$\Zt$ depend on the renormalization scale $\mu=L^{-1}$ which, as before, has been identified with the inverse
of $L$, the linear extent of the spatial volume. To obtain renormalized correlation functions in perturbation theory
one may e.g.~adopt the minimal subtraction of logarithms scheme~\cite{Sint:1997jx} (with $\mu=L^{-1}$).
However, one must then still allow for finite renormalizations, as otherwise the continuum
relations between correlation functions will not hold in general. More precisely, to renormalize consistently
with the expected continuum relations derived in Section~2, one may start and renormalize a given field minimally but
allow for finite parts in the renormalization of its chirally transformed counterpart.

Given these definitions, fixing the renormalized parameters $\gr$ and $\mr=0$
amounts to tuning the bare parameters according to 
\begin{equation}
  g_0^2 =\gr^2 +\rmO(\gr^4),\qquad
  m_0   = \mcr^{(1)} \gr^2 +\rmO(\gr^4),
\end{equation}
and, up to higher orders in the coupling, the boundary counterterm coefficients are set to
\begin{equation}
 \zf   = 1 + \zf^{(1)}\gr^2,\qquad
  d_s  = \half + d_s^{(1)}\gr^2,\qquad
 \ctt  = 1 + \ctt^{(1)}\gr^2,\qquad
 \ct   = 1.
\end{equation}
Note that, to the order considered, the gluonic boundary counterterm $\propto \ct$
enters the fermionic correlation functions only at tree-level via the gluon propagator.
In order to determine its one-loop value for the $\chi$SF we have also computed a gluonic observable,
namely the SF coupling constant at one-loop order (cf.~Section~\ref{sec:coupling}).
Except for this calculation we stay with vanishing background gauge field and
thus only require $\csw$ to be set at tree-level, i.e.~$\csw=\csw^{(0)}=1,0$, for O($a$) improved and
unimproved Wilson fermions, respectively.

\begin{figure}[!t]
\centering
\includegraphics[clip=true,scale=0.54]{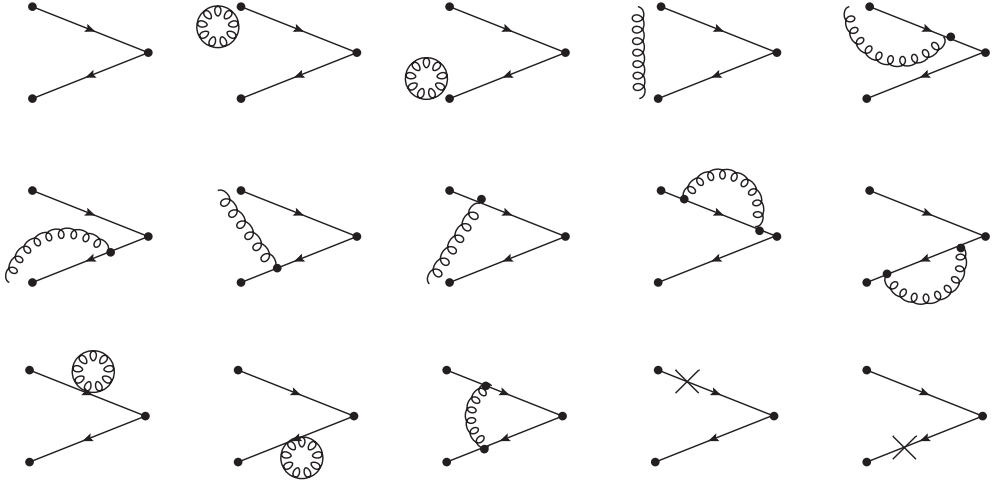}
\caption{The set of tree-level and one-loop diagrams contributing to the
boundary-to-bulk correlation functions $\gx^{f_{1}f_{2}}$
and $\ly^{f_{1}f_{2}}$. Fermion propagators are represented by continuous lines,
while curly lines represent the gluon propagator. Fermionic counterterms insertions are
represented by a cross on a fermion line. Gluon lines not starting from a fermion line
originate from the explicit time like link variables in the fermionic
boundary fields $\zeta$ and $\zetabar$, Eqs.~(\ref{eq:zeta}),(\ref{eq:zetabar}).}
\label{fig:diag_gX_lY}
\end{figure}

\begin{figure}[!t]
\centering
\includegraphics[clip=true,scale=0.54]{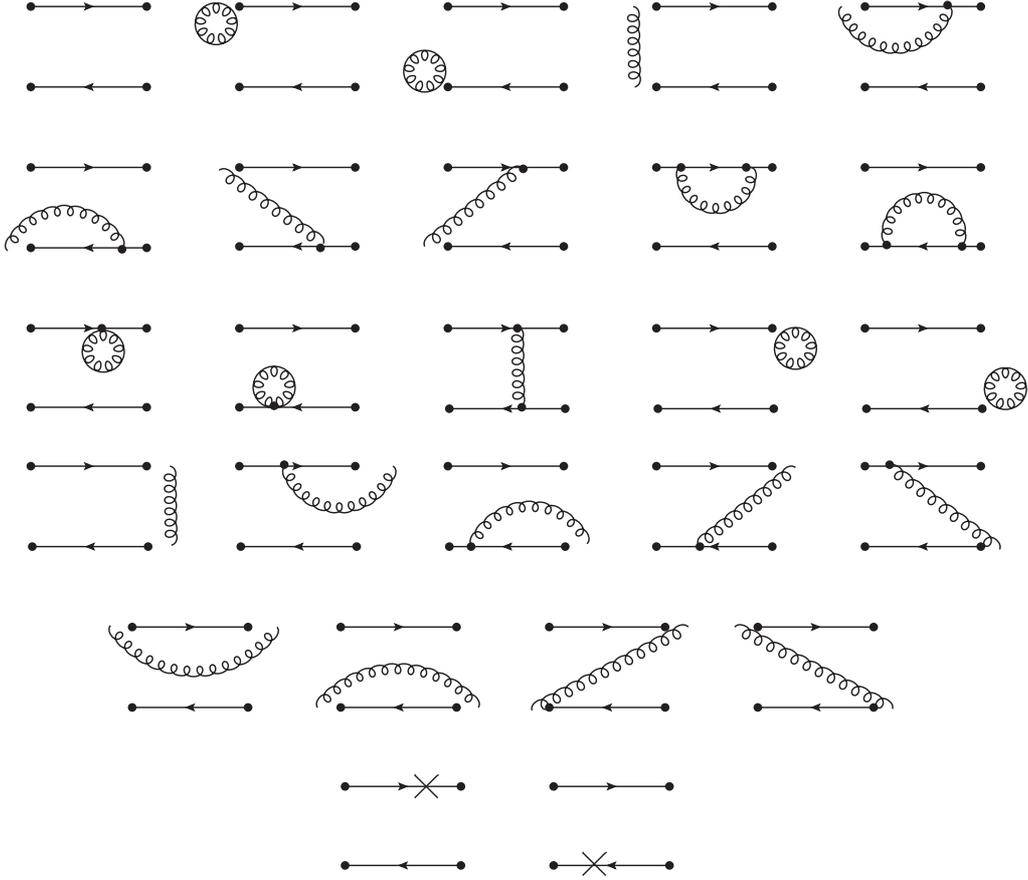}
\caption{The set of tree-level and one-loop diagrams contributing to the
boundary-to-boundary correlation functions $g_{1}^{f_{1}f_{2}}$ and $l_{1}^{f_{1}f_{2}}$.}
\label{fig:diag_g1_l1}
\end{figure}

\begin{figure}[!t]
\centering
\includegraphics[clip=true,scale=0.54]{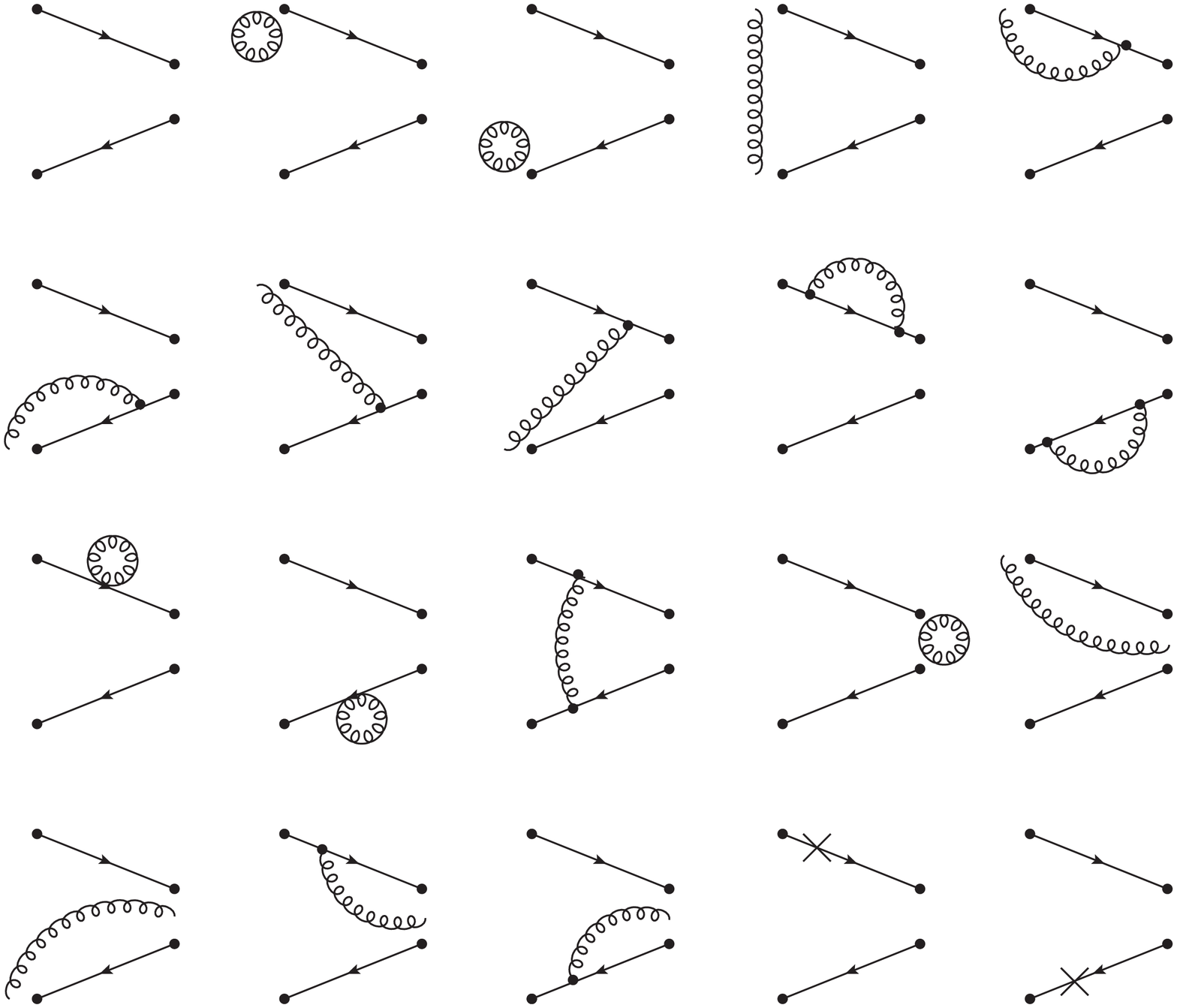}
\caption{The set of tree-level and one-loop diagrams contributing to the boundary-to-bulk
correlation functions involving the point-split vector current, i.e.,
$g_{\rm \widetilde V}^{f_{1}f_{2}}$ and $l_{\rm \widetilde V}^{f_{1}f_{2}}$. Note that each diagram
in the figure represents the two terms forming the point-split current (\ref{eq:Vtilde}).
The two fermion lines do not meet at the vertex due to the point-split nature of the current, 
and gluons lines may originate from the gauge links that appear in the operator.}
\label{fig:diag_gVt_lVt}
\end{figure}

We are now ready to expand the renormalized correlation functions in Eq.~(\ref{eq:ren_g})
in powers of $\gr^2$. Defining the  expansion coefficients of the renormalized and O($a$) improved 
correlation functions by
\begin{equation}
   [\gx]_{\rm R}(x_0) =  \gx^{(0)}(x_{0}) + \gr^{2}\,\gx^{(1)}(x_{0}) + \rmO(\gr^4),\qquad
    [g_{1}]_{\rm R} = g^{(0)}_{1} + \gr^{2}\, g^{(1)}_{1} +\rmO(\gr^4),
\end{equation}
the one-loop coefficients take the form,
\begin{eqnarray}
 \gx^{(1)}(x_{0}) &=& \sum_{n}g_{{\rm X},n}^{(1)}(x_{0}) +\mcr^{(1)}  g_{{\rm X};m_0}^{(0)}(x_{0}) 
  + \left(Z_{\rm X}^{(1)}+2 Z_{\zeta}^{(1)}\right)\gx^{(0)}(x_{0}) \nonumber\\
&& +\, z_{f}^{(1)}g_{{\rm X};z_{f}}^{(0)}(x_{0})+  d_{s}^{(1)} g_{{\rm X};d_{s}}^{(0)}(x_{0})
+ \bar d_{s}^{(1)}g_{{\rm X};\bar d_{s}}^{(0)}(x_{0})
+ a\,c_{\rm X}^{(1)} g_{\delta {\rm X}}^{(0)}(x_{0})\,,
\label{eq:gX_1loop} \\[1ex]
g^{(1)}_{1} &=& \sum_{n}g_{1,n}^{(1)} +\mcr^{(1)} g_{1;m_0}^{(0)}+ 4Z_{\zeta}^{(1)}g_{1}^{(0)}
+ z_{f}^{(1)}g_{1;z_{f}}^{(0)}+d_{s}^{(1)}g_{1;d_{s}}^{(0)} + \bar d_{s}^{(1)}g_{1;\bar d_{s}}^{(0)}\,.
\label{eq:g1_1loop}
\end{eqnarray}
Note that, for the sake of readability, we have left out the flavour indices on all terms of these equations, and
we have defined the counterterm contributions for $\gx$,
\begin{eqnarray}
 g_{{\rm X};m_0}^{f_{1}f_{2}(0)} &=&
   \left.\dfrac{\partial}{\partial m_0}\gx^{f_{1}f_{2}(0)}\right\vert_{m_0=0}\,,\\
 g_{{\rm X};z_{f}}^{f_{1}f_{2}(0)} &=&
   \left.\dfrac{\partial}{\partial \zf}\gx^{f_{1}f_{2}(0)}\right\vert_{\zf=1}\,,\\
 g_{{\rm X};d_{s}}^{f_{1}f_{2}(0)} &=&
  \left.\dfrac{\partial}{\partial d_s}\gx^{f_{1}f_{2}(0)}\right\vert_{d_s=1/2}\,,\\
 g_{{\rm X};\bar d_{s}}^{f_{1}f_{2}(0)} &=&
   \left.\dfrac{\partial}{\partial \bar{d}_s}\gx^{f_{1}f_{2}(0)}\right\vert_{\bar{d}_s=0}\,,
\end{eqnarray}
and similarly for $g_1$. The correlation functions $g_{\delta {\rm X}}$ refer to the bulk O($a$) counterterms
$\delta X$ associated with some of the fermion bilinear fields $X$ (cf.~Eqs.~(\ref{eq:imp_bil_list})).
We have assumed that their respective coefficients $\cx$ vanish at tree-level, i.e.~$\cx^{(0)}=0$,
which is known to be the case for the local bilinears (cf. Appendix~\ref{app:bilinear}).
Analogous expansions are obtained for the correlation functions $[\ly]_{\rm R}$ and $[l_{1}]_{\rm R}$,
and also for the standard SF functions (with the obvious modifications).
The sums over $n$ in (\ref{eq:gX_1loop}) and (\ref{eq:g1_1loop}) run over the set of all those diagrams 
containing a gluon line (see Figures \ref{fig:diag_gX_lY}, \ref{fig:diag_g1_l1} and \ref{fig:diag_gVt_lVt}).
For later use we give the sum of these diagrams a separate name,
\begin{equation}
 \gx^{(1,a)} =  \sum_n g_{{\rm X},n}^{(1)},\qquad  g_1^{(1,a)} =  \sum_n g_{{1},n}^{(1)},
\end{equation}
and analogously for all other correlation functions. As said,
the terms with subscripts ``$m_0$", ``$z_{f}$'', ``$d_{s}$'' and ``$\bar d_{s}$'', indicate the
contributions due to insertions of the counterterms proportional to these coefficients. Diagrammatically 
these are represented by  crosses on the fermion lines. Note that we have included the counterterm $\propto \bar d_{s}^{(1)}$ 
for completeness of notation, although this counterterm has been omitted in our calculation.
While the $\bar{d}_s$ counterterm is correctly implemented at tree level ($\bar{d}_s^{(0)}=0$, cf.~\cite{Sint:2010eh}),
in the following we omit the one-loop counterterm, effectively setting $\bar{d}_s^{(1)}=0$ in
Eqs.~(\ref{eq:gX_1loop}),(\ref{eq:g1_1loop}), and all other correlators.
The reason this can be done consistently is that, by the mechanism of automatic O($a$) improvement,
it only contributes O($a^2$) effects to any of the $P_5$-even correlation functions.
Its inclusion would however be required for the study of O($a$) improvement for the  $P_5$-odd correlation functions,
which is beyond the scope of this work.

\subsection{The numerical calculation and checks performed}

All terms appearing in (\ref{eq:gX_1loop}) and (\ref{eq:g1_1loop}) are functions of
$a/L$ that can be evaluated numerically by inserting the explicit time-momentum representation of
the vertices and propagators into the expressions of each diagram. To this end, we have produced
a FORTRAN program for the numerical evaluation of Feynman diagrams both in the standard and chirally
rotated SF. Numerical results for each diagram and counterterm have been compared against previous
calculations~\cite{Weisz:private} in the case of the standard SF, finding agreement up to rounding errors.
For the $\chi$SF we have checked all diagrams for the $\gx$ and $\ly$ correlators by an independent FORTRAN program,
excluding the ones involving the point-split vector current. A check for the latter has been performed by comparing ratios of 
correlators to Monte Carlo simulations at small values of the bare coupling, $g_0^2$, 
cf.~Appendix~\ref{app:MC}. Further confidence in the correctness of our code is gained by the perfect agreement
with results in the literature for the current normalization
constants (cf.~Section~\ref{sec:Ren_bil}).
We have numerically checked gauge parameter independence for all correlators on small lattices and then performed
all subsequent calculations in the Feynman gauge (setting the gauge parameter $\lambda_0=1$),
in which the gluon propagator for the plaquette action is diagonal. This allows
for a considerable speed-up in the numerical computation. A technical point worth noting
is that we calculated  the fermion propagator for fixed spatial momentum by numerical
matrix inversion, as the available analytic result assumes $d_s=1$, whereas
the correct tree-level value is $d_s^{(0)}=1/2$~\cite{Sint:2010eh}. While it would have
been possible to calculate an approximate fermion propagator analytically by single or double
insertion of the boundary counterterm, we refrained from doing this as it would prevent a
direct comparison with non-perturbative data at finite lattice spacing.

In the remainder of this section we determine the one-loop parameters of
the lattice action, $\mcr^{(1)}$, $\zf^{(1)}$ and $d_s^{(1)}$ from this data,
and $\ct^{(1)}$ is quoted from a separate calculation of the fermion determinant following the
lines of ref.~\cite{Sint:1995ch}, as described in Section~\ref{sec:coupling}.

\subsection{Determination of $\mcr^{(1)}$ and $z_{f}^{(1)}$}\label{sec:mc_zf}

The determination of $\mcr$ and $z_{f}$ is done by solving simultaneously the 
system of equations consisting of the conditions (\ref{eq:cond_mc}) (with flavours $f_1f_2 = ud$)
and (\ref{eq:cond_zf_gA}).
Expanding these equations to order $\gr^2$, we obtain
\begin{equation}
\begin{split}
0 = &\,\tilde{\partial}_{0}\ga^{ud(1,a)}+\mcr^{(1)}\tilde{\partial}_{0}g_{{\rm A};m_0}^{ud(0)} 
+ z_{f}^{(1)}\tilde{\partial}_{0} g_{{\rm A};z_f}^{ud(0)}+d_{s}^{(1)}\tilde{\partial}_{0}g_{{\rm A};d_s}^{ud(0)}\\
&\,+ \bar d_{s}^{(1)}\tilde{\partial}_{0} g_{{\rm A};\bar{d}_s}^{ud(0)} + a \ca^{(1)}\partial_0^\ast\partial_0 \gp^{ud(0)},
\label{eq:cond_mc_1loop}
\end{split}
\end{equation}
and
\begin{equation}
0= \ga^{ud(1,a)}+\mcr^{(1)}g_{{\rm A};m_0}^{ud(0)} + z_{f}^{(1)}g_{{\rm A};z_f}^{ud(0)}+d_{s}^{(1)}g_{{\rm A};d_s}^{ud(0)}
+\bar d_{s}^{(1)}g_{{\rm A};\bar{d}_s}^{ud(0)} + a \ca^{(1)} \tilde{\partial}_0\gp^{ud(0)},
\label{eq:cond_zf_1loop}
\end{equation}
where we always assume $x_0=T/2$, and $T=L$. The determination of $\mcr^{(1)}$ and $z_{f}^{(1)}$ 
becomes particularly simple when choosing $\theta=0$. Indeed, for this choice, 
the contributions of the counterterms proportional
to $d_{s}$, $\bar d_{s}$  and $\ca$ vanish. Moreover, for $\theta=0$, the contribution of 
the counterterm proportional to $z_{f}^{(1)}$ in (\ref{eq:cond_zf_1loop}) 
is  constant in $x_{0}$, and hence the derivative $\tilde{\partial}_{0}g_{{\rm A};z_f}^{ud(0)}$ in (\ref{eq:cond_mc_1loop})
vanishes. The determination of $\mcr^{(1)}$ thus becomes independent of $z_{f}^{(1)}$ in this case.
For a given lattice size in the range $L/a\in[6,48]$ we then solve the 2 equations
and obtain the series
\begin{eqnarray}
  \mcr^{(1)}(a/L) &=& - \dfrac{\tilde{\partial}_{0}\ga^{ud(1,a)}(L/2)}{\tilde{\partial}_{0}g_{{\rm A};m_0}^{ud(0)}(L/2)}\,,
  \label{eq:mc1_L}\\
  z_f^{(1)}(a/L)  &=&
  -\dfrac{ \ga^{ud(1,a)} (L/2) + \mcr^{(1)}(a/L) g_{{\rm A};m_0}^{ud(0)}(L/2) }{g_{{\rm A};z_f}^{ud(0)}(L/2)}.
  \label{eq:zf1_L}
\end{eqnarray}
From these, we extrapolate to the asymptotic values
\begin{equation}
  \mcr^{(1)} = \lim_{a/L\rightarrow 0} \mcr^{(1)}(a/L),\qquad
   z_f^{(1)} = \lim_{a/L\rightarrow 0} z_f^{(1)}(a/L),
  \label{eq:mc1_zf1}
\end{equation}
following the blocking method described in~\cite{Bode:1999sm}.
The values obtained in this way are collected  in Table~\ref{tab:mc1_zf1} for the fundamental representation
of the gauge group\footnote{Values of  $a\mcr^{(1)}$ and $z_{f}^{(1)}$ for a representation $R$
can be obtained from the numbers quoted in Table  \ref{tab:mc1_zf1} by replacing 
$C_{\rm F}\rightarrow C_{2}(R)$. For the symmetric, antisymmetric, and adjoint representations one has 
$C_{2}(R)=2C_{\rm F}(N+2)/(N+1)$, $2C_{\rm F}(N-2)/(N+1)$ and $N$, respectively.}.
We reproduce the values of $\mcr^{(1)}$ available in the 
literature~\cite{GonzalezArroyo:1982ts,Stehr:1982en,Wholert_desy},
as expected, since these asymptotic results only depend on the regularization 
of the bulk action, and are hence unaffected by the choice of boundary conditions. 
This is a further strong check on the correctness of our calculation.
The values for $z_{f}^{(1)}$, instead, have been calculated here for the first time,
cf.~Table~\ref{tab:mc1_zf1}.

\begin{table}[htb!]
\begin{center}
\begin{tabular}{|c|c|c|}
\hline
        &  $a\mcr^{(1)} $       &     $z_{f}^{(1)}$   \\
    \hline
 $  \csw^{(0)}=1 $  &  $  -0.2025565(3)\times C_{\rm F} $  &  $  0.16759(1)\times C_{\rm F} $  \\
 $  \csw^{(0)}=0 $  &  $  -0.32571(2)\times C_{\rm F} $   &  $  0.3291(2)\times C_{\rm F} $  \\
 
\hline 
\end{tabular}
\caption{Results for $a\mcr^{(1)}$ and $z_{f}^{(1)}$ with and without the clover term, for the fundamental
representation of the gauge group.}
\label{tab:mc1_zf1}
\end{center}
\end{table}

In order to check the correctness of the determination of $z_{f}^{(1)}$, we recompute it 
using the following  alternative renormalization conditions (again for $x_0=T/2$, $T=L$ and $\theta=0$),
\begin{equation}
 \gp^{uu'}=0,\qquad \lv^{ud}=0 \qquad \textrm{and} \qquad \lt^{uu'}=0\,,
\label{eq:zf_more_cond}
\end{equation}
and the same solution for $\mcr^{(1)}$ as before. In each case we obtained an
asymptotic  value of $z_{f}^{(1)}$ consistent with those in Table \ref{tab:mc1_zf1}. 

\begin{figure}[!t]
\centering
\includegraphics[clip=true,scale=0.57]{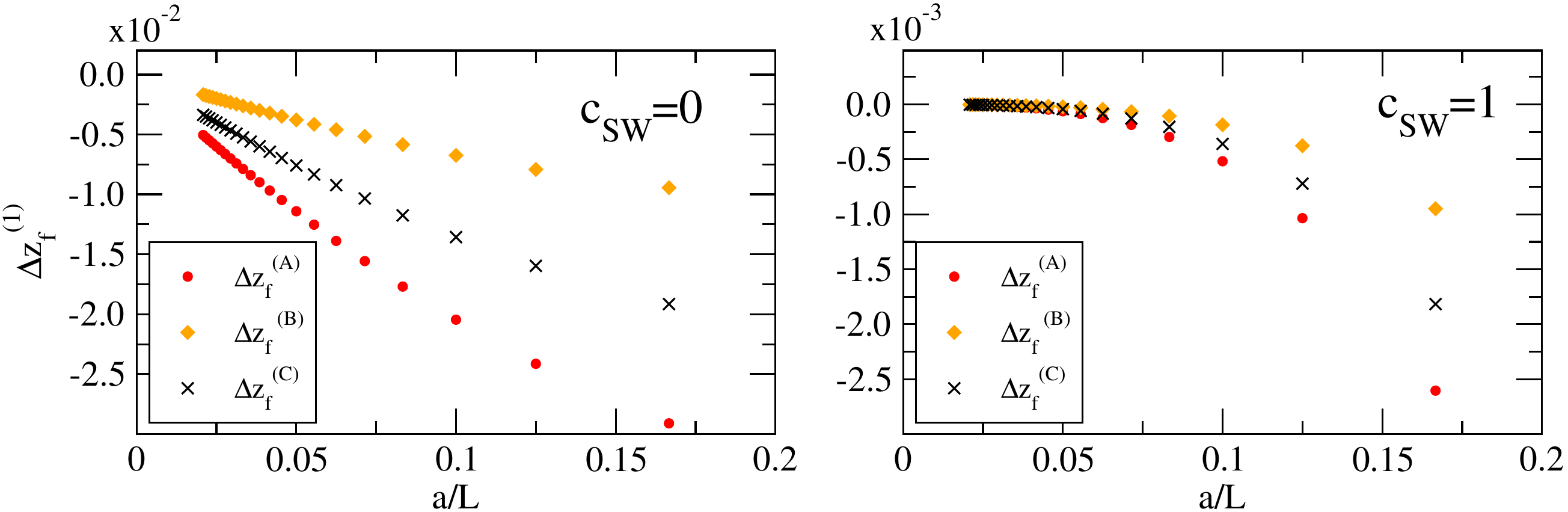}
\caption{Differences in the value of $z_{f}^{(1)}$ at finite lattice spacing obtained with the different tuning
conditions given in Eqs.~(\ref{eq:cond_zf_gA}) and (\ref{eq:zf_more_cond}) (all data for $\theta=0$ and $\Cf=4/3$).}
\label{fig:zf_conditions}
\end{figure}

Finally, we calculated the differences $\Delta z_{f}^{(1)}(a/L)$ at finite lattice spacing between 
$z_{f}^{(1)}(a/L)$ obtained using the condition (\ref{eq:cond_zf_gA}), and that obtained with 
the conditions (\ref{eq:zf_more_cond}), i.e.,
\begin{eqnarray}
        \Delta z_{f}^{(A)} &=& \left.z_{f}^{(1)}\right|_{\gp^{uu'}=0}-\left.z_{f}^{(1)}\right|_{\ga^{ud}=0},\nonumber \\
        \Delta z_{f}^{(B)} &=& \left.z_{f}^{(1)}\right|_{\lt^{uu'}=0}-\left.z_{f}^{(1)}\right|_{\ga^{ud}=0},\\
        \Delta z_{f}^{(C)} &=& \left.z_{f}^{(1)}\right|_{\lv^{ud}=0} -\left.z_{f}^{(1)}\right|_{\ga^{ud}=0}.\nonumber 
\label{eq:zf_diff}
\end{eqnarray}
These are displayed in Figure~\ref{fig:zf_conditions}. For $\theta = 0$ the only source of cutoff
effects in these differences comes from the bulk action, and is completely eliminated by the 
clover term.  Hence, for $\csw^{(0)}=1$ the differences $\Delta z_{f}^{(1)}$ behave as an O($a^2$)
effect, in contrast to $\csw^{(0)}=0$ for which they behave linearly in $a$, up to  possible 
logarithmic corrections.

\subsection{Determination of $d_{s}^{(1)}$}\label{5_ds}

The determination of the 1-loop boundary improvement coefficient $d_{s}^{(1)}$ can be obtained by 
requiring the absence of O($a$) effects at O($\gr^{2}$) in some $P_{5}$-even observable. 
Following a strategy similar to the one used in~\cite{Luscher:1996vw} 
for the extraction of the boundary improvement coefficient $\ctt^{(1)}$, 
we consider the ratio
\begin{equation}
R_{\rm P}(\theta,a/L) = 
\left.{\left[\gp^{ud}(x_0;\theta,a/L)\right]_{\rm R}\over\left[\gp^{ud}(x_{0};0,a/L)\right]_{\rm R}}\right|_{x_{0}=T/2} 
= R_{\rm P}^{(0)}(\theta,a/L)\left(1 + \gr^2\, r_{\rm P}^{(1)}(\theta,a/L) + \rmO(\gr^4)\right),
\label{eq:ds_ratio}
\end{equation}
which has a finite continuum limit, and the tree level ratio, $R_{\rm P}^{(0)}(\theta,a/L)$, is O($a$) improved.
The one-loop ratio $r_{\rm P}^{(1)}(\theta,a/L)$ can then be expanded in $a/L$
\begin{equation}
\begin{split} 
r_{\rm P}^{(1)}(\theta,a/L) = & \left(\left.{\gp^{ud(1,a)}\over \gp^{ud(0)}}\right|_{\theta} - 
\left.{\gp^{ud(1,a)}\over \gp^{ud(0)}}\right|_{\theta=0} \right) 
+\mcr^{(1)}\left(\left.{g_{{\rm P};m_0}^{ud(0)}\over \gp^{ud(0)}}\right|_{\theta} -
\left.{g_{{\rm P};m_0}^{ud(0)}\over \gp^{ud(0)}}\right|_{\theta=0} \right) \\
 & + z_{f}^{(1)}\left(\left.{g_{{\rm P};z_f}^{ud(0)}\over \gp^{ud(0)}}\right|_{\theta} -
 \left.{g_{{\rm P};zf}^{ud(0)}\over \gp^{ud(0)}}\right|_{\theta=0} \right) +  d_{s}^{(1)}\left.{g_{{\rm P};d_s}^{ud(0)}\over \gp^{ud(0)}}\right|_{\theta}, \\
 = &\; r_{\rm P}^{(1)}(\theta,0) + \frac{a}{L}\left(r_1 +    d_{s}^{(1)}\frac{L}{a}\left.{g_{{\rm P};d_s}^{ud(0)}\over \gp^{ud(0)}}\right|_{\theta}\right) +   \rmO(a^{2}),
\label{eq:ds1}
\end{split}
\end{equation}
where the constant $r_1$ is the coefficient of the O($a$) effect in $r_{\rm P}^{(1)}(\theta,a/L)$ in the absence of the $d_s$-counterterm. Hence,
the condition that $r_{\rm P}^{(1)}(\theta,a/L)$ be O($a$) improved leads to the equation
\begin{equation}
    d_{s}^{(1)} = -  r_1 \times \left[\lim_{a/L\rightarrow 0}
    \left(\dfrac{L}{a}\left.{g_{{\rm P};d_s}^{ud(0)}\over\gp^{ud(0)}}\right|_{\theta}\right) \right]^{-1} \,.
\end{equation}
We have analysed the sequence of values for $L/a=6,8,\ldots,48$ with the blocking procedure of ref.~\cite{Bode:1999sm}.
Besides $\theta=0.5$ we have produced further data for the set of values $\theta=0.1,0.25,0.75$ and $1.0$.
In the case of the O($a$) improved data $\csw^{(0)}=1$ we also considered analogous ratios to Eq.~(\ref{eq:ds_ratio}) using
$l_{\rm A}^{ud}(x_0)$ and the boundary-to-boundary correlation functions $g_1^{ud}$ and $l_1^{ud}$.
Consistent numerical results were obtained and we quote
\begin{equation}
d_{s}^{(1)}=\left\{\begin{array}{cl}
 -0.0006(3)\times C_{\rm F},\qquad &\csw^{(0)}=1\,,\\
 -0.0184(5)\times C_{\rm F},\qquad &\csw^{(0)}=0\,. \\
\end{array} \right.
\label{eq:ds1_value}
\end{equation}
Note that this consistency indirectly verifies automatic O($a$) improvement,
as it demonstrates the irrelevance at O($a$) of both the counterterm proportional to $\bar d_{s}$ (which was omitted)
and of the SW-term in the case of the unimproved Wilson fermion data.

\subsection{Determination of $\ct^{(1)}$}
\label{sec:ct}

In order to obtain the complete set of $\chi$SF action parameters to order $g_0^2$,
we would also like to compute the one-loop coefficient,
\begin{equation}
  \ct^{(1)} = \ct^{(1,0)} + \Nf \ct^{(1,1)},
\end{equation}
for the lattice $\chi$SF regularization. However, $\ct$ multiplies a gluonic counterterm,
so that the fermionic correlation functions at one-loop order
are only sensitive to its tree-level value, $\ct^{(0)}=1$.
We thus consider a gluonic observable, the SF coupling, $\bar{g}^2(L)$, defined as
the response coefficient to a chromo-electric background field in ref.~\cite{Luscher:1992an}.
Expanding in the bare coupling,
\begin{equation}
  \bar{g}^2(L) = g_0^2  + p_1(L/a) g_0^4 + \rmO(g_0^6),
  \label{eq:gbarsq}
\end{equation}
the logarithmically divergent one-loop coefficient, $p_{1}(L/a)$, decomposes into a purely gluonic, and a
fermionic contribution,
\begin{equation}
 p_{1}(L/a) = p_{1,0}(L/a) + \Nf\, p_{1,1}(L/a).
 \label{eq:p1}
\end{equation}
For gauge groups SU(2) and SU(3) the gluonic coefficient $p_{1,0}$ was first computed
in~\cite{Luscher:1992an, Luscher:1993gh} and the fermionic
part,  $p_{1,1}$,  in ref.~\cite{Sint:1995ch}, for fermions in the fundamental representation
and with standard SF boundary conditions. Given the nature of these calculations with
a non-trivial gauge background field, it is not obvious how these results depend
on the number of colours, $N$, and the fermion representation.
This dependence has been worked out in ref.~\cite{Hietanen:2014lha} where the results are given
for general $N$ and SU($N$) group constants. In particular the gluonic coefficient,
first computed for SU(3) in ref.~\cite{Luscher:1993gh}, takes the form,
\begin{equation}
  \ct^{(1,0)}=-0.08900(5) = \left [-0.0316483(4)\times N + \dfrac{0.017852(13)}{N} \right]_{N=3},
\end{equation}
and is, to this order, independent of the fermion regularization.
The analysis of $p_{1,1}(L/a)$ nicely illustrates some of the main points of this paper and
is left to Section~\ref{sec:coupling}. We here just quote the result
of this analysis for fermions in the fundamental representation,
\begin{equation}
\ct^{(1,1)}=\left\{\begin{array}{cll}
-0.006610(5),\qquad& \text{$\chi$SF},  &\csw^{(0)}=0 ,\\
\phantom{-} 0.006890(5), \qquad& \text{$\chi$SF}, &\csw^{(0)}=1 ,\\
\phantom{-} 0.019141(2), \qquad& \phantom{\chi}\text{SF},       &\csw^{(0)}=1.\\
\end{array} \right.
\label{eq:ct11}
\end{equation}
The value for the standard SF is in perfect agreement with ref.~\cite{Sint:1995ch}.
According to ref.~\cite{Hietanen:2014lha}, for a general fermion representation $R$ these
numbers need to be scaled by $T(R)/T(F)$, where $T(R)$ refers to the normalisation of the trace of two (hermitian) 
SU($N$)-generators  in the representation $R$.\footnote{$T(R)=1/2$, $(N+2)/N$, $(N-2)/N$, and $N$ for 
the fundamental, symmetric, antisymmetric, and adjoint representations, respectively.}

\section{Perturbative tests}
\label{sec:PTtests}

Having determined the action parameters to O($g_0^2$) we may now 
test the theoretical expectations discussed in Sect.~\ref{sec:theory} to 
this order in perturbation theory. This section describes our tests of the boundary conditions,
the mechanism of automatic O($a$) improvement, the restoration of flavour symmetry
and a direct comparison between SF and $\chi$SF observables.

\subsection{Boundary conditions}

On the lattice boundary conditions are not so much imposed as implicitly encoded by
the structure of the action near the boundary. Testing whether the boundary
conditions are satisfied (up to cutoff effects) is therefore not trivial.
Considering the first ratios of Eq.~(\ref{eq:rgxlyminus}) at $x_0=T/2$,
we expand perturbatively,
\begin{equation}
 R_{\rm X,-}^{g,f_1f_2} = R_{\rm X,-}^{g,f_1f_2 (0)} + \gr^2 R_{\rm X,-}^{g,f_1f_2 (1)} + \rmO(\gr^{4}),
\label{eq:ratio_mp_pt}
\end{equation}
with the tree-level and one-loop terms given by
\begin{equation}
   R_{\rm X,-}^{g,f_1f_2 (0)}  =\dfrac{g_{{\rm X},-}^{f_1f_2(0)}}{\gx^{f_1f_2(0)}}\,,\qquad
   R_{\rm X,-}^{g,f_1f_2 (1)} = R_{\rm X,-}^{g,f_1f_2 (0)}\left\{
     \dfrac{g_{{\rm X},-}^{f_1f_2(1)}}{g_{{\rm X},-}^{f_1f_2(0)}} - \dfrac{\gx^{f_1f_2(1)}}{\gx^{f_1f_2(0)}}\right\}\,.
\label{eq:ratios_minusplus}
\end{equation}
Analogous expressions are obtained for other ratios in Eq.~(\ref{eq:rgxlyminus}),
and for the corresponding ratios of standard SF correlation functions.

\begin{figure}[!t]
\centering
\includegraphics[clip=true,scale=0.65]{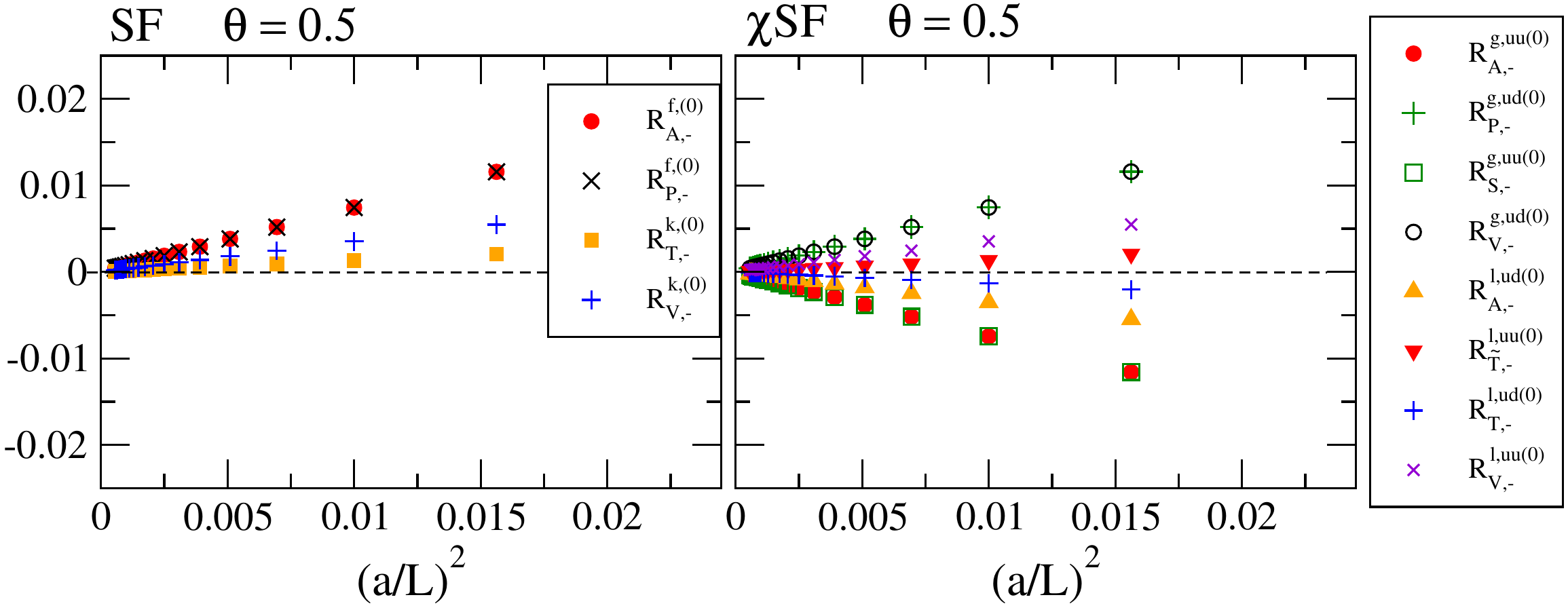}
\caption{Tree-level ratios (\ref{eq:ratios_minusplus}) between correlation functions defined with reverted projectors and
correct projectors, respectively. Ratios for both standard SF (left panel) and $\chi$SF (right panel) 
boundary conditions are shown for $\theta=0.5$.}
\label{fig:even_van_tree}
\end{figure}

Using these definitions we compute the tree-level and one-loop terms in (\ref{eq:ratio_mp_pt}) for all the
$P_5$-even boundary-to-bulk correlation functions, for $\csw^{(0)}=1$ and
for $\theta=0,0.5$, and their standard SF counterparts.
The tree-level ratios vanish exactly when $\theta=0$, both in the $\chi$SF and in the standard SF. 
For $\theta=0.5$ instead, the tree-level ratios are non-zero
at finite lattice spacing, and vanish at a rate of O($a^{2}$), cf.~Figure \ref{fig:even_van_tree}. 
We find that the size of the cutoff effects in both set-ups is comparable at tree-level. Note that the
tree-level correlators do not depend on $\csw^{(0)}$, due to our choice of trivial gauge background field.

\begin{figure}[t]
\centering
\includegraphics[clip=true,scale=0.58]{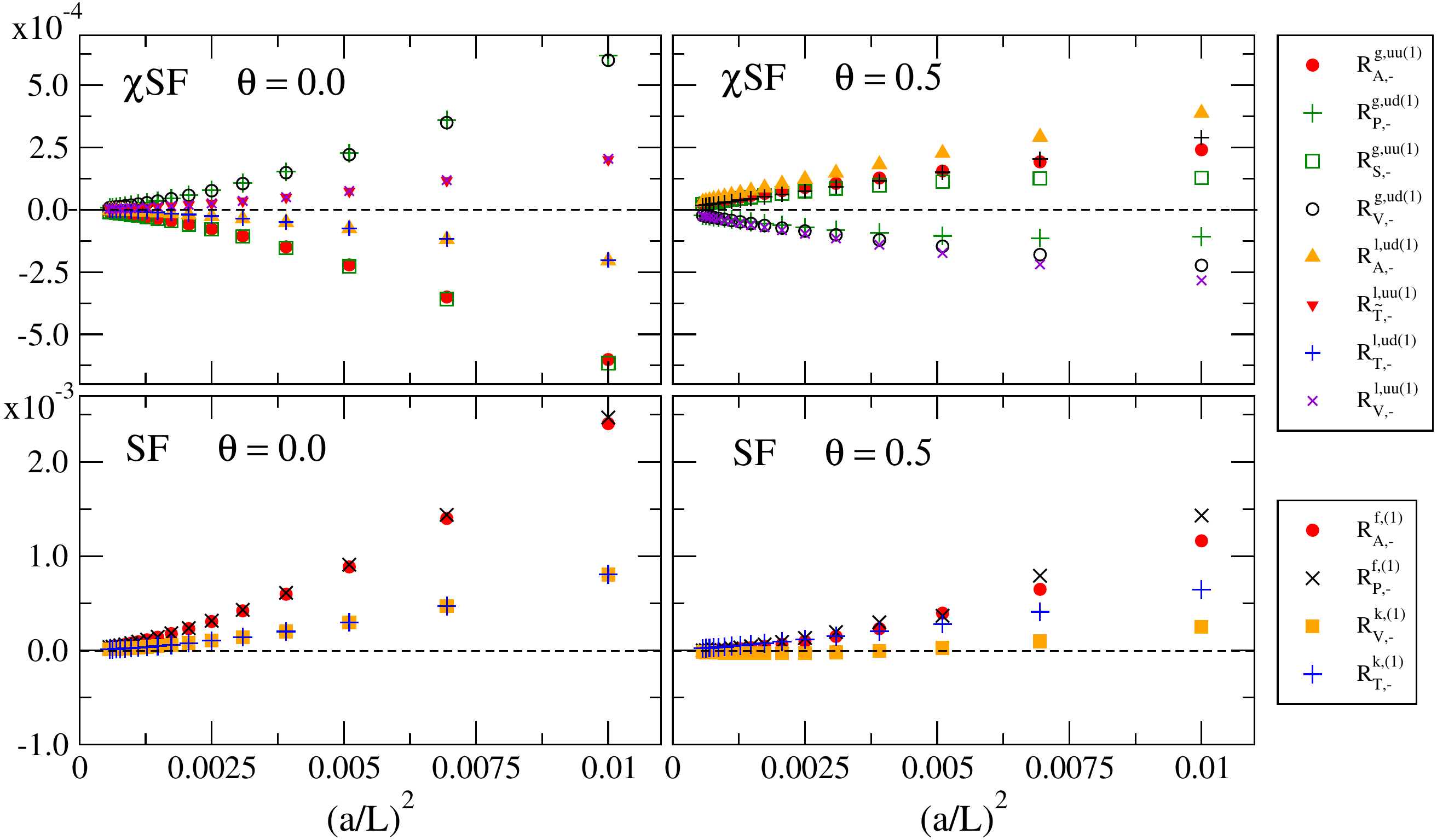}
\caption{One-loop ratios (\ref{eq:ratios_minusplus}) between correlation functions defined with reverted and correct projectors,
respectively, for $\csw=1$ and both $\theta=0$ and $0.5$. The factor $\Cf=4/3$ is included.
The ratios for the $\chi$SF are displayed in the  upper panels,
while those for the SF are shown in the lower panels (note the scale difference). }
\label{fig:even_van}
\end{figure}

In order to evaluate the same ratios at one-loop order, we insert the series $\mcr^{(1)}(a/L)$ and $z_f^{(1)}(a/L)$
obtained from $g^{ud}_{\rm A}$ at finite $L/a$ and for $\theta=0$.
The convergence to the continuum limit of the ratios is displayed in
Figure~\ref{fig:even_van}. We note that the ratios are very
small for the $\chi$SF already at the coarsest lattices,
both for $\theta=0$ and $0.5$. In the first case, cutoff effects are particularly suppressed,
and seem to approach zero faster than O($a^2$)
(top-left panel of Figure~\ref{fig:even_van}), whereas the data for $\theta=0.5$
shows the O($a^2$) continuum approach that one might have expected 
(top-right panel of Figure~\ref{fig:even_van}). For the standard SF the ratios at one-loop,
although still small, are an order of magnitude larger than their $\chi$SF counterparts (see bottom panels
of Figure~\ref{fig:even_van}).
In summary, we note that all the ratios considered approach zero in the continuum limit, at least with a rate
of O($a^2$). This confirms that the boundary conditions are correctly implemented to one-loop order
of perturbation theory.

\subsection{Automatic O($a$) improvement}
\label{sec:test_autoOa}

As explained in Subsection~\ref{subsect:automatic} we may test automatic
O($a$) improvement either by confirming the O($a^2$) continuum approach
of $P_5$-even observables, or by showing that the associated bulk O($a$) counterterm contributions,
or, more generally, $P_5$-odd correlations, are pure O($a$) effects.
Several examples of the former will appear below, where the absence of cutoff effects linear in $a$ is observed.
We here focus on the $P_5$-odd correlations functions, which are the ones
translating to $\fs,\fv$ or $\ka,\ktt$ according to our dictionary of Section~2.
Among those we omit the ones which vanish identically, Eq.~(\ref{eq:exact0}),
which leaves us with non-trivial tests of automatic O($a$) improvement to be performed for
\begin{equation}
  g_{\rm A}^{ud}, g_{\rm P}^{uu'}, l_{\rm V}^{ud}, l_{\rm T}^{uu'},
  \label{eq:P5odd}
\end{equation}
as well as the derivatives
\begin{equation}
  \tilde\partial_0 g_{\rm P}^{uu'}, \tilde\partial_0 l_{\rm V}^{ud}, \tilde\partial_0 l_{\rm T}^{uu'},
\end{equation}
which also appear as O($a$) counterterms to the $P_5$-even correlation functions
$g_{\rm A}^{uu'}$, $l_{\rm T}^{ud}$ and $l_{\rm V}^{uu'}$, respectively (cf.~Appendix~\ref{app:bilinear}).

\begin{figure}[!t]
\centering
\includegraphics[clip=true,scale=0.53]{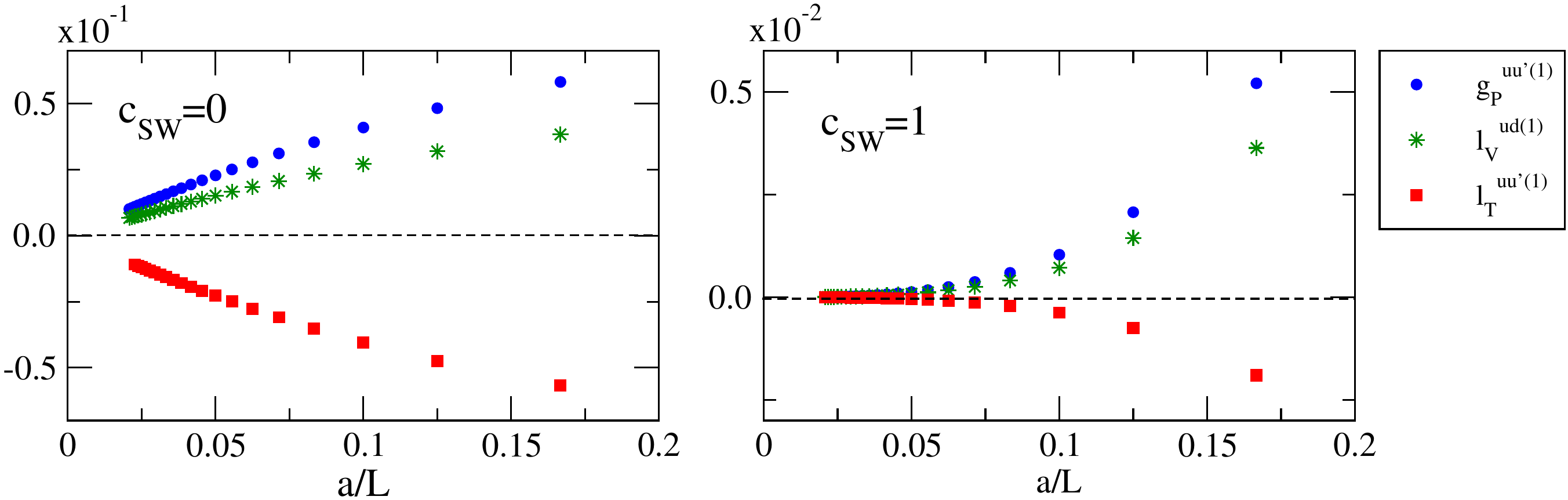}
\caption{Vanishing $P_{5}$-odd correlation functions at one-loop order, calculated for $\csw=0$ (left panel), $\csw=1$ (right panel) and
for $\theta=0$. The series for $\mcr^{(1)}(a/L)$ (\ref{eq:mc1_L}) and $z_f^{(1)}(a/L)$ (\ref{eq:zf1_L}) have been inserted and
$\Cf=4/3$ (note the scale difference between the panels).}
\label{fig:odd_van_1}
\end{figure}

We first choose data at $\theta=0$, set $x_0=T/2$ and insert the series Eqs.~(\ref{eq:mc1_L}),(\ref{eq:zf1_L})
for $\mcr^{(1)}$ and $z_f^{(1)}$. For $\theta=0$, all $P_5$-odd correlation functions
at tree-level vanish identically already at finite lattice spacing.
At one-loop order, we focus on the correlation functions in Eq.~({\ref{eq:P5odd}}), where
$g_{\rm A}^{ud}(T/2)=0$ holds by definition, as this is our tuning condition for $z_f$.
The remaining ones are shown in Figure~\ref{fig:odd_van_1}
for both $\csw^{(0)}=0$ (left panel) and $\csw^{(0)}=1$ (right panel).
While non-zero at finite lattice spacing, all these $P_5$-odd correlation functions do
indeed vanish in the continuum limit, as expected from automatic O($a$) improvement.
To understand the faster continuum approach in the case of $\csw^{(0)}=1$, we note
that with $\theta=0$ the counterterm insertions
$\propto \ca^{(1)},\cv^{(1)},\cT^{(1)}$  vanish,
\begin{equation}
   \tilde{\partial}_0 \gp^{(0)}(x_0)\vert_{\theta=0} =
   \tilde{\partial}_0 \lt^{(0)}(x_0)\vert_{\theta=0} =
   \tilde{\partial}_0 \lv^{(0)}(x_0)\vert_{\theta=0} = 0\,,
\end{equation}
and similarly the contributions $\propto \bar{d}_s^{(1)}$,
\begin{equation}
  g_{{\rm X};\bar{d}_s}^{(0)}(x_0)\vert_{\theta=0} = l_{{\rm Y};\bar{d}_s}^{(0)}(x_0)\vert_{\theta=0}=0.
\end{equation}
The same holds for the $d_s$-counterterm. However, both this and the $\ct$-counterterm
are $P_5$-even so that their contribution would anyway be at most an O($a^2$) effect.
Hence, the only relevant counterterm for O($a$) improvement of these observables
is the Sheikholeslami-Wohlert term and its inclusion thus changes the rate of the approach to the continuum limit
from O($a$) to O($a^2$). As an aside we remark that this observation could be used to determine $\csw^{(0)}$
and thus provides a perturbative example for the kind of O($a$) improvement conditions that can be obtained 
from the $\chi$SF.

\begin{figure}[!t]
\centering
\includegraphics[clip=true,scale=0.53]{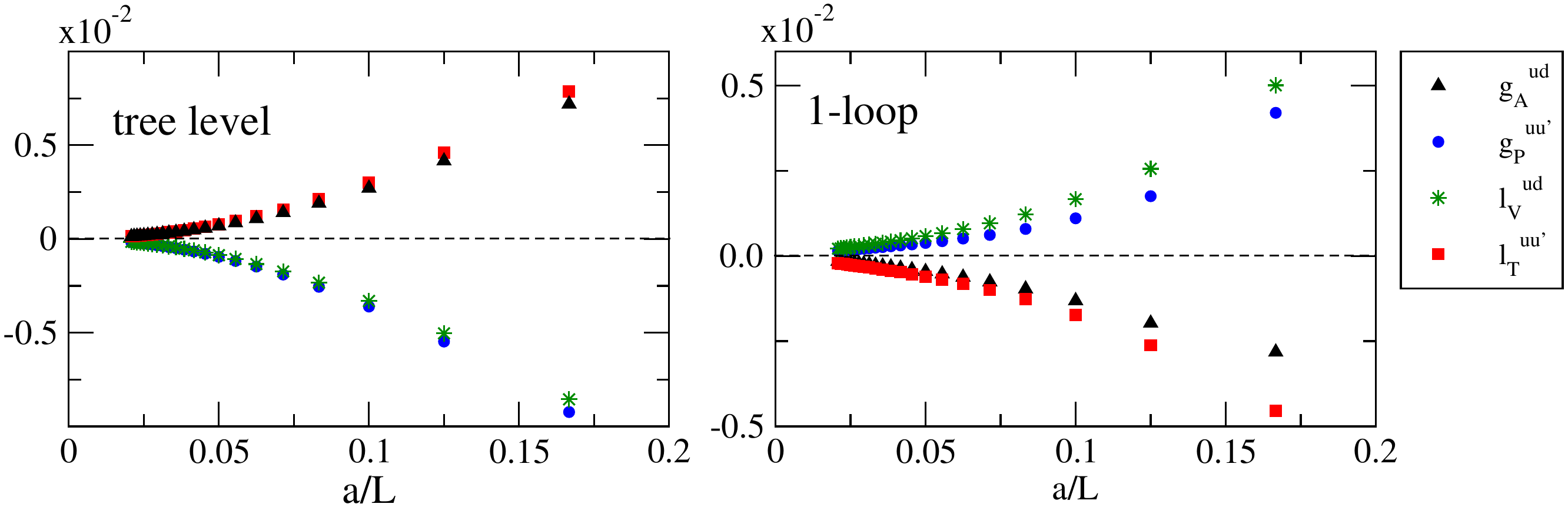}
\caption{Vanishing $P_{5}$-odd correlation functions for $\theta=0.5$
both at tree-level (left panel) and at one-loop order with $\csw^{(0)}=1$ (right panel).
The series for $\mcr^{(1)}(a/L)$ (\ref{eq:mc1_L}) and $z_f^{(1)}(a/L)$ (\ref{eq:zf1_L}) have been
used and the group factors have been set to $N=3$ and $\Cf=4/3$, respectively.}
\label{fig:odd_van_2}
\end{figure}

Passing to data for $\theta=0.5$ and $\csw^{(0)}=1$, the $P_5$-odd correlation functions are found to vanish in the continuum
limit, both at the tree- and one-loop level, with a rate of O($a$) as should be expected (cf.~Figure~\ref{fig:odd_van_2}).
In this case the vanishing of $g_{\rm A}^{ud}(T/2)$ is non-trivial as we obtain $z_f$ from data at $\theta=0$.
In conclusion, we confirm that $P_5$-odd observables are indeed pure lattice artefacts,
and confirm that automatic O($a$) improvement works out as theoretically expected.

\subsection{Flavour symmetry restoration}

\begin{figure}[!t]
\centering
\includegraphics[clip=true,scale=0.55]{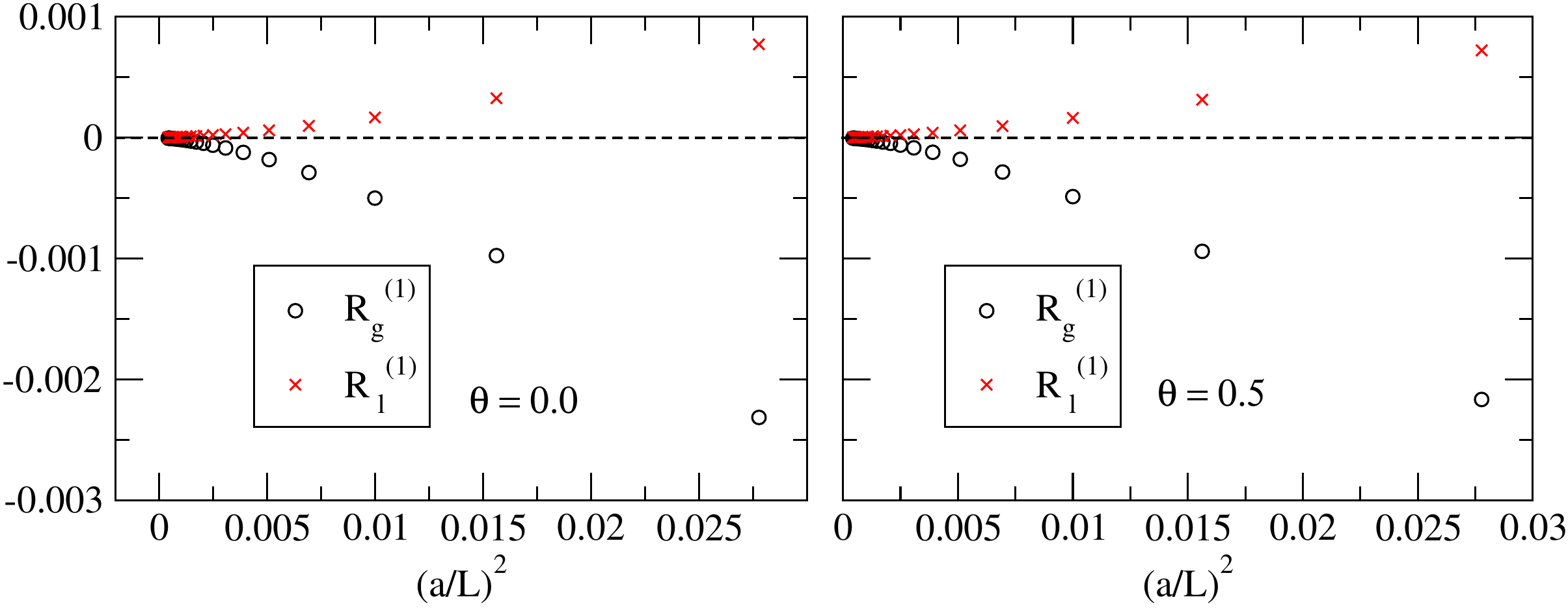}
\caption{One-loop ratios $R_{g}^{(1)}$ and $R_{l}^{(1)}$, Eq.~(\ref{eq:Rgl01}),
as a function of $(a/L)^{2}$ for $c_{\rm sw}^{(0)}=1$.
The factor $\Cf=4/3$ has been included.}
\label{fig:R_g_l}
\end{figure}

In order to check if flavour symmetry is restored in the continuum limit,
we consider the relations between boundary-to-boundary correlation functions with different flavour
content. Taking the ratios in Eq.~(\ref{eq:ratflav}) and expanding them to order $\gr^2$, 
\begin{equation}
    \label{eq:ratios}
    R_{g,l} = R_{g,l}^{(0)} + \gr^{2}\,R_{g,l}^{(1)} + {\rm O}(\gr^{4}),
\end{equation}
we should find that the tree-level coefficients,
\begin{equation}
 R_{g}^{(0)}={g_{1}^{uu'(0)}\over g_{1}^{ud(0)}},\qquad
 R_{l}^{(0)}={l_{1}^{uu'(0)}\over l_{1}^{ud(0)}},
\end{equation}
approach unity, whereas the one-loop coefficients,
\begin{equation}
 R_{g}^{(1)}=R_{g}^{(0)}\left\{{g_{1}^{uu'(1)}\over g_{1}^{uu'(0)}}-{g_{1}^{ud(1)}\over g_{1}^{ud(0)}}\right\},\qquad
 R_{l}^{(1)}=R_{l}^{(0)}\left\{{l_{1}^{uu'(1)}\over l_{1}^{uu'(0)}}-{l_{1}^{ud(1)}\over l_{1}^{ud(0)}}\right\},
 \label{eq:Rgl01}
\end{equation}
should vanish in the continuum limit.
Computing these coefficients for $\csw^{(0)}=1$ and $0$ and for $\theta=0$ and $0.5$, 
we find that the ratios at tree-level are exactly $R_{g}^{(0)}=R_{l}^{(0)}=1$
for all values of $L/a$ and independently of $\theta$. The one-loop coefficients $R_{g}^{(1)}$ and $R_{l}^{(1)}$ 
are non-zero at finite lattice spacing, but vanish as $a/L\rightarrow 0$, thus confirming 
the restoration of flavour symmetry. The counterterm insertions proportional to $d_s^{(1)}$
vanish exactly in this ratio rendering this counterterm irrelevant not only at O($a$)
(as expected from the discussion in~Subsect.~~\ref{subsect:flavour_restoration}) but
to all orders in $a$. Somewhat surprisingly, the same statement holds for the counterterm insertions
proportional to $\mcr^{(1)}$ and $z_f^{(1)}$, so that the choice of the critical mass or the
precise definition of $z_f$ become irrelevant, too. The results for the coefficients $R_{g}^{(1)}$ and $R_{l}^{(1)}$ are displayed
in Figure~\ref{fig:R_g_l} for $\csw^{(0)}=1$. The behaviour for both values of $\theta$ is very similar and the continuum limit
is approached at an even faster rate than the expected O($a^2$).

\subsection{Direct comparison SF vs.~$\chi$SF}
\label{5_univ}

As explained in Sect.~\ref{subsec:scaledep}
the bare fermionic boundary source fields being different represents an obstacle when directly
comparing fermionic correlation functions between the SF and $\chi$SF. We are thus led
to consider (double) ratios where the boundary source renormalization factors $Z_\zeta$ are cancelled
separately for SF and $\chi$SF observables, e.g.
\begin{equation}
 R_{\rm A}=\Bigg[{{[\ga^{uu'}]_{\rm R}}\over{\sqrt{[g_{1}^{uu'}]_{\rm R}}}}\Bigg]
 \times\Bigg[{{[\fa]_{R}}\over{\sqrt{[f_{1}]_{R}}}}\Bigg]^{-1},\qquad
R_{\rm P}=\Bigg[{{[\gp^{ud}]_{\rm R}}\over{\sqrt{[g_{1}^{ud}]_{\rm R}}}}\Bigg]
\times\Bigg[{{[\fp]_{\rm R}}\over{\sqrt{[f_{1}]_{\rm R}}}}\Bigg]^{-1}\,,
\label{eq:univ_pt}
\end{equation}
where we have suppressed the $x_0$-dependence.
Such ratios are expected to approach 1 in the continuum limit, and similar ratios could be obtained
from the $k$- and $l$-functions, with vector and tensor bilinears. In fact, up to a tree-level factor,
all these double ratios correspond to ratios between $Z$-factors defined in SF schemes, cf.~Eq.~(\ref{eq:RatioZp}).
Since the bare fermion bilinear operators and the bulk lattice regularization
here are taken to be the same for SF and $\chi$SF, the renormalization factors must be equal up to cutoff effects.
For these effects to be reduced to O($a^2$) full Symanzik improvement of the action and fields is
required on the SF side. Note that this requirement imposes the use of the improved action also for the $\chi$SF.
Furthermore, one needs to implement boundary O($a$) improvement for the $\chi$SF by tuning $d_s$ and $\ct$.
Automatic O($a$) improvement of the $\chi$SF then ensures that
the bulk O($a$) counterterms to the fields as well as the $P_5$-odd boundary
counterterm $\propto \bar{d}_s$ do not contribute at O($a$) and may be omitted.

\begin{figure}[!t]
\centering
\includegraphics[clip=true,scale=0.55]{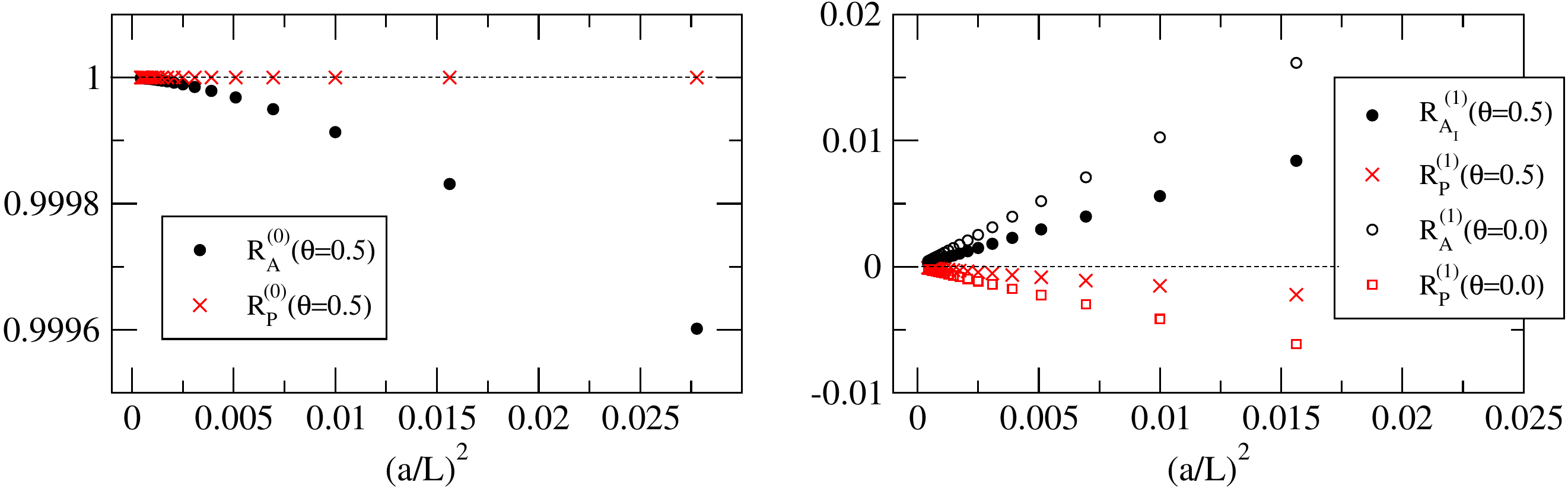}
\caption{Ratios $R_{\rm A}$ and $R_{\rm P}$ at tree-level (left panel) and 1-loop (right panel)
calculated for $\csw^{(0)}=1$ and $\Cf=4/3$. }
\label{fig:univ}
\end{figure}

To study the continuum approach for $R_{\rm A}$ and $R_{\rm P}$ to O($\gr^{2}$),
we expand the ratios in the coupling,
\begin{equation}
 R_{\rm X} = R_{\rm X}^{(0)}+\gr^{2}\,R_{\rm X}^{(1)}+{\rm O}(\gr^{4}),
\end{equation}
with the tree-level terms given by
\begin{equation}
 R_{\rm X}^{(0)}={\gx^{(0)}\over\sqrt{g_{1}^{(0)}}}\cdot{\sqrt{f_{1}^{(0)}}\over \fx^{(0)}},
\end{equation}
and the 1-loop terms,
\begin{equation}
   R_{\rm X}^{(1)}=R_{\rm X}^{(0)}\left\{{\gx^{(1)}\over \gx^{(0)}}-{\fx^{(1)}\over \fx^{(0)}}
  -{1\over2}\left({g_{1}^{(1)}\over g_{1}^{(0)}}-{f_{1}^{(1)}\over f_{1}^{(0)}}\right)
 \right\}.
\end{equation}
Looking at data for $x_0=T/2$ and $\theta=0$, the tree-level coefficients $R_{\rm A}^{(0)}$ and $R_{\rm P}^{(0)}$ are exactly 1
even at finite $L/a$. For $\theta\neq0$, $R_{\rm P}^{(0)}$ is still exactly 1, whereas
$R_{\rm A}^{(0)}$ shows a small deviation from 1 which apparently vanishes even faster than O($a^2$)
(see left panel of Figure \ref{fig:univ}). The one-loop terms $R_{\rm A}^{(1)}$ and $R_{\rm P}^{(1)}$ calculated
at $\csw^{(0)}=1$, $\theta=0$ and $\theta=0.5$ are  displayed in the right panel of Figure \ref{fig:univ}.
Again we have inserted the finite $a/L$ estimates of $\mcr^{(1)}$ (\ref{eq:mc1_L}) and of $z_f^{(1)}$ (\ref{eq:zf1_L}).
Boundary O($a$) improvement by the $d_s$- and $\ctt$-counterterms, respectively, has been implemented.
Furthermore, for $\theta=0.5$, the correlation function $f^{(1)}_{\rm A}$ receives 
a contribution from the operator improvement counterterm proportional to $\ca$, which vanishes for 
$\theta=0$. We thus also consider $f^{(1)}_{\rm A}$  with the improved axial current $A_{\rm I}$
and label the corresponding ratio of correlation functions as $R_{\rm A_{I}}^{(1)}$.
In all cases considered, the one-loop ratios $R_{\rm X}^{(1)}$ converge to 0, thus confirming the expectation of universality.
Furthermore, the convergence rate is found to be O$(a^2)$ provided O($a$) improvement is correctly implemented at the boundaries
and in the bulk for the action and the SF correlation functions. Again, this indirectly
confirms automatic O($a$) improvement, as the omitted $P_5$-odd counterterms $\propto \bar{d}_s$ and $\propto \ca$
on the $\chi$SF side are not required.

\section{Applications based on universality}
\label{sec:Ren_bil}
In this section we now assume universality and demonstrate
the determination of scale independent renormalization factors
like $\Za$ or $\Zv$, which are traditionally obtained from chiral
and flavour Ward identities, respectively.
We then take another look at SF schemes for the pseudo-scalar and
tensor densities, and study both the renormalization factors  and
the associated step-scaling functions.

\subsection{Scale-independent renormalization factors}\label{subsec:SI_ren}

We now consider the ratios of Subsect.~\ref{subsec:rat_finiteZ}, which should yield
the scale independent factors $\Za$ and $\Zv$
and the scale independent ratios $\Zp/\Zs$ and $\Zt/\Ztt$, up to cutoff effects
of order $a^2$. Taking for example $R^{g}_{\rm V\tilde{V}}$, Eq.~(\ref{eq:RgVVt}),
we write the perturbative expansion,
\begin{equation}
   R^{g}_{\rm V\widetilde{V}} = R^{g(0)}_{\rm V\widetilde{V}} + \gr^2  R^{g(1)}_{\rm V\widetilde{V}} +\rmO(\gr^4)\,.
\end{equation}
We set $x_0=T/2$ and $T=L$ and then expect the tree-level term to approach unity
with O($a^2$) corrections and we find this
is indeed the case. Focusing on the one-loop contribution, we simplify notation by
writing
\begin{equation}
   R^{g(1)}_{\rm V\widetilde{V}} = \Zv^{g(1)}(L/a),
\end{equation}
and similarly for the other estimators of Subsect.~\ref{subsec:rat_finiteZ}, including
those which yield ratios of $Z$-factors, e.g.
\begin{equation}
   R^{g(1)}_{\rm PS} = [\Zp/\Zs]^{(1)}(L/a),
\end{equation}
and the superscript $g$ or $l$ referring to the $\gx$ or $\ly$ correlation functions
is only used when a confusion is possible. Note that, besides the $P_5$-odd $\bar{d}_s$-counterterm, we
also omit the $d_s$-counterterm at one-loop order: for $\theta=0$ it vanishes exactly, however, in general
it is expected to be irrelevant for the O($a$) improvement of such ratios and will at most cause
additional O($a^2$) effects (cf. Section~4).
We have verified this expectation explicitly by studying the combination of $d_s$-counterterm insertions
entering the one-loop $Z$-factors. In the case of the vector current normalization constants
this combination is even found to  vanish exactly.

Following Symanzik's analysis of cutoff effects, one expects that the asymptotic behaviour
for $a/L\rightarrow0$ is described by,
\begin{equation}
  \Zx^{(1)}(L/a)\sim\sum_{n=0}^{\infty}\left[r_{{\rm X},n}+s_{{\rm X},n}\ln(L/a)\right](a/L)^{n}.
  \label{eq:ZX_1_as_expansion}
\end{equation}
The coefficient $r_{\rm X,0}$ defines the finite asymptotic value $\Zx^{(1)}$.
For scale independent renormalization constants, the coefficient multiplying
the logarithmic divergence must be zero i.e.~$s_{\rm X,0}=0$. All
subsequent coefficients in Eq.~(\ref{eq:ZX_1_as_expansion})
describe the cutoff effects in $\Zx^{(1)}(L/a)$. The term linear in $a/L$ should be absent
according to the discussion in Subsect.~\ref{subsec:rat_finiteZ} regarding the boundary O($a$) effects.
The term proportional to $s_{{\rm X},1}$ is $0$ provided that O($a$) effects are absent in the bulk.

We obtain the first asymptotic coefficients in (\ref{eq:ZX_1_as_expansion}) following the blocking 
procedure described in~\cite{Bode:1999sm}. For all cases we confirm 
that the coefficients $s_{{\rm X},0},r_{{\rm X},1}$ and $s_{{\rm X},1}$
are compatible with zero up to at least 5 decimal digits. Assuming these to be zero in the subsequent
analysis, we can then easily extract the asymptotic values $r_{{\rm X},0}$.
The results  are collected in Table \ref{table:Z_1loop}.

\begin{table}[htb!]
\begin{center}
\begin{tabular}{|l|c|c|}
\hline
        &  $c_{\rm sw}=1$       &     $c_{\rm sw}=0$   \\
\hline 
$ \Za^{(1)} $  &  $  -0.116458(2) $  & $-0.133375(2)$   \\
$ \Zv^{(1)} $  &  $  -0.129430(2) $  & $-0.174085(2)$   \\
$  \left[\Zp/\Zs\right]^{(1)} $  &  $-0.025944(3)$  &  $-0.081420(3)$  \\
\hline
\end{tabular}
\caption{One-loop values of the scale-independent renormalization factors of fermion bilinears
for O($a$) improved and unimproved Wilson fermions in QCD ($\Cf=4/3$). Values for general $N$ can be obtained by
multiplying the quoted numbers by $(3/4)\times C_{\rm F}$.}
\label{table:Z_1loop}
\end{center}
\end{table}

Within the quoted errors the asymptotic values $\Zv^{(1)}$ and $\Za^{(1)}$ calculated using the $g$- and the $l$-functions are
in agreement with each other. We also found agreement with the
literature~\cite{Gabrielli:1990us,Martinelli:1982mw,Martinelli:1983be,Meyer:1983ds,Groot:1983ng}
for all renormalization factors, indicating that the method described in
Subsection~\ref{subsec:rat_finiteZ} for defining finite renormalization constants is well-founded.

\subsubsection{Lattice artefacts}

Next, we consider the cutoff effects in the finite renormalization factors to 
O($g_{0}^{2}$) in perturbation theory. At tree-level and one-loop order 
we define the difference between a given renormalization constant at finite lattice spacing
and its asymptotic value, i.e.,
\begin{equation}
 \delta Z^{(i)}_{\rm X}(L/a)=Z_{\rm X}^{(i)}(L/a)-Z_{\rm X}^{(i)},\quad i=0,1.
\end{equation}
In view of non-perturbative applications we will focus on the case of O($a$) improved Wilson
fermions and set $\csw^{(0)}=1$.

At tree-level, all renormalizaton constants are unity, $\Zx^{(0)}=1$. For the particular choice of $\theta=0$
this is also true  at finite lattice spacing, i.e.~$\Zx^{(0)}(L/a)=1$, and hence the cutoff
effects vanish exactly, $\delta\Zx^{(0)}(L/a)=0$ for all $a/L$.
For $\theta=0.5$, the tree-level cutoff effects $\delta\Zv^{g(0)}$ and $\delta\Za^{g(0)}$
are numerically around $0.01$ for $L/a=6$ and vanish at a rate $\propto a^2$.
In all other cases (including $\delta\Zv^{l(0)}$ and $\delta\Za^{l(0)}$) the cutoff effects are numerically much smaller
and also vanish at a higher rate than the expected O($a^2$).

\begin{figure}[!ht]
\centering
\includegraphics[clip=true,scale=0.54]{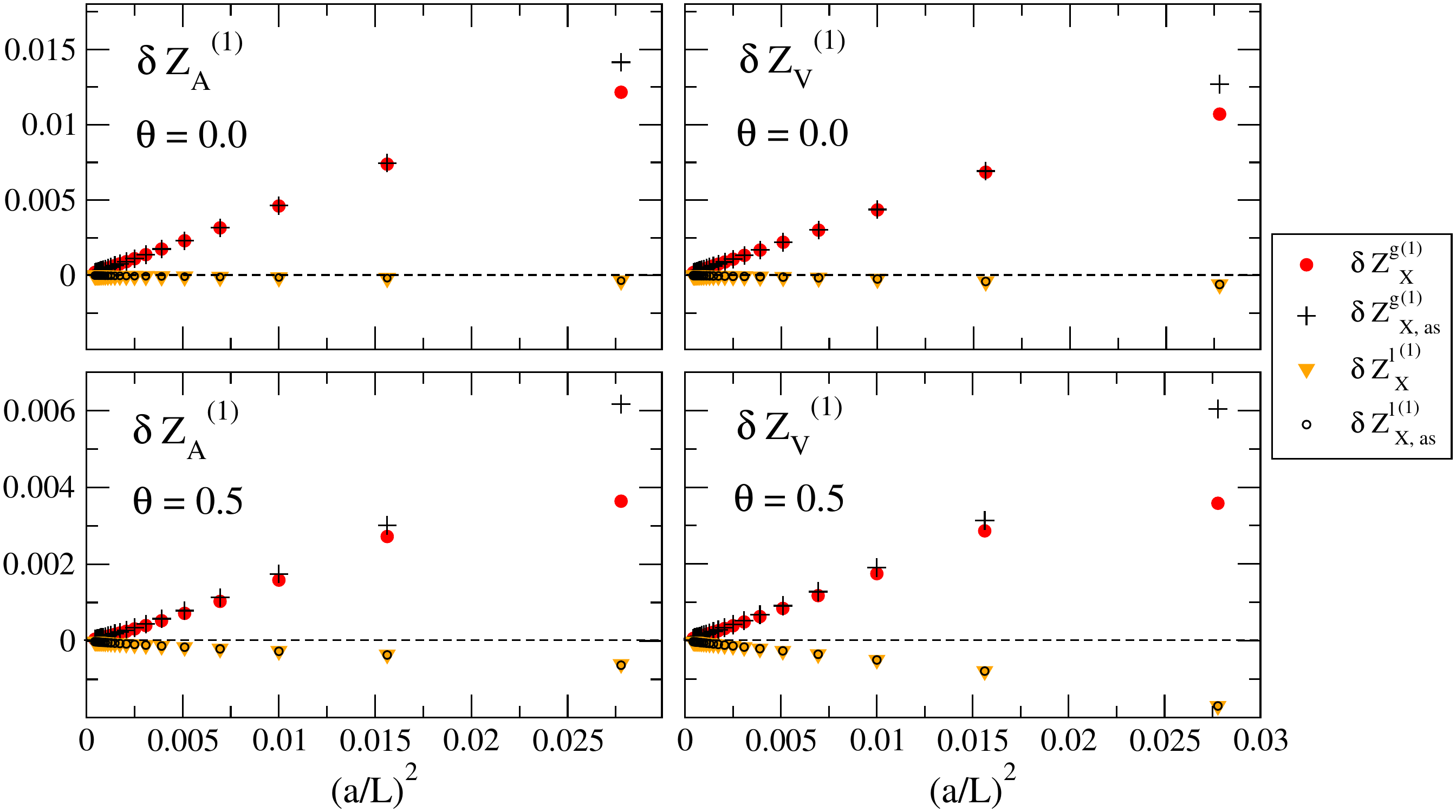}
\caption{Cutoff effects in $Z_{\rm V}^{(1)}(L/a)$ and $Z_{\rm A}^{(1)}(L/a)$ 
computed using the different definitions in Eqs.~(\ref{eq:ZAgdef})-(\ref{eq:ZlAV}), for $\csw^{(0)}=1$ and $\Cf=4/3$.}
\label{fig:delta_Z_1loop}
\end{figure}

The one-loop cutoff effects in $\Zv$ and $\Za$ are shown in Figure \ref{fig:delta_Z_1loop} for
$\theta=0$ and $0.5$. We study the cutoff effects obtained by using
the asymptotic values of $\mcr^{(1)}$ and $z_{f}^{(1)}$ (cf.~Table~\ref{tab:mc1_zf1}) in the expansions of $\Zv$ and $\Za$,
and also those obtained using the values $\mcr^{(1)}(a/L)$ and $z_{f}^{(1)}(a/L)$ at finite $L/a$ and for $\theta=0$
from Eqs.~(\ref{eq:mc1_L}),(\ref{eq:zf1_L}).
The latter are denoted $\delta \Zx^{(1)}$, whereas the former are labelled $\delta Z_{\rm X, as}^{(1)}$. 
The qualitative picture is similar to that observed at
tree-level\footnote{However, differently to the tree-level case, cutoff effects at one-loop are non-zero even
if $\theta=0$.}.  Cutoff effects associated to the definitions $\Zv^{l}$ and $\Za^{l}$ are always
very small even for the smallest lattices, in contrast to the definitions $\Zv^{g}$ and $\Za^{g}$ where we observe
considerably larger but still rather small effects. An interesting observation is that the insertion of the mass counterterm
causes an O(1) effect on $\Zv^{g}$ and $\Za^{g}$, whereas it is suppressed by a
further power of $a/L$ for $\Zv^{l}$ and $\Za^{l}$. The O(1) behaviour is expected
since the insertion of the $P_5$-odd mass counterterm into the $P_5$-even observables 
combines a power of $a/L$ with a linear divergence $\propto L/a$.
What comes as a surprise is the above mentioned additional O($a/L$) suppression, 
which is also seen for the ratio of tensor densities and
in the pseudo-scalar to scalar ratio. Similarly, regarding the $z_f$-counterterm 
we find that its insertion combines to an O($a^2$) effect in all cases,
except for the vector current where it vanishes exactly. Finally, we recall that the $d_s$-counterterm 
vanishes exactly at $\theta=0$, whereas
for $\theta=0.5$ its contributions are at least of O($a^2$)
and numerically insignificant in all cases, due also to the smallness
of $d_{s}^{(1)}$ [cf.~Eq.~(\ref{eq:ds1_value})]. Regarding $P_5$-odd counterterms, we
find no sign of an O($a$) contamination due to the omission of either the $\bar{d}_s$-counterterm or the 
bulk counterterms to the currents.
In conclusion, in all cases cutoff effects vanish proportionally to $(a/L)^{2}$, 
nicely confirming the theoretical expectations expressed in Section~4.


\subsection{Scale-dependent renormalization factors}

Here we compute to one-loop order in perturbation theory the scale-dependent
renormalization factors $\Zp$ and $\Zt$ in SF schemes, defined by the renormalization
conditions,
Eqs.~(\ref{eq:Zp}) and (\ref{eq:tensor_ren}). Again we focus on the O($a$) improved
action with $\csw^{(0)}=1$ and we first insert the the series Eqs.~(\ref{eq:mc1_L}),(\ref{eq:zf1_L})
for $\mcr^{(1)}$ and $z_f^{(1)}$. Expanding both $\Zp$ and $\Zt$ in the bare coupling,
\begin{equation}
  \Zx(g_0^2,L/a)=1+\sum_{k=1}^{\infty}\Zx^{(k)}(L/a)g_0^{2k},
 \qquad {\rm X=P,T},
\end{equation}
their one-loop coefficients, $\Zp^{(1)}(L/a)$ and $\Zt^{(1)}(L/a)$,
have an asymptotic expansion analogous to Eq.~(\ref{eq:ZX_1_as_expansion}), with the
finite parts and the coefficients of the logarithmic divergences given by 
\begin{align}
&r_{{\rm P},0} = z_{\rm P}^{(1)}(\theta),
\qquad s_{{\rm P},0} = - d_{0}= -{6 C_{\rm F}\over(4\pi)^{2}},\label{eq:sP0}\\
&r_{{\rm T},0} = z_{\rm T}^{(1)}(\theta),
\qquad s_{{\rm T},0} =  \gamma_{\rm T}^{(0)}= {2 C_{\rm F}\over(4\pi)^{2}}\label{eq:sT0}.
\end{align}
Here $-d_0$ and $\gamma_{\rm T}^{(0)}$ are the universal one-loop anomalous dimensions
of the pseudoscalar and tensor density, respectively.
One then expects the coefficients $r_{{\rm X},1}$ and $s_{{\rm X},1}$
to vanish provided that O($a$) lattice artefacts
are absent due to both boundary O($a$) improvement ($d_s^{(1)}$ and $\ct^{(0)}=1$),
and automatic O($a$) improvement.

We extract the first asymptotic coefficients in (\ref{eq:ZX_1_as_expansion}) 
for $\Zp^{(1)}(L/a)$ and $\Zt^{(1)}(L/a)$ in the way described in Subsect.~\ref{subsec:SI_ren}.
Note that we here omit the $d_s$ counterterm: its contribution vanishes in
all cases considered except for $\Zt^{(1)}(L/a)$ at $\theta=0.5$, where its contribution
is so small as to be below our resolution for the O($a$) coefficient
$r_{{\rm T},1}$ and can be safely neglected.
We then confirm that for all cases the coefficients $r_{{\rm X},1}$ and $s_{{\rm X},1}$
are compatible with zero to least 4 decimal digits. For the $\theta=0.5$ data and
to this level of precision we may therefore exclude contributions at O($a$)
from the omitted $\bar{d}_s$-counterterm, as well as from the bulk O($a$)
counterterm $\propto \cT$ in the case of the tensor density,
thereby providing further evidence for automatic O($a$) improvement.

The coefficients $s_{{\rm P},0}$ and $s_{{\rm T},0}$ agree with their theoretically expected values in
Eqs.~(\ref{eq:sP0}) and (\ref{eq:sT0})  to about 5 decimal digits. With
this confirmation we set these to their expected values
and proceed to extract the asymptotic coefficients $z_{\rm P}^{(1)}$
and $z_{\rm T}^{(1)}$, which we collect in Table \ref{table:Z_PT_1loop}. The values
of $z_{\rm P}^{(1)}$ and $z_{\rm T}^{(1)}$ obtained here are in perfect agreement 
with the results found in ref.~\cite{Sint:1998iq} and~\cite{Fritzsch:2015lvo}, 
respectively.

\begin{figure}[!ht]
\centering
\includegraphics[clip=true,scale=0.65]{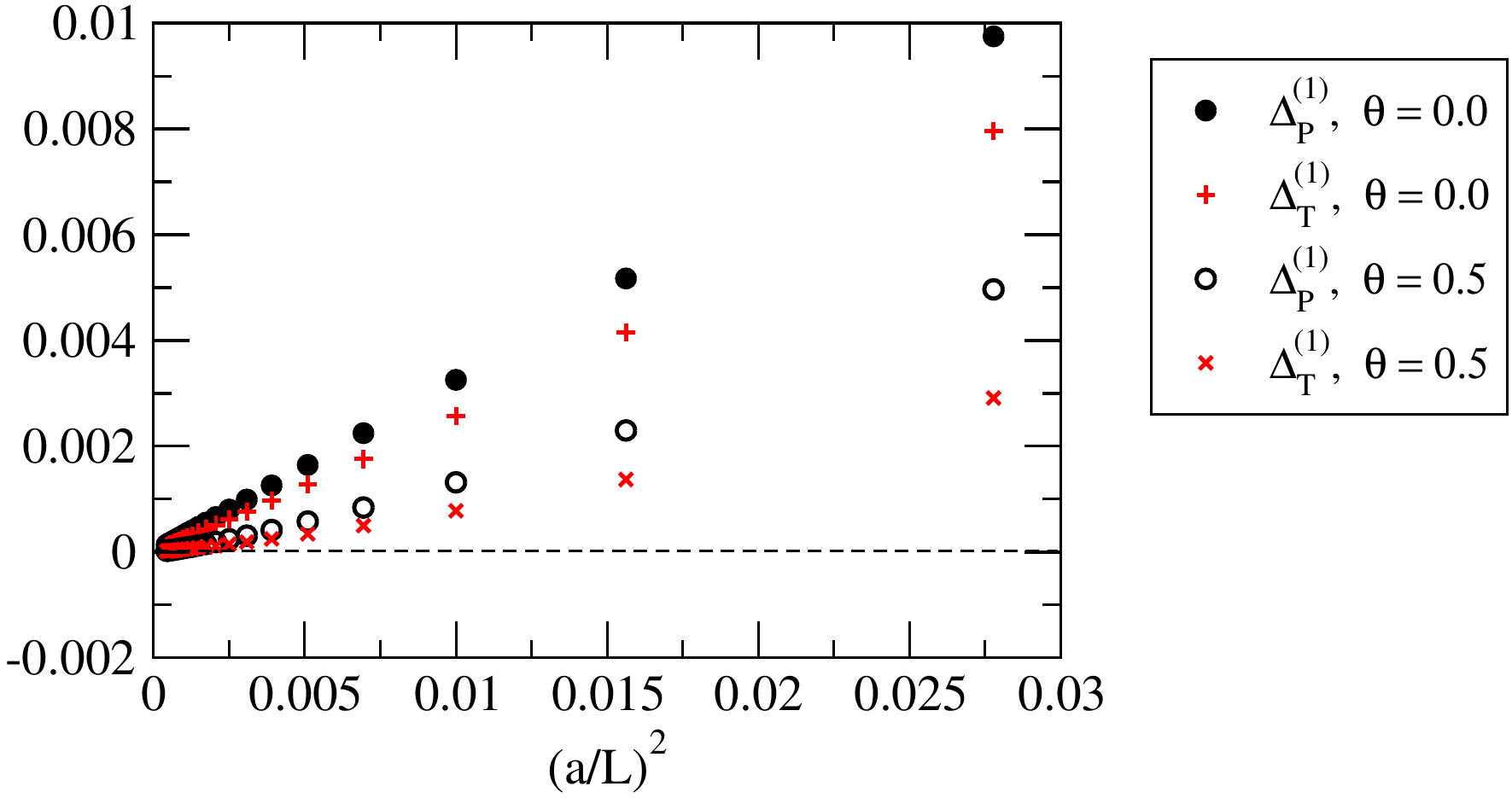}
\caption{Convergence to the continuum limit of the subtracted one-loop coefficients $\Delta_{\rm P}^{(1)}$ and
 $\Delta_{\rm T}^{(1)}$, Eq.~(\ref{eq:Delta_X}), with $\Cf=4/3$.}
\label{fig:ZP_ZT}
\end{figure}

To study the convergence to the continuum we define the subtracted one-loop renormalization constants
\begin{equation}
   \Delta_{\rm X}^{(1)} = Z_{\rm X}^{(1)}(L/a) - z_{\rm X}^{(1)}(\theta) - s_{\rm X,0} \ln(L/a)\,, \qquad
   \text{${\rm X=P,T}$},
 \label{eq:Delta_X}
\end{equation}
where we have now inserted the asymptotic values $\mcr^{(1)}$ and $z_f^{(1)}$ from table~\ref{tab:mc1_zf1}.
Figure~\ref{fig:ZP_ZT} clearly shows the O($a^2$) behaviour of the data, with cutoff effects
being largest for $\theta=0$.

\begin{table}[htb!]
\begin{center}
\begin{tabular}{|l|c|c|}
\hline
        &  $\theta = 0 $       &     $\theta=0.5$   \\
\hline 
$ z_{\rm P}^{(1)} $  &  $  -0.119542(1)\times \Cf $  & $-0.092815(1) \times \Cf$   \\
$ z_{\rm T}^{(1)} $  &  $  -0.019852(1)\times \Cf $  & $-0.06270(1) \times \Cf$   \\
\hline
\end{tabular}
\caption{One-loop results for the finite parts of the scale dependent renormalization 
factors $Z_{\rm P}$ and $Z_{\rm T}$, for $\csw^{(0)}=1$ and $\theta=0$ and $0.5$. }
\label{table:Z_PT_1loop}
\end{center}
\end{table}

\subsubsection{Lattice artefacts in the step scaling functions}

\begin{figure}[!ht]
\centering
\includegraphics[clip=true,scale=0.65]{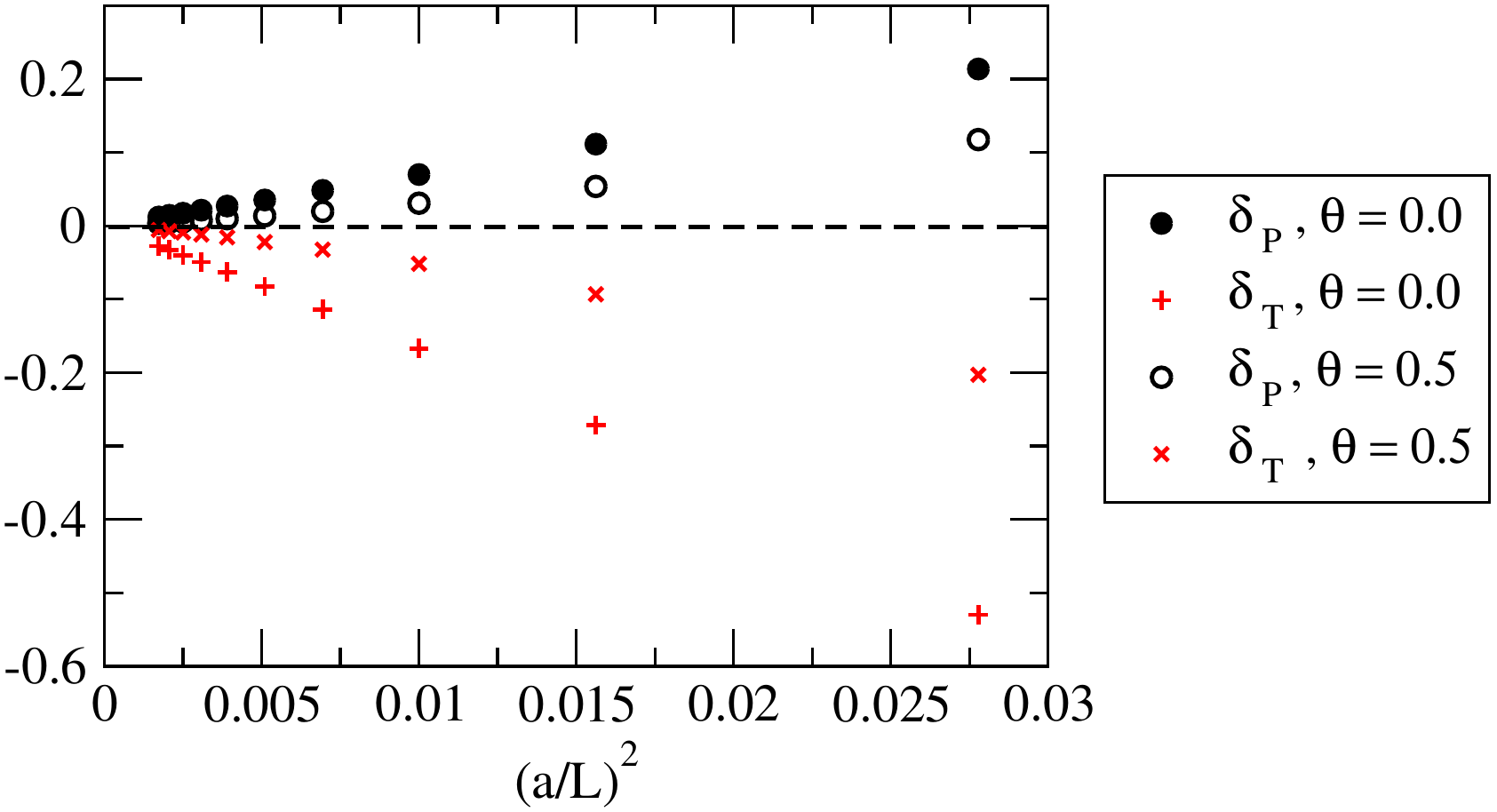}
\caption{One-loop cutoff effects in the step scaling functions $\Sigma_{\rm P}$ and 
$\Sigma_{\rm T}$, for $\theta=0$ and $0.5$.}
\label{fig:delta_k}
\end{figure}

For further illustration we look at the respective step-scaling functions
for $\Zp$ and $\Zt$ (cf.~Subsection~\ref{subsec:scaledep}),
\begin{equation}
   \Sigma_{\rm X}(u,a/L) = \left.\frac{\Zx(g_0^2,2L/a)}{\Zx(g_0^2,L/a)}\right\vert_{u=\bar{g}^2(L)}= 1 + k_{\rm X}(L/a)\times u 
   + {\rm O}(u^2),
\end{equation}
where
\begin{equation}
 k_{\rm X}(L/a) = Z_{\rm X}^{(1)}(2L/a) - Z_{\rm X}^{(1)}(L/a).
\end{equation}
Taking the continuum limit at order $u$ the preceding discussion of the respective $Z$-factors
implies the results $k_{\rm P}(\infty)=-d_{0}\ln(2)$ and $k_{\rm T}(\infty)=\gamma_{\rm T}^{(0)}\ln(2)$.
To study the approach to these continuum values we define the relative cutoff effects by
\begin{equation}
 \delta_{\rm P}(a/L) ={k_{\rm P}(L/a)\over k_{\rm P}(\infty)}-1,
 \qquad \textrm{and} \qquad
 \delta_{\rm T}(a/L) ={k_{\rm T}(L/a)\over k_{\rm T}(\infty)}-1.
\end{equation}
These coefficients are shown in figure \ref{fig:delta_k} for
$\theta = 0$ and $0.5$. Note that we have used the asymptotic values of $m_{\rm cr}^{(1)}$ and $z_f^{(1)}$,
and we have again omitted the vanishing or (in the case of the tensor density) numerically very small
$d_s$-counterterm contributions. In all cases the convergence to the continuum limit
is dominated by $(a/L)^{2}$ effects already at intermediate lattice sizes.
Lattice artefacts turn out to be smaller for $\theta =0.5$
than for $\theta = 0$. This difference is particularly pronounced for $\Sigma_{\rm T}$, for
which cutoff effects are quite large at $\theta =0$.
Note that a similar observation was made for the cutoff effects in $\Sigma_{\rm P}$
when calculated in the standard SF \cite{Sint:1998iq}. 

\section{The standard SF coupling and $\ct$ to one-loop order}
\label{sec:coupling}

We here consider the SF coupling as introduced in~\cite{Luscher:1992an}.
Apart from the calculation of the gluonic counterterm $\propto\ct$ to order $g_0^2$,
this provides yet another confirmation of universality and automatic
O($a$) improvement. With boundary O($a$) improvement in place
we also compare the residual lattice effects in the $\chi$SF
regularized step scaling functions to the standard SF.
In this section we restrict attention to lattice QCD i.e.~we assume $N=3$ and
fermions in the fundamental representation.

\subsection{Analysis of the fermionic one-loop coefficient $p_{1,1}(L/a)$}

Taking the expansion of the renormalized SF coupling in $g_0^2$, Eq.~(\ref{eq:gbarsq}),
as starting point, the fermionic coefficient $p_{1,1}(L/a)$ can be calculated
as in ref.~\cite{Sint:1995ch}, for a given lattice resolution $L/a$ using a recursive
evaluation of the determinant for fixed spatial momentum and colour component.
The necessary modifications due to $\chi$SF boundary conditions are described
in Appendix~\ref{app:p11}. We have written 2 independent FORTRAN codes
implementing both SF and $\chi$SF boundary conditions. Perfect agreement (up to rounding
errors) was found between both codes using double precision arithmetic.
One of the codes was then used to produce data for $p_{1,1}(L/a)$ 
in  quadruple (128 bit) precision arithmetic, for  $\theta=\pi/5$,
both for $\csw=0$ and $\csw=1$ and for a range of lattice sizes up to $L/a=64$.
We have used the asymptotic tree-level values for the
fermionic action parameters $\zf^{(0)}=1$ and $\mcr^{(0)}=0$ and $d_s^{(0)}=1/2$,
and $\ctt^{(0)}=1$. The gluonic action parameter is set to $\ct = 1 + g_0^2 \ct^{(1)}$,
with the fermionic contribution, $\ct^{(1,1)}$, as free parameter, to be determined by this calculation.
For $a/L\rightarrow 0$ one then expects the data to show the asymptotic behaviour,
\begin{equation}
   p_{1,1}(L/a)\sim\sum_{n=0}^{\infty}\left(r_{n}+s_{n}\ln(L/a)\right)(a/L)^{n}.
\end{equation}
The logarithmic divergence must be cancelled by the coupling renormalization,
implying that its coefficient, $s_0$, must be given in terms of the one-loop $\beta$-function.
Using the notation 
\begin{equation}
   b_0 = b_{0,0} + \Nf b_{0,1}, \qquad
   b_{0,0} =  \dfrac{11 N}{48 \pi^2},\qquad
   b_{0,1} = -\dfrac{1}{24\pi^2},
\end{equation}
one expects to find~\cite{Sint:1995ch}
\begin{equation}
  s_0 = 2 b_{0,1} = -\dfrac{1}{12\pi^2} \approx -0.008443431966\,.
\end{equation}
We extracted the asymptotic coefficients of $p_{1,1}$ from the numerical results following the method
described in~\cite{Bode:1999sm}. We first confirmed the expected value for $s_0$ for all
data sets with a relative precision better than  $1$ in $10^{4}$. Then we subtracted $s_0\ln(L/a)$
from the data using the analytically expected coefficient for $s_0$. This improves
the attainable precision for the analysis of the remaining coefficients.
The coefficient $r_0$ depends on the details of the chosen renormalization scheme for the SF coupling,
such as the choice of $\theta$, the aspect ratio $T/L$ or
the parameters of the background gauge field. Its value also depends on the regularization
through the bare coupling used in the expansion~(\ref{eq:gbarsq}).
This regularization dependence disappears once the bare coupling is replaced e.g.~by the $\msbar$ coupling
(cf.~\cite{Sint:1995ch}). For $r_0$ we find complete agreement with ref.~\cite{Sint:1995ch},
with comparable precision,
\begin{equation}
   r_0\bigr\vert_{\text{$\chi$SF},\,\theta=\pi/5,\,\csw^{(0)}=1} = -0.0346649(1),\qquad
   r_0\bigr\vert_{\text{$\chi$SF}\,,\theta=\pi/5,\,\csw^{(0)}=0} = -0.0098682(1),
\end{equation}
and similarly for data at $\theta=0$, thereby completely confirming the expectation regarding universality.

The coefficients $r_1$ and $s_1$ are relevant for O($a$) improvement. In particular, with the standard
SF, $s_1$ was found to vanish only for $\csw^{(0)}=1$, and is therefore related
to bulk O($a$) improvement. For the $\chi$SF we thus expect that automatic O($a$) improvement implies $s_1=0$, independently
of $\csw$. Indeed we find that for all our $\chi$SF data sets  $|s_1| < 10^{-4}$, thus confirming the expectation.

Finally, the coefficient $r_1$ is related to boundary O($a$) effects. From the $\theta=\pi/5$ data set with $\csw=1$,
we obtain
\begin{equation}
   r_1\bigr\vert_{\text{$\chi$SF},\,\theta=\pi/5,\,\csw^{(0)}=1} =  -2\ct^{(1,1)} + 0.01378(1).
  \label{eq:r1}
\end{equation}
Requiring the absence of O($a$) effects in the SF coupling at one-loop order
means $r_1=0$, and thus determines $\ct^{(1,1)}$. Note that this result must be
independent of $\theta$ or other kinematical parameters.
We have checked that the result~(\ref{eq:r1}) is reproduced within errors with data at $\theta=0$.

For the $\chi$SF data with $\csw=0$ the corresponding result is
\begin{equation}
   r_1\bigr\vert_{\text{$\chi$SF},\,\theta=\pi/5,\,\csw^{(0)}=0} =  -2\ct^{(1,1)} -0.01322(1)\,,
  \label{eq:r1csw0}
\end{equation}
independently of $\theta$. Note that this is in contrast to the standard SF where $r_1$ is found
to be $\theta$-dependent, indicating that boundary O($a$) improvement in the standard SF
cannot be achieved separately from bulk O($a$) improvement. As our data shows, with the $\chi$SF this is
indeed possible. More abstractly, this is due to the fact that
$P_5$-parity distinguishes the even O($a$) boundary counterterms ($\propto \ct,d_s$) from
the odd bulk O($a$) counterterm $\propto\csw$.

\subsection{Residual cutoff effects in the step-scaling function}

\begin{figure}[!t]
\centering
\includegraphics[clip=true,scale=0.6]{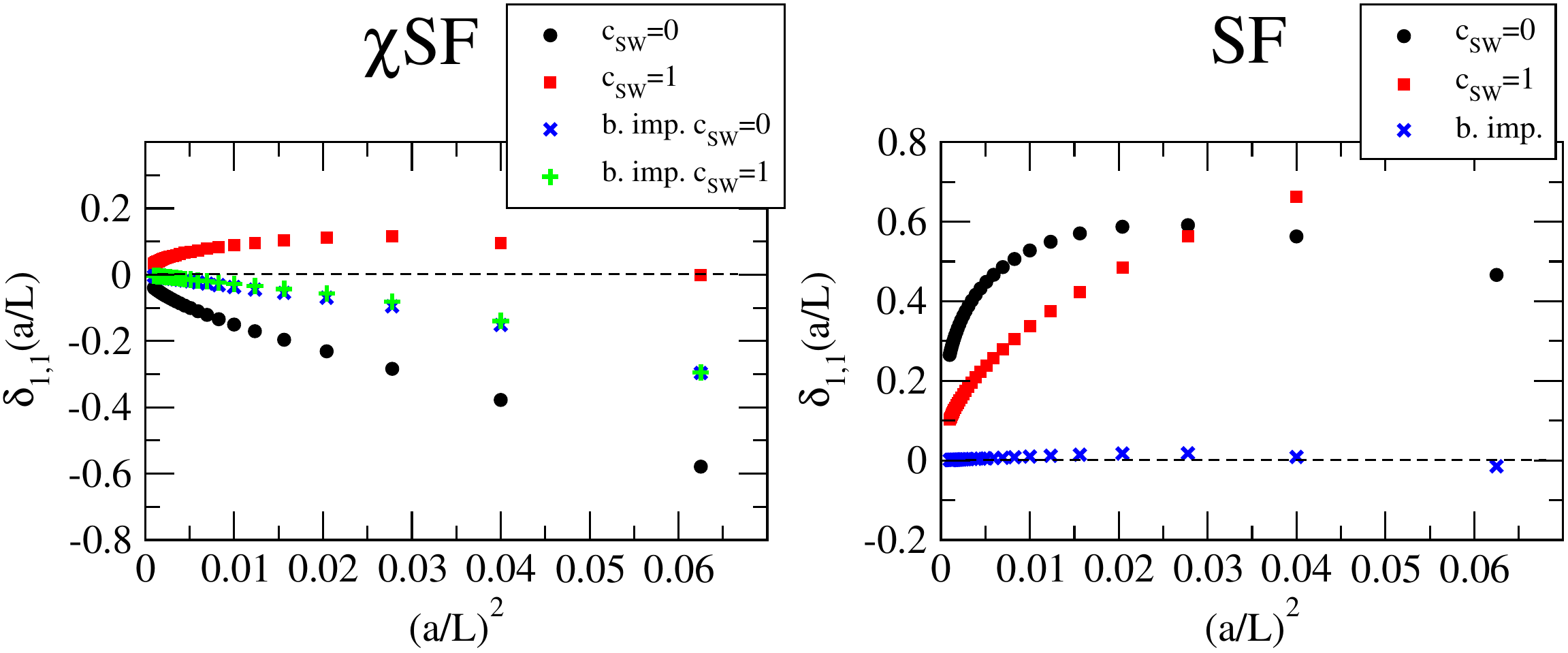}
\caption{Cutoff effects $\delta_{1,1}$, Eq.~(\ref{eq:del11}), for the $\chi$SF (left panel)
and the standard SF (right panel). For both the SF and $\chi$SF we show the results with and without
clover term. The legend ``b.~imp.'' refers to $\ct^{(1,1)}$ being set to the correct values, Eqs.~(\ref{eq:ct11}),
otherwise it is set to zero.}
\label{fig:su3_imp_XSF_SF}
\end{figure}

In non-perturbative applications the scale evolution of the SF coupling can be traced with the help
of the step scaling function (SSF)~\cite{Luscher:1991wu},
\begin{equation}
 \sigma(u)=\left.\overline{g}^{2}(2L)\right|_{u=\overline{g}^{2}(L)},
\label{eq:ssf_cont}
\end{equation}
which relates the value $u$ of the coupling $\overline{g}^{2}$ at a scale $L$ to its value at a scale $2L$. The
lattice version $\Sigma(u,L/a)$ of the step scaling function depends on the details of the regularization
and converges to (\ref{eq:ssf_cont}) in the continuum limit,
\begin{equation}
 \sigma(u)=\lim_{a/L\rightarrow 0}\Sigma(u,a/L).
\end{equation}
Both continuum and lattice versions of the SSF are expanded in perturbation theory as,
\begin{equation}
 \sigma(u)=u+\sigma_{1}u^{2}+{\rm O}(u^{3}),\qquad
 \Sigma(u,a/L)=u+\Sigma_{1}(a/L)u^{2}+{\rm O}(u^{3}),
\end{equation}
with the 1-loop terms given by
\begin{equation}
 \sigma_{1}=2b_{0}\ln(2),\qquad
\Sigma_{1}(a/L)=p_{1}(2L/a)-p_{1}(L/a).
\end{equation}
We would like to monitor the size of the lattice artefacts in the fermionic contribution to the SSF.
Isolating the part $\propto \Nf$,
\begin{equation}
  \Sigma_{1}(L/a) = \Sigma_{1,0}(L/a) +\Nf \Sigma_{1,1}(L/a),
\end{equation}
and analogously for $\sigma_1$, their relative difference,
\begin{equation}
 \delta_{1,1}(a/L)= \dfrac{\Sigma_{1,1}(L/a)-\sigma_{1,1}}{\sigma_{1,1}},
 \label{eq:del11}
\end{equation}
is shown in Figure \ref{fig:su3_imp_XSF_SF} for different levels of improvement.
For the $\chi$SF (Figure \ref{fig:su3_imp_XSF_SF}, left panel), the cutoff effects are asymptotically
O($a^{2}$) once $\ct^{(1,1)}$ is fixed to the correct value (\ref{eq:ct11}).
Note that boundary O($a$) effects are very different between $\csw=0$ or $1$. Somewhat surprisingly,
once these are removed by including the respective values for $\ct^{(1,1)}$, the remaining cutoff effects are quite
similar for $\csw=0$ and $\csw=1$.
For the standard SF (Figure \ref{fig:su3_imp_XSF_SF}, right panel), cutoff effects are essentially zero
after O($a$) improvement is implemented in the bulk and at the boundaries. This smallness of the remaining
cutoff effects seems to be an accident for this particular choice of background
field and kinematical parameters. 

\section{Conclusions}
In this paper we have defined a complete set of boundary-to-bulk and boundary-to-boundary  
correlation functions with both $\chi$SF and standard SF boundary conditions. 
Universality allows to establish a dictionary between both sets which should be applicable 
to appropriately renormalized correlation functions. We have discussed
renormalization and Symanzik O($a$) improvement in terms of these correlation functions. 
We have then formulated a few theoretical expectations, from the restoration of
$\chi$SF boundary conditions, flavour and parity symmetry, to automatic O($a$) improvement,
all of which follow from the assumption of a universal continuum limit. 
We have thus provided the framework for applications and checks of the 
$\chi$SF both in perturbation theory and beyond. 

We have then carried out the perturbative expansion in order to test 
the theoretical expectations to one-loop order. Based on numerical data for a range 
of lattice sizes from $L/a=6$ to $L/a=48$ (for both
SF and $\chi$SF with and without the SW-term), we have first calculated the action counterterm coefficients 
$\mcr$, $z_f$ and $\ct,d_s$ to order $g_0^2$.
The critical mass $\mcr$ and the renormalization constant  $z_{f}$ are required to restore
physical parity and flavour symmetries which are broken at finite lattice spacing.
Their determination is thus a pre-condition for any further tests regarding the 
continuum limit. The counterterms with coefficients $\ct,d_s$ remove O($a$) effects 
originating from the time boundaries (analogous to $\ct,\ctt$ in the standard SF).

Having determined the action to this order we have performed the following tests:
first, we have confirmed that the correct boundary conditions are implemented on the lattice. 
This was done by reversing the projectors in the boundary sources such as to project on 
the expected Dirichlet components of the fermionic boundary fields. 
The modified correlation functions were then seen to vanish in the continuum limit,
with O($a^2$) corrections. For comparison we also looked at the corresponding SF correlation functions, 
where comparable if larger cutoff effects are observed. Secondly, we have verified 
that flavour symmetry is restored in the 
continuum limit. This has been done by checking that ratios of boundary-to-boundary correlation functions
with different flavour content converge to unity, such that the continuum relations (\ref{eq:dictf1}),(\ref{eq:dictk1})
are satisfied. We then studied ratios of boundary-to-bulk correlation functions which should also approach unity, provided
the fermion bilinear operators in the bulk are correctly renormalized. This was confirmed and reproduced
a number of results from the literature for ratios of fermion bilinear renormalization constants.
Next, we have confirmed the universality between the SF and $\chi$SF set-ups by comparing 
renormalization constants for the pseudoscalar and tensor densities in SF schemes.
Finally, we have checked that the mechanism of automatic O($a$) improvement 
works as expected. This was done directly, by observing that a set of $P_{5}$-odd correlation functions 
vanish with a rate of O($a$), and indirectly by observing the absence of O($a$) terms
in $P_5$-even observables, the cancellation of which would require the O($a$) bulk counterterms.
In summary, the perturbative study fully confirms all theoretical expectations and lends further
support to the $\chi$SF framework.

With the $\chi$SF firmly established as a new tool, we would like
to give a short outlook on current and future applications. With automatic O($a$) improvement in place, 
any bulk O($a$) effect in physical observables vanishes without the need to tune 
either the $\csw$ coefficient in the action or any of the operator improvement coefficients. 
This last property is particularly appealing when studying the renormalization of complicated operators 
such as 4-fermion or higher-twist operators, where the non-perturbative determination 
of improvement coefficients is difficult or impractical. A project to determine the step-scaling
functions for a complete set of 4-quark operators in lattice QCD is currently 
in progress~\cite{Brida:2015b,Brida:2015c}.
In this context we remark that,  in practice, it seems advantageous to include the clover term in the action, 
as it drastically reduces the O($a$) ambiguity in the critical mass, 
even if the axial current in the PCAC relation remains unimproved.
This in turn renders the tuning of $z_f$ easier and higher order cutoff effects seem strongly reduced,
even though the qualitative asymptotic behaviour is expected to remain unchanged. 
This feature has been observed before in the quenched approximation~\cite{Sint:2010xy} and 
is now confirmed by our perturbative study.

In forthcoming non-perturbative studies~\cite{nosotros1:2015,nosotros2:2015} we will present 
further non-perturbative tests of the $\chi$SF and, 
in particular, results for the non-singlet current normalization constants, 
$\Za$ and $\Zv$ (for preliminary results in $\Nf=2$ lattice
QCD cf.~\cite{Brida:2014zwa}).
In this context, perturbation theory allows us to make an informed choice of the parameters 
and to perturbatively eliminate cutoff effects from the numerical simulation data. 

As a further promising application of the $\chi$SF we envisage
the determination of the bulk O($a$) improvement coefficients $\csw$, $\ca,\cv,\cT$. 
Convenient improvement conditions can be obtained by requiring some $P_5$-odd observables
to vanish exactly (besides the one used to determine $z_f$). 
A systematic investigation along these lines both in perturbation theory and
non-perturbatively is left to future work.

\section*{Acknowledgments}
The authors acknowledge initial support by the Research Executive
Agency (REA) of the European Union under Grant Agreement number PITN-GA-2009-238353
(ITN STRONGnet). M.D.B.~has been partially supported by the Irish Research Council
through the award of an "embark" scholarship. Research by S.S. is funded 
by SFI under grant 11/RFP/PHY3218. The numerical data have been produced
using computer resources at the Trinity Centre for High Performance Computing and the 
Irish Centre for High End Computing. The authors are grateful to both
centres for their support.

\appendix

\newcommand\numberthis{\addtocounter{equation}{1}\tag{\theequation}}

\section{Fermion bilinears}
\label{app:bilinear}

We refer to appendix A of ref.~\cite{Luscher:1996sc} for our conventions on the Euclidean
$\gamma$-matrices. An over-complete set of fermion bilinear operators is then given by
\begin{align*}
 V_{\mu}^{f_{1}f_{2}}(x)&=\overline{\psi}_{f_{1}}(x)\gamma_{\mu}\psi_{f_{2}}(x) ,
 & A_{\mu}^{f_{1}f_{2}}(x)&=\overline{\psi}_{f_{1}}(x)\gamma_{\mu}\gamma_{5}\psi_{f_{2}}(x), \\
  S^{f_{1}f_{2}}(x)&=\overline{\psi}_{f_{1}}(x)\psi_{f_{2}}(x) ,
 & P^{f_{1}f_{2}}(x)&=\overline{\psi}_{f_{1}}(x)\gamma_{5}\psi_{f_{2}}(x),\\
  T_{\mu\nu}^{f_{1}f_{2}}(x)&=i\overline{\psi}_{f_{1}}(x)\sigma_{\mu\nu}\psi_{f_{2}}(x) ,
 & \widetilde{T}_{\mu\nu}^{f_{1}f_{2}}(x)&=i\overline{\psi}_{f_{1}}(x)\gamma_{5}\sigma_{\mu\nu}\psi_{f_{2}}(x),
 \numberthis \label{eq:bilinear_list}
\end{align*}
where explicit flavour indices are used instead of the usual labeling through
the generators of the flavour group. Over-completeness follows from the fact that
only 6 of the tensor densities $T_{\mu\nu}$ and $\widetilde{T}_{\mu\nu}$
are independent due to the identity,
\begin{equation}
   \gamma_5\sigma_{\mu\nu} = -\frac12\varepsilon_{\mu\nu\rho\sigma}\sigma_{\rho\sigma},
\end{equation}
with the totally antisymmetric $\varepsilon$-tensor normalized by $\varepsilon_{0123}=1$.

In order to achieve on-shell O($a$) improvement a single counterterm is
needed for the bilinear operators in (\ref{eq:bilinear_list}), so that,
\begin{equation}
 X_{\rm I}^{f_{1}f_{2}}(x) = X^{f_{1}f_{2}}(x) + ac_{\rm X}(g_0) \delta X^{f_{1}f_{2}}(x).
\end{equation}
Here $X^{f_{1}f_{2}}$ is any bilinear operator, while the corresponding O(a) counterterms 
$\delta X^{f_{1}f_{2}}(x)$ are given by,
\begin{align*}
 \delta V_{\mu}^{f_{1}f_{2}}(x)&=\tilde{\partial}_{\nu}T_{\mu\nu}^{f_{1}f_{2}}(x), &
 \delta A_{\mu}^{f_{1}f_{2}}(x)&=\tilde{\partial}_{\mu}P^{f_{1}f_{2}}(x), \\
 \delta S^{f_{1}f_{2}}(x)&=0, & \delta P^{f_{1}f_{2}}(x)&=0, \\
 \delta T_{\mu\nu}^{f_{1}f_{2}}(x)&=\tilde{\partial}_{\mu}V_{\nu}^{f_{1}f_{2}}(x)-
 \tilde{\partial}_{\nu}V_{\mu}^{f_{1}f_{2}}(x), &
 \delta \widetilde{T}_{\mu\nu}^{f_{1}f_{2}}(x)&= -\varepsilon_{\mu\nu\rho\sigma}
 \tilde\partial_\rho V^{f_1f_2}_\sigma(x).
 \numberthis\label{eq:imp_bil_list}
\end{align*}
The coefficients multiplying the O($a$) counterterms are functions of the bare coupling $g_{0}$.
In perturbation theory these read:
\begin{equation}
   c_{\rm X}(g_0) = c^{(0)}_{\rm X} + c^{(1)}_{\rm X} g_0^2 + {\rm O}(g_0^4),
\end{equation}
where their tree-level values $c^{(0)}_{\rm X}$ are zero, while the 1-loop values are 
given by~\cite{Luscher:1996vw,Sint:1997jx,Sint:1997dj}
\begin{equation}
 c_{\rm V}^{(1)}=-0.01225(1)\times \Cf,\quad c_{\rm A}^{(1)}=-0.005680(2)\times \Cf,
 \quad c_{\rm T}^{(1)}=c_{\rm \widetilde{T}}^{(1)}=0.00896(1)\times \Cf.
\end{equation}
In the standard SF, the boundary bilinear operators are 
 \begin{align*}
 \mathcal{O}_{5}^{f_{1}f_{2}} &=a^{6}\sum_{\mathbf{y,z}}\overline{\zeta}_{f_{1}}(\mathbf{y})P_{+}\gamma_{5}\zeta_{f_{2}}(\mathbf{z}), &
 \mathcal{O}_{5}^{'f_{1}f_{2}} &=a^{6}\sum_{\mathbf{y,z}}\overline{\zeta}'_{f_{1}}(\mathbf{y})P_{-}\gamma_{5}\zeta'_{f_{2}}(\mathbf{z}),\\
 \mathcal{O}_{k}^{f_{1}f_{2}} &=a^{6}\sum_{\mathbf{y,z}}\overline{\zeta}_{f_{1}}(\mathbf{y})P_{+}\gamma_{k}\zeta_{f_{2}}(\mathbf{z}),&
 \mathcal{O}_{k}^{'f_{1}f_{2}} &=a^{6}\sum_{\mathbf{y,z}}\overline{\zeta}'_{f_{1}}(\mathbf{y})P_{-}\gamma_{k}\zeta'_{f_{2}}(\mathbf{z}),
 \numberthis \label{eq:bound_bil_SF}
 \end{align*}
where $\mathcal{O}_{5}$ and $\mathcal{O}_{k}$ are the bilinears at $x_{0}=0$, while $\mathcal{O}'_{5}$
and $\mathcal{O}'_{k}$ are the bilinears at $x_{0}=T$. 

Given these definitions, the boundary bilinear operators for the $\chi$SF depend on the flavour
structure and are given by
 \begin{align*}
 \mathcal{Q}_{5}^{uu'} &=a^{6}\sum_{\mathbf{y,z}}\overline{\zeta}_{u}(\mathbf{y})\gamma_{0}\gamma_{5}Q_{-}\zeta_{u'}(\mathbf{z}),&
 \mathcal{Q}_{k}^{uu'} &=a^{6}\sum_{\mathbf{y,z}}\overline{\zeta}_{u}(\mathbf{y})\gamma_{k}Q_{-}\zeta_{u'}(\mathbf{z}),  \\
 \mathcal{Q}_{5}^{dd'} &=a^{6}\sum_{\mathbf{y,z}}\overline{\zeta}_{d}(\mathbf{y})\gamma_{0}\gamma_{5}Q_{+}\zeta_{d'}(\mathbf{z}),&
 \mathcal{Q}_{k}^{dd'} &=a^{6}\sum_{\mathbf{y,z}}\overline{\zeta}_{d}(\mathbf{y})\gamma_{k}Q_{+}\zeta_{d'}(\mathbf{z}),  \\
 \mathcal{Q}_{5}^{ud\phantom{'}} &=a^{6}\sum_{\mathbf{y,z}}\overline{\zeta}_{u}(\mathbf{y})\gamma_{5}Q_{+}\zeta_{d}(\mathbf{z}),&
 \mathcal{Q}_{k}^{ud\phantom{'}} &=a^{6}\sum_{\mathbf{y,z}}\overline{\zeta}_{u}(\mathbf{y})\gamma_{0}\gamma_{k}Q_{+}\zeta_{d}(\mathbf{z}),  \\
 \mathcal{Q}_{5}^{du\phantom{'}} &=a^{6}\sum_{\mathbf{y,z}}\overline{\zeta}_{d}(\mathbf{y})\gamma_{5}Q_{-}\zeta_{u}(\mathbf{z}),&
 \mathcal{Q}_{k}^{du\phantom{'}} &=a^{6}\sum_{\mathbf{y,z}}\overline{\zeta}_{d}(\mathbf{y})\gamma_{0}\gamma_{k}Q_{-}\zeta_{u}(\mathbf{z}),
 \numberthis \label{eq:bound_bil_zero}
 \end{align*}
for the boundary at $x_{0}=0$, and
 \begin{align*}
  \mathcal{Q}_{5}'^{uu'} &=-a^{6}\sum_{\mathbf{y,z}}\overline{\zeta}'_{u}(\mathbf{y})\gamma_{0}\gamma_{5}Q_{+}\zeta'_{u'}(\mathbf{z}),&
  \mathcal{Q}_{k}'^{uu'} &=\phantom{-}a^{6}\sum_{\mathbf{y,z}}\overline{\zeta}'_{u}(\mathbf{y})\gamma_{k}Q_{+}\zeta'_{u'}(\mathbf{z}),  \\
  \mathcal{Q}_{5}'^{dd'} &=-a^{6}\sum_{\mathbf{y,z}}\overline{\zeta}'_{d}(\mathbf{y})\gamma_{0}\gamma_{5}Q_{-}\zeta'_{d'}(\mathbf{z}),&
  \mathcal{Q}_{k}'^{dd'} &=\phantom{-}a^{6}\sum_{\mathbf{y,z}}\overline{\zeta}'_{d}(\mathbf{y})\gamma_{k}Q_{-}\zeta'_{d'}(\mathbf{z}),  \\
  \mathcal{Q}_{5}'^{ud\phantom{'}} &=\phantom{-}a^{6}\sum_{\mathbf{y,z}}\overline{\zeta}'_{u}(\mathbf{y})\gamma_{5}Q_{-}\zeta'_{d}(\mathbf{z}),&
  \mathcal{Q}_{k}'^{ud\phantom{'}} &=-a^{6}\sum_{\mathbf{y,z}}\overline{\zeta}'_{u}(\mathbf{y})\gamma_{0}\gamma_{k}Q_{-}\zeta'_{d}(\mathbf{z}),  \\
  \mathcal{Q}_{5}'^{du\phantom{'}} &=\phantom{-}a^{6}\sum_{\mathbf{y,z}}\overline{\zeta}'_{d}(\mathbf{y})\gamma_{5}Q_{+}\zeta'_{u}(\mathbf{z}),&
  \mathcal{Q}_{k}'^{du\phantom{'}} &=-a^{6}\sum_{\mathbf{y,z}}\overline{\zeta}'_{d}(\mathbf{y})\gamma_{0}\gamma_{k}Q_{+}\zeta'_{u}(\mathbf{z}),
  \numberthis \label{eq:bound_bil_T}
 \end{align*}
for the boundary at $x_{0}=T$.

\section{One-loop contribution to the SF coupling from fermions in the $\chi$SF}
\label{app:p11}

We present a few details on the perturbative calculation of the coefficient $p_{1,1}(L/a)$
in Eqs.~(\ref{eq:gbarsq}) and (\ref{eq:p1}) with $\chi$SF boundary conditions.
The discussion follows very closely Appendix~A of ref.~\cite{Sint:1995ch},
where the coefficient $p_{1,1}(L/a)$ was calculated for the standard SF.
The reader will be assumed to be familiar with this
reference, as we will adopt much of the notation from there without further notice
(in particular we use lattice units $a=1$ and $t=x_0$ for Euclidean time).

For definiteness we assume a doublet with $\Nf=2$ flavours. One then has
\begin{equation}
   p_{1,1} = \frac{1}{2k} \left.\frac{\partial}{\partial\eta}
   \ln \det \left({\cal D}_W +\delta {\cal D}_W + m_0\right) \right\vert_{U_\mu(x) = V_\mu(x)}\,,
\label{eq:p11chiSF}
\end{equation}
where ${\cal D}_W + \delta {\cal D}_W$ is the $\chi$SF Dirac operator including the counterterms,
Eqs.~(\ref{eq:wilson_dirac_chiSF}),(\ref{eq:deltaDW}),
$V_\mu(x)$ denotes the Abelian background field which depends on the parameters
$\eta$ and $\nu$, which are set to zero after differentiation by $\eta$. Finally,  $k$ is the
tree-level normalization constant which ensures the correct normalization
of the SF coupling (cf.~\cite{Luscher:1993gh,Sint:1995ch}).
The large determinant in Eq.~(\ref{eq:p11chiSF}) can be reduced to subsectors
of fixed spatial momentum $\bfp$, colour $n_c$ and flavour $f$,
such that
\begin{equation}
 p_{1,1}(L/a)=\frac{1}{2k}\sum_{n_{c}=1}^{3}\sum_{f=u,d}\sum_{\bf{p}}
 \left.\frac{\partial}{\partial\eta}\ln\det {{\mathcal{D}}}^{(f)}(n_c,\bf{p})\right\vert_{\eta=\nu=0}.
\label{eq:p11_start}
\end{equation}
The flavour structure can be further reduced to the up-type determinant by recalling from ref.~\cite{Sint:2010eh}
that
\begin{equation}
   {\cal D}_W^{(u)} = \gamma_5 \left({\cal D}_W^{(d)}\right)^\dagger \gamma_5,
\end{equation}
so that their determinants are complex conjugate to each other. Moreover, we anticipate that both
determinants are real when taken in the Abelian background fields, so that we can omit the modulus
and obtain:
\begin{equation}
 p_{1,1}(L/a)=\frac{1}{k}\sum_{n_{c}=1}^{3}\sum_{\bf{p}}
 \left.\frac{\partial}{\partial\eta}\ln\det {\mathcal{D}}^{(u)}(n_c,\bf{p})\right\vert_{\eta=\nu=0}.
\label{eq:p11_start1}
\end{equation}
The task is thus reduced to many evaluations of (the $\eta$-derivative of) the determinant of ${\mathcal{D}}^{(u)}$
for fixed colour and spatial momentum, which corresponds to a matrix of size $4(T+1)\times4(T+1)$.
This is most efficiently done by setting up a recursion relation in Euclidean time, following refs.~\cite{Luscher:1992an,Sint:1995ch}.
The starting point is an eigenvalue equation for a hermitian operator, which requires us to temporarily remain
in 2-flavour space and consider:
\begin{equation}
   \left( \gamma_5\tau^1 {\mathcal{D}} -\mu\right) f(t) = 0\,.
   \label{eq:EVeq}
\end{equation}
The reduced operator $\mathcal{D} =\diag\left(\mathcal{D}^{(u)},\mathcal{D}^{(d)}\right)$
acts on eigenfunctions $f(t)$ as a finite difference operator in Euclidean time,
\begin{equation}
 \left(\mathcal{D} f\right)(t)=- P_{-}f(t+1) + h(t)f(t) - P_{+}f(t-1)\,,
 \label{eq:rec_allt}
\end{equation}
where we have extended the functions $f(t)$ beyond the interval $[0,T]$ by setting
\begin{equation}
 f(-1)=-i\gamma_{0}\gamma_{5}\tau^3 f(0),\qquad f(T+1)=i\gamma_{0}\gamma_{5}\tau^3 f(T),
\label{eq:bound_f}
\end{equation}
and $f(t) =0$ for $t<-1$ and $t>T+1$. In the notation of \cite{Sint:1995ch}, the function $h(t)$ is given by
\begin{equation}
\begin{split}
 h(t)=  1 & + m_{0} + i\tilde{q}_{k}(t)\gamma_{k} + \frac12 \sum_{k=1}^{3}\hat{q}_{k}(t)^{2}
     -(1-\delta_{t,0} -\delta_{t,T})\frac12 \csw\gamma_{0}\gamma_{k}p_{0k}\\
   & +\left(\delta_{t,0} + \delta_{t,T}\right)\Bigl\{
       (z_{f}-1)+(d_{s}-1)\bigl( i\tilde{q}_{k}(t)\gamma_{k} + \frac{1}{2}\sum_{k=1}^{3}\hat{q}_{k}(t)^{2}  \bigr)\Bigr\}\,,
\end{split}
\end{equation}
where summation over repeated spatial indices is assumed.
Note that, at the boundaries $t=0$ and $t=T$, the function $h(t)$ contains the contribution
coming from the boundary counterterms and the term proportional to $\csw$ is absent (cf.~Section~3).

To obtain a first order recursion we now reformulate~\cite{Sint:1995ch},
\begin{equation}
 F(t)=P_{-}f(t)+P_{+}f(t-1),\qquad 0\leq t\leq T+1\,,
\end{equation}
and, as a consequence of Eq.~(\ref{eq:bound_f}), $F(t)$ satisfies the boundary conditions
\begin{equation}
 \tilde{Q}_{+}F(0)=0, \qquad
 \tilde{Q}_{-}F(T+1)=0.
 \label{eq:bound_F}
\end{equation}
The eigenvalue equation (\ref{eq:EVeq}) now takes the form of a first order recursion relation,
\begin{equation}
  F(t+1) = A(t) F(t),
\end{equation}
with 
\begin{equation}
\begin{split}
 A(t) =-a(t)^{-1}&\left\{  P_{-}\left[\mu^{2}-a(t)^{2}+\mu\gamma_{5}\tau^{1}
   \left(c_{k}(t)\gamma_{k}-b_{k}(t)\gamma_{k}+1\right)\right.\right.\\
  & \quad\quad \left.+ c_{k}(t)\gamma_{k}\left(b_{j}(t)\gamma_{j}-1\right)\right]\\
  & \quad +P_{+} \left.\left[b_{k}(t)\gamma_{k}-\mu\gamma_{5}\tau^1-1\right]\right\}.
\end{split}
\end{equation}
The coefficients $a(t)$, $b_{k}(t)$ and $c_{k}(t)$ are scalar functions of $t$ given by
 \begin{align}
  a(t) & =1+m_{0}+\frac{1}{2}\sum_{k=1}^{3}\hat{q}_{k}(t)^{2}\nonumber\\
  &\qquad+\left(\delta_{t,0}+\delta_{t,T}\right)
     \Bigl\{(z_{f}-1)+(d_{s}-1)\frac{1}{2}\sum_{k=1}^{3}\hat{q}_{k}(t)^{2}\Bigr\}\,,\\
  b_{k}(t) & =i\left[1+\left(\delta_{t,0}+\delta_{t,T}\right)
    (d_{s}-1)\right]\tilde{q}_{k}(t) - \frac{1}{2}\left(1-\delta_{t,0}-\delta_{t,T}\right)\csw p_{0k}\,,\\
  c_{k}(t) & =i\left[1+\left(\delta_{t,0}+\delta_{t,T}\right)
    (d_{s}-1)\right]\tilde{q}_{k}(t) + \frac{1}{2}\left(1-\delta_{t,0}-\delta_{t,T}\right)\csw p_{0k}\,.
 \end{align}
After $T+1$ steps one arrives at $F(T+1)$ which depends linearly on $F(0)$,
through
\begin{equation}
 F(T+1)=M^{\tilde{Q}}(\mu)F(0)\,, \qquad   M^{\tilde{Q}}(\mu) = A(T)A(T-1)...A(0)\,.
\label{eq:recursion}
\end{equation}
The boundary conditions (\ref{eq:bound_F}) then imply
\begin{equation}
   \det\left(M^{\tilde{Q}}_{--}(\mu)\right) = 0\,,
\end{equation}
where $M^{\tilde{Q}}_{--}(\mu)$ is the matrix $\tilde{Q}_-M^{\tilde{Q}}(\mu)\tilde{Q}_-$ reduced to the subspace (of dimension $2\times 2$)
defined by the projectors $\tilde{Q}_-$.
Taking into account the dimensionality of the matrices and following the reasoning of ref.~\cite{Luscher:1992an}, the
characteristic polynomial of $\gamma_5\tau^1\mathcal{D}$ is given by
\begin{equation}
 \det\left(\gamma_5\tau^1 \mathcal{D}-\mu\right)=\det\left( M^{\tilde{Q}}_{--}(\mu) \prod_{t=0}^{t=T} a(t)\right).
 \label{eq:charpoly}
 \end{equation}
In practice it is slightly inconvenient to choose a representation of the $\gamma$-algebra
where $\tilde{Q}_\pm$ are diagonal. Using diagonal $\gamma_0$ instead, one may perform a unitary rotation,
\begin{equation}
 U^{\dag}\tilde{Q}_\pm U=P_\pm,\quad \textrm{  where  } \quad U=(1-i\tau^{3}\gamma_{5})/\sqrt{2},
\end{equation}
and use the $P_\pm$ projectors. More precisely, Eq.~(\ref{eq:recursion}) reads,
\begin{equation}
   U^{\dag} F(T+1) =  U^{\dag} M^{\tilde{Q}}(\mu) U U^{\dag} F(0)\,,
\end{equation}
and the boundary conditions for $U^{\dag} F$ are now given in terms of the $P_\pm$ projectors,
\begin{equation}
  P_+ U^\dag F(0) = U^\dag \tilde{Q}_+F(0) =0\,,\qquad P_- U^\dag F(T+1) = U^\dag \tilde{Q}_-F(T+1) = 0\,.
\end{equation}
Hence, if we define
\begin{equation}
   M^P(\mu) = U^{\dag} M^{\tilde{Q}}(\mu) U\,,
\end{equation}
we conclude
\begin{equation}
   \det\bigl( M^{\tilde{Q}}_{--}(\mu)\bigr) =  \det\bigl( M^P_{--}(\mu)\bigr)\,,
\end{equation}
where on the RHS the restriction is now to the subspace defined by the $P_-$ projector.
At this point one may set $\mu=0$ and $M^P_{--}(0)$ becomes flavour diagonal.
The final result may be written in the form
\begin{equation}
\frac{\partial}{\partial\eta} \ln \det{\mathcal{D}^{(u)}}
= \Tr\left\{\left(M^{(u)}\right)^{-1}\frac{\partial}{\partial\eta} M^{(u)}\right\}\,,
\end{equation}
where the matrix $M^{(u)}$ is given by
\begin{equation}
   M^{(u)} = \frac12 \Bigl([1+i\gamma_5]B(T)B(T-1)\cdots B(0)[1-i\gamma_5]\Bigr)_{--}\,,
\end{equation}
with
\begin{equation}
 B(t) = P_{-} a(t)^{2} + P_+ \left\{1- b_k(t)\gamma_k\right\} + c_k(t)\gamma_k \left\{1- b_j(t)\gamma_j\right\}\,.
\end{equation}
Note that the matrix $M^{(u)}$ and its $\eta$-derivative (or any other derivative, here generically denoted by ``prime")
can be generated by the coupled recursion,
\begin{equation}
   G(t+1) = B(t) G(t), \qquad G'(t+1) = B'(t)G(t) + B(t)G'(t)\,,
\end{equation}
starting with $G(0)=(1-i\gamma_5)P_-$ and $G'(0)=0$.

\section{Perturbation theory versus Monte Carlo data at large $\beta$}
\label{app:MC}

In order to further corroborate the perturbative results we obtained for the finite
renormalization constants $Z_{\rm V}$ and $Z_{\rm A}$ involving the point-split current,
we decided to compare the determinations with results from Monte Carlo simulations
at small values of the bare coupling $g_0$. To this end, we performed simulations at
a fixed lattice size $L/a=8$, and for 25 different values of $\beta=6/g_0^2$, in the
range $\beta\in[50:1200]$. We note that at O($g_0^2$) only gluonic loops appear in 
the perturbative expansion of the fermionic correlation functions entering the 
definition of $Z_{\rm A,V}$ (cf. Figure \ref{fig:diag_gX_lY} and \ref{fig:diag_gVt_lVt}). 
This allowed us to simply generate pure SU(3) gauge-field configurations on which we
measured the relevant fermionic correlators. 

For the comparison to be meaningful the Monte Carlo determinations need to mimic exactly
the perturbative computations. This means that the lattice set-up, as well as the values
for the bare parameters and improvement coefficients need to be the same in the two 
computations. We therefore set $\rho=T/L=1$ and $\theta=0$. For the bare parameters 
we took: $m_0(g_0)=m^{(1)}_{\rm cr}(a/L)g_0^2$ and $z_f(g_0)=1+z_f^{(1)}(a/L)g^2_0$,
where $m^{(1)}_{\rm cr}(a/L)$ and $z_f^{(1)}(a/L)$ were given by Eqs. 
(\ref{eq:mc1_L}),(\ref{eq:zf1_L}) for $L/a=8$. Finally, for the improvement 
coefficients we considered their asymptotic values up to the relevant order
of perturbation theory. Specifically, we set the boundary improvement coefficients
$d_s(g_0) = \frac{1}{2} + d_s^{(1)}g_0^2$, $c_t=1$, and $\bar{d}_s=0$
(cf. Section \ref{sec:PTsetup}), while for the bulk improvement coefficients we 
chose $c_{\rm sw}=1$, and $c_{\rm A}=c_{\rm V}=c_{\rm \widetilde{V}}=0$.

\begin{figure}[!hpbt]
\centering
\includegraphics[clip=true,scale=0.54]{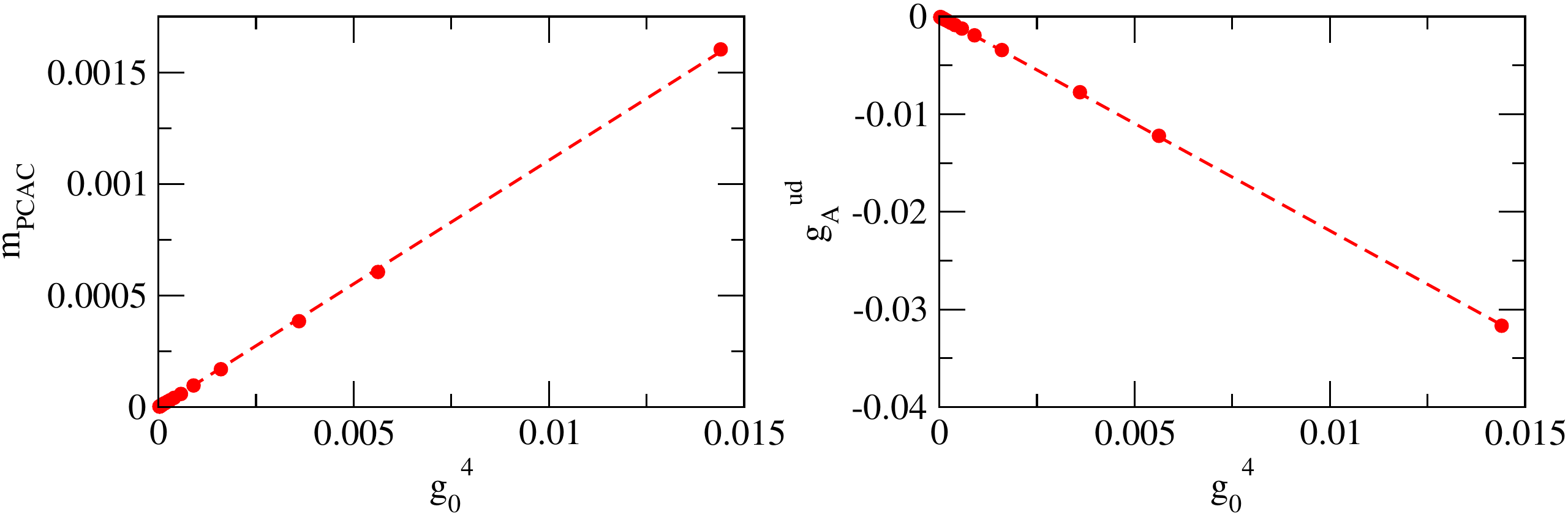}
\caption{Values for the PCAC quark-mass $m_{\rm PCAC}$ (\ref{eq:mPCAC}) and 
for $g_{\rm A}^{ud}$, as a function of $g^4_0$, obtained from Monte Carlo 
simulations at $L/a=8$. Both quantities have been measured in the middle of 
the lattice i.e. for $x_0=T/2$.}
\label{fig:MCrencond}
\end{figure}

In order to confirm that the bare parameters $m_0(g_0)$ and $z_f(g_0)$ were chosen properly, 
we checked whether the conditions (\ref{eq:cond_mc}) and (\ref{eq:cond_zf_gA}) were realized
up to O($g_0^4$) corrections. To this end, we looked at the quantities,
\begin{equation}
 \label{eq:mPCAC}
 m_{\rm PCAC}(x_0) = {\tilde{\partial}_0 g_{\rm A}^{ud}(x_0)\over 2g_{\rm P}^{ud}(x_0)}
 \quad\text{and}\quad g_{\rm A}^{ud}(x_0),
\end{equation}
which are show in Figure \ref{fig:MCrencond}. As we can see, the data are very well described by a
pure O($g_0^4$) effect over the whole range of $g_0$ we investigated. The renormalization
conditions are then satisfied up to 1-loop order in perturbation theory.

\begin{table}[htb]
\centering
\caption{Comparison between the 1-loop coefficients $Z_{\rm X}^{(1)}(L/a)$ of $Z_{\rm X}$, 
${\rm X=V,A}$, for both the $l$ and $g$ definitions, as obtained from perturbation theory 
(PT) and Monte Carlo simulations (MC) at $L/a=8$.}
\begin{tabular}{ccc}
\toprule
$Z_{\rm X}^{(1)}(L/a)$  & PT & MC \\
\midrule
$Z_{\rm V}^{g(1)}$ & $-0.122586$ & $-0.122596(19)$ \\
$Z_{\rm V}^{l(1)}$ & $-0.129838$ & $-0.129822(12)$ \\
$Z_{\rm A}^{g(1)}$ & $-0.109076$ & $-0.109074(22)$ \\
$Z_{\rm A}^{l(1)}$ & $-0.116640$ & $-0.116645(10)$ \\
\bottomrule
\end{tabular}
\label{tab:PTvsMC}
\end{table}

\begin{figure}[!hptb]
\centering
\includegraphics[clip=true,scale=0.54]{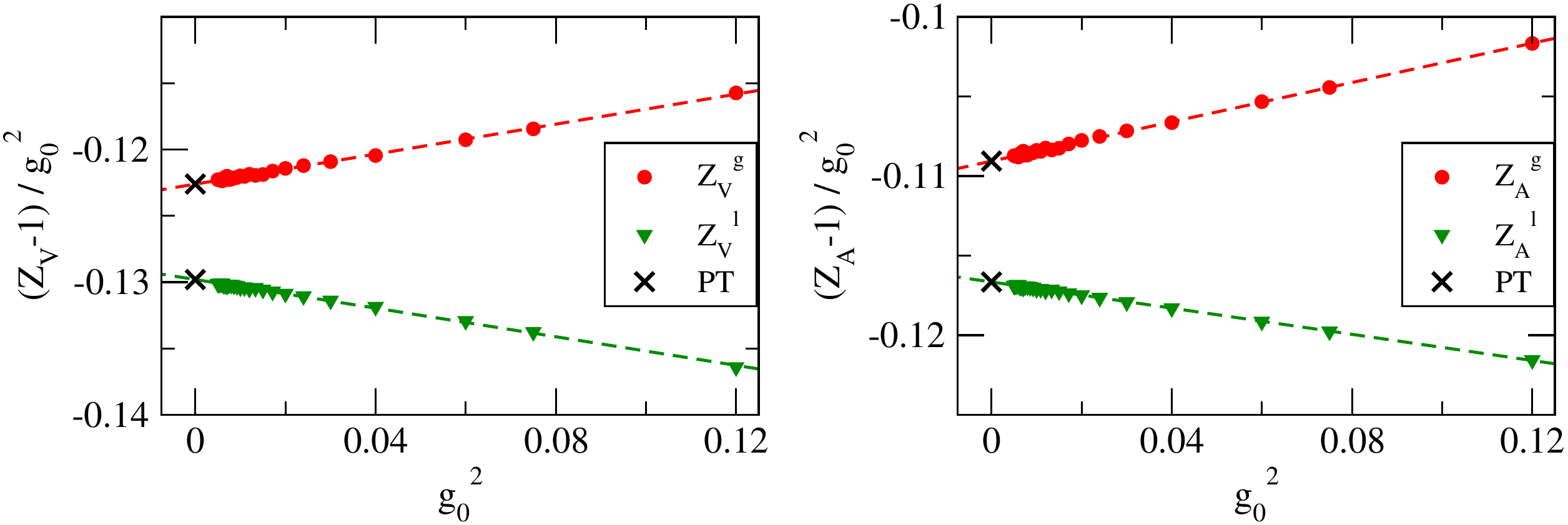}
\caption{Results for the two definitions of $Z_{\rm V}$, $Z^{g}_{\rm V}$ and $Z^{l}_{\rm V}$, 
and of $Z_{\rm A}$, $Z^{g}_{\rm A}$ and $Z^{l}_{\rm A}$, as a function of $g^2_0$, 
obtained from Monte Carlo simulations at $L/a=8$. The perturbative results, PT, for 
$Z_{\rm X}^{(1)}(L/a)$, ${\rm X=V,A}$, are also shown.}
\label{fig:MCrenfact}
\end{figure}

In Figure \ref{fig:MCrenfact}, instead, we present the results for the two definitions of 
$Z_{\rm V}$ (left panel), and $Z_{\rm A}$ (right panel). Specifically, after verifying that
$Z_{\rm V}$ and $Z_{\rm A}$ extrapolated correctly to 1 for $g_0\to0$, we looked at 
$(Z_{\rm X}-1)/g_0^2$, ${\rm X=V,A}$, in order to extract the 1-loop coefficients 
$Z_{\rm X}^{(1)}(L/a)$ to be compared with perturbation theory. 
As we can see from the figure, there is nice agreement between the perturbative and Monte 
Carlo determinations. We note that $Z_{\rm X}^{(1)}(L/a)$ was obtained from the Monte Carlo
data by considering a linear fit of $(Z_{\rm X}-1)/g_0^2$ with respect to $g_0^2$, including
all but the largest value of $g_0$ we simulated. For completeness, we collected in Table 
\ref{tab:PTvsMC} the results from 1-loop perturbation theory and the results of the 
extrapolations of the Monte Carlo data.

\bibliographystyle{JHEP}
\bibliography{references}

\end{document}